\pdfoutput=1

\documentclass[british]{article}

\usepackage{lmodern}

\usepackage[T1]{fontenc}
\usepackage[utf8]{inputenc}
\usepackage{booktabs}
\usepackage{url}
\usepackage{amsmath}
\usepackage{amssymb}
\usepackage{graphicx}
\usepackage{esint}
\usepackage{tcolorbox}
\usepackage[normalem]{ulem}

\usepackage{tikz} 

\makeatletter

\usepackage{jcappub}
\numberwithin{equation}{section}

\usepackage{fullpage}
\usepackage{array}
\allowdisplaybreaks

\@ifundefined{showcaptionsetup}{}{%
 \PassOptionsToPackage{caption=false}{subfig}}
\usepackage{subfig}
\makeatother

\usepackage[UKenglish]{babel}
\usepackage{listings}
\usepackage{enumerate}

\newcommand{\hiclass}{{\tt hi\_class} }
\newcommand{\hiclassx}{{\tt hi\_class}}
\newcommand{\aK}{\alpha_\text{K}}
\newcommand{\aB}{\alpha_\text{B}}
\newcommand{\aT}{\alpha_\text{T}}
\newcommand{\aM}{\alpha_\text{M}}
\newcommand{\OmDE}{\Omega_\text{DE}}
\newcommand{\Mpl}{M_\text{Pl}}

\begin{document}

\title{\hiclassx: Background Evolution, Initial Conditions and Approximation Schemes}

\subheader{\includegraphics[width=0.3\textwidth]{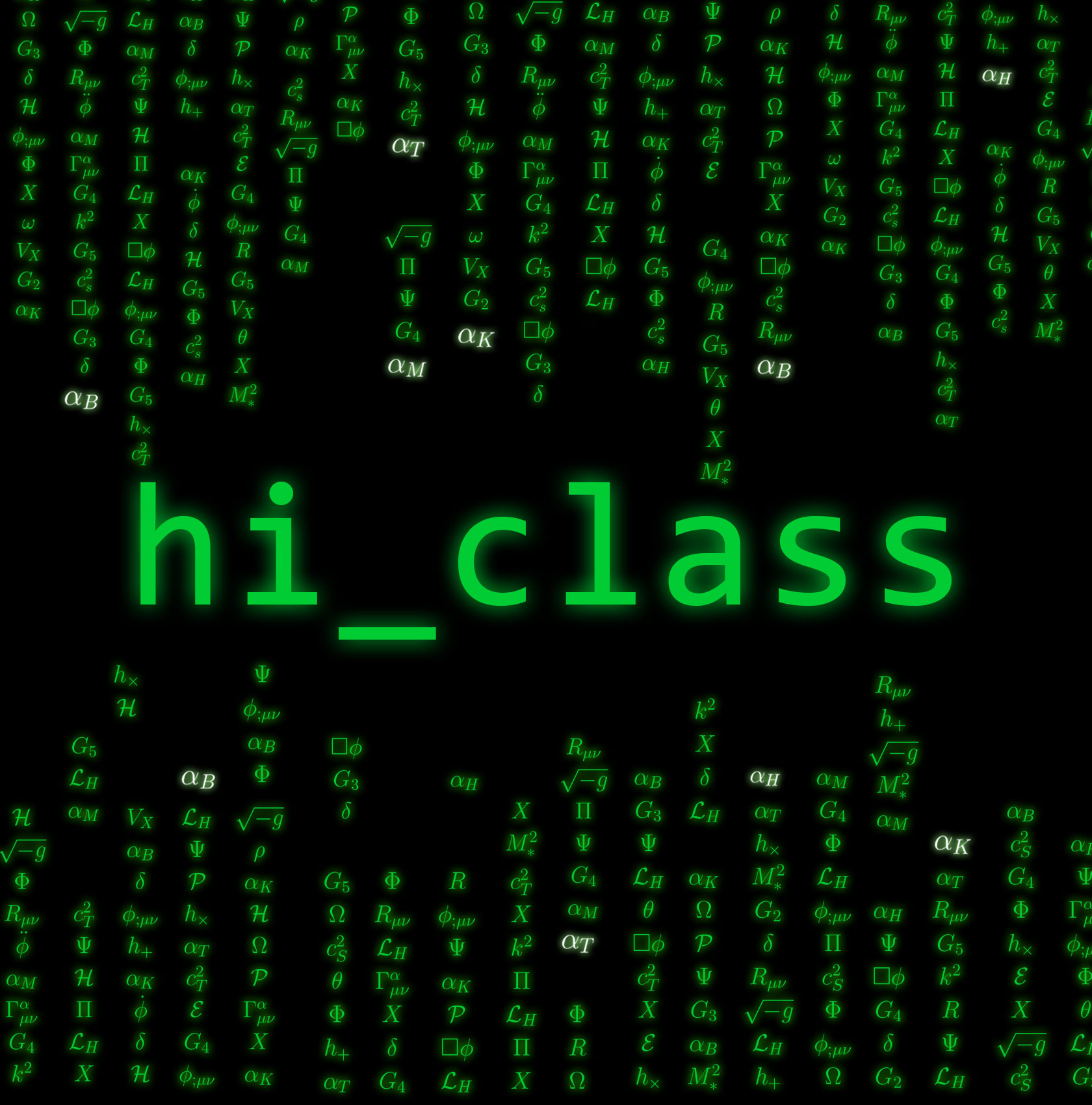} }

\author[a]{Emilio Bellini,}
\author[b]{Ignacy Sawicki}
\author{and}
\author[c,d,e]{Miguel Zumalacárregui}

\affiliation[a]{Astrophysics, University of Oxford, \\ Denys Wilkinson Building, Keble Road, Oxford OX1 3RH, United Kingdom}

\affiliation[b]{CEICO, Institute of Physics of the Czech Academy of Sciences,\\Na Slovance 2, 182 21 Praha 8, Czech Republic}

\affiliation[c]{Berkeley Center for Cosmological Physics, \\ LBNL and University of California at Berkeley, \\
Berkeley, California 94720, USA}
\affiliation[d]{Institut de Physique Th\' eorique, Universit\'e  Paris Saclay 
CEA, CNRS, \\
91191 Gif-sur-Yvette, France}
\affiliation[e]{Max Planck Institute for Gravitational Physics (Albert Einstein Institute), \\
Am Mühlenberg 1, D-14476 Potsdam-Golm, Germany}

\emailAdd{emilio.bellini@physics.ox.ac.uk}
\emailAdd{ignacy.sawicki@fzu.cz}
\emailAdd{miguelzuma@berkeley.edu}

\abstract{
Cosmological datasets have great potential to elucidate the nature of dark energy and test gravity on the largest scales available to observation.
Theoretical predictions can be computed with \hiclass  (\url{www.hiclass-code.net}), an accurate, fast and flexible code for linear cosmology,
incorporating a wide range of dark energy theories and modifications to general relativity.
We introduce three new functionalities into \hiclassx:
(1) Support for models based on covariant Lagrangians, including a constraint-preserving integration scheme for the background evolution 
and a series of worked-out examples: Galileon, nKGB, quintessence (monomial, tracker) and Brans-Dicke.
(2) Consistent initial conditions for the scalar-field perturbations in the deep radiation era,
identifying the conditions under which modified-gravity isocurvature perturbations may grow faster than adiabatic modes leading to a loss of predictivity.
(3) An automated quasi-static approximation scheme allowing order-of-magnitude improvement in computing performance without sacrificing accuracy for wide classes of models.
These enhancements bring the treatment of dark energy and modified gravity models to the level of detail comparable to software tools restricted to standard $\Lambda$CDM cosmologies.
The \hiclass code is publicly available (\url{https://github.com/miguelzuma/hi_class_public}), ready to explore current data and prepare for next-generation experiments. 
}

\maketitle

\section{Introduction}

Universal and attractive, gravity is the main force shaping our cosmos.
Despite the successful conceptual description of the universe's evolution and a wide range of phenomena on terrestrial and astrophysical scales, Einstein's General Relativity (GR) is tightly linked to a wide variety of open questions in fundamental physics. 
A better understanding of gravity may shed light on problems such as the initial conditions for the universe, the smallness of the cosmological constant, the nature of space-time singularities and the unification of gravitational and quantum-mechanical phenomena.
Many among the unsolved problems in physics are connected, directly or indirectly, with the nature of gravity.%
\footnote{\url{https://en.wikipedia.org/wiki/List_of_unsolved_problems_in_physics}}

The nature of Dark Energy (DE) and Dark Matter (DM) stand out among these open questions. On one hand, there exists overwhelming observational support for their existence, making up 69 and 26\% of the Universe's energy density today \cite{Aghanim:2018eyx}. 
On the other hand, all the evidence collected so far relies on their gravitational effects, making the inferred abundance and properties contingent on our understanding of gravity.
And while Einstein's theory has been exquisitely validated in terrestrial, solar and astrophysical systems \cite{Will:2014kxa}, our only evidence for DM and DE stems from observations at galactic and cosmological scales on which most tests of GR rely on assumptions about these dark components. Whether or not new gravitational physics are necessary to explain cosmological data, testing GR is necessary to 
transform observations of the universe into advances in fundamental physics.

The minimal working model of the universe, known as $\Lambda$CDM, assumes the validity of GR on all scales. In it, late-time cosmic acceleration is provided by a cosmological constant ($\Lambda$) and cold dark matter (CDM) drives structure formation.
This remarkably simple model can be tested by a variety of methods, including observations of the large-scale structure (LSS) of the universe, the cosmic microwave background (CMB) and transient phenomena, including type Ia supernovae (SNe) and gravitational waves (GWs) (cf.~Refs.~\cite{Weinberg:2012es,Ade:2015rim,Alam:2016hwk,Scolnic:2017caz,Abbott:2018xao}). 
While most datasets are compatible, tensions have started to emerge when independent observables are interpreted in the simple $\Lambda$CDM model. 
These discrepancies typically appear between the expected properties of the late universe based on observations that probe the early universe (e.g.~CMB) and direct probes of this late-universe.
These tensions include the cosmological expansion rate today \cite{Riess:2019cxk,Freedman:2019jwv,Wong:2019kwg}, the amplitude of cosmological perturbations (e.g.\ through weak lensing \cite{Hildebrandt:2016iqg,Joudaki:2016kym,Troxel:2017xyo,Hikage:2018qbn,Joudaki:2019pmv} or cluster counts \cite{Ilic:2019pwq}) and problems on small scales \cite{Bullock:2017xww}.
The persistence of these discrepancies has motivated the exploration of possible solutions based on extensions of the standard model involving new physics beyond $\Lambda$, CDM or GR (see Refs.~\cite{Bernal:2016gxb,Renk:2017rzu,Belgacem:2017cqo,Poulin:2018zxs,Poulin:2018cxd,Vattis:2019efj,Raveri:2019mxg,Agrawal:2019lmo,Lin:2019qug,Pan:2019hac,DiValentino:2019ffd,Kreisch:2019yzn,Smith:2019ihp,Knox:2019rjx} for some examples).

The ambitious program to test GR and alternatives to $\Lambda$CDM is progressing steadily since the discovery of the accelerated expansion of the universe \cite{Perlmutter:1998np,Riess:1998cb}.
In the near future an array of observational campaigns will improve vastly on the quantity and quality of data and provide new opportunities to test $\Lambda$CDM and its foundations.
Surveys such as the extended Baryon Oscillation Spectroscopic Survey \cite{Dawson:2015wdb}, the Dark Energy Spectroscipic Instrument \cite{Levi:2019ggs}, the EUCLID Satellite \cite{Laureijs:2011mu} and the Large Synoptic Survey Telescope \cite{Abell:2009aa} will map the universe's expansion history and matter distribution with unprecedented accuracy using complementary techniques. 
These upcoming experiments will provide new opportunities to test fundamental physics, likely producing a detection of neutrino mass \cite{Font-Ribera:2013rwa} and a $10-100\times$ precision improvement in testing deviations from GR \cite{Alonso:2016suf}.
In addition, GW observations from upcoming LIGO Virgo and Kagra observing runs \cite{Aasi:2013wya}, the Laser Interferometer Space Antenna \cite{Baker:2019nia} and Next-generation Earth-based Observatories \cite{GW3G} will provide novel ways to test gravity and DE \cite{Gair:2012nm,Ezquiaga:2018btd,Belgacem:2019pkk,Sathyaprakash:2019yqt}.

Optimal exploitation of current and upcoming data requires accurate predictions across a wide range of scenarios.
Addressing the vast landscape of alternatives to GR and DE models (see Refs.~\cite{Clifton:2011jh,Koyama:2015vza,Ishak:2018his,Heisenberg:2018vsk,Ezquiaga:2018btd} for recent reviews) requires new conceptual and computational frameworks flexible enough to encompass specific instantiations.
A unifying paradigm was realized by the re-discovery of Horndeski's theory \cite{Horndeski,Deffayet:2011gz,Kobayashi:2011nu}, a very general action for scalar-tensor gravity that encompasses a large fraction of theories of gravity, including many proposed models for DE and inflation.
In parallel, the exploration of model independent descriptions of gravity led to the formulation of the effective (field) theory of DE (EFT-DE) \cite{Gubitosi:2012hu,Bloomfield:2012ff} (see \cite{Gleyzes:2014rba,Frusciante:2019xia} for reviews), which can be matched to concrete models or used as an model-independent parameterization of deviations from GR in cosmology.
In order to test these scenarios against data, their predictions need to be computed efficiently, fast enough to sample the parameter space of each model.

The theoretical advances in DE and tests of GR have been translated into flexible tools to compute cosmological predictions. 
The \hiclass code \cite{Zumalacarregui:2016pph} provides a publicly available tool to study the phenomenology of Horndeski theories in the linear regime of cosmological perturbations.
This paper introduces new features available in the second version of the code.

\begin{tcolorbox}
\textbf{Code usage:}
The use of the public version of \hiclass is free to the scientific
community but conditional on the inclusion of references to at least
this article, the first \hiclass paper \cite{Zumalacarregui:2016pph} and the original CLASS paper \cite{Blas:2011rf}.
\end{tcolorbox}

\section{Horndeski's Theory and the \hiclass code}

The purpose of the \hiclass code is to provide
a fast, flexible and accurate tool to compute predictions and enable tests of general gravity and dark energy theories, with the level of detail and control available to standard cosmology.
\hiclass is based on the Cosmic Linear Anisotropy Solving System (CLASS), a modern and flexible Einstein-Boltzmann solver for linear cosmological perturbations \cite{Lesgourgues:2011re,Blas:2011rf}.
Most CLASS features are available in \hiclassx, including neutrino and DM properties \cite{Lesgourgues:2011rh}, general primordial power spectra, including isocurvature modes and the computation of galaxy number counts \cite{DiDio:2013bqa}. Features not available in the current \hiclass version are non-flat universes \cite{Lesgourgues:2013bra}, Newtonian gauge and non-linear calculations consistent with deviations from GR.
\hiclass can be interfaced with any code that supports CLASS, including MontePython \cite{Audren:2012wb,Brinckmann:2018cvx}, CosmoSIS \cite{Zuntz:2014csq}, FalconIC \cite{Valkenburg:2015dsa}, the Core Cosmology Library \cite{Chisari:2018vrw} and Cobaya\footnote{\url{https://cobaya.readthedocs.io/en/latest/}}.

\hiclass is easy to use and to modify and is publicly available to the scientific community.%
\footnote{The code lives on \url{https://github.com/miguelzuma/hi_class_public}, resources are available on \url{www.hiclass-code.net}.}
The structure of the code and the modifications with respect to the CLASS code are described in detail in the first \hiclass paper \cite{Zumalacarregui:2016pph}.
\hiclass can be used to explore different aspects of gravitational theories and their cosmological implications \cite{Bellini:2015oua,Renk:2016olm,Lorenz:2017iez,Burrage:2019afs,Melville:2019wyy}, test models against current data \cite{Bellini:2015xja,Renk:2017rzu,Garcia-Garcia:2018hlc,SpurioMancini:2019rxy} or in preparation for future experiments \cite{Alonso:2016suf,Mancini:2018qtb,Garcia-Garcia:2019ees}.%
\footnote{For a complete and updated list of projects using \hiclass see \url{www.hiclass-code.net}.}
\hiclass has been cross-validated against other codes both implementing a general EFTDE framework (e.g.~EFTCAMB \cite{Hu:2013twa}, COOP \cite{Huang:2015srv}) and ones dealing with an approximation to it or specific models of gravity  (e.g.\ DASh \cite{Kaplinghat:2002mh}, GalCAMB \cite{Barreira:2014jha}, EoS\_class \cite{Pace:2019uow} or EFCLASS \cite{Arjona:2019rfn}) reaching the level of accuracy needed by next-generation cosmological experiments \cite{Bellini:2017avd}. 
Currently the publicly available version of the \hiclass code incorporates Horndeski's scalar-tensor theory, a very general overarching framework that incorporates a large class of DE and beyond GR theories.

Horndeski's theory is the most general theory of gravity featuring a metric and a scalar field in four space-time dimensions subject to the requirements of Lorentz invariance, locality and being described by second order equations of motion~\cite{Horndeski,Deffayet:2011gz,Kobayashi:2011nu}. 
The theory is given by the following action
\begin{equation}
S[g_{\mu\nu},\phi]=\int\mathrm{d}^{4}x\,\sqrt{-g}\left[\sum_{i=2}^{5}\frac{1}{8\pi G_{\text{N}}}{\cal L}_{i}[g_{\mu\nu},\phi]\,+\mathcal{L}_{\text{m}}[g_{\mu\nu},\psi_{M}]\right]\,,\label{eq:action}
\end{equation}
where the four Lagrangians $\mathcal{L}_{i}$ 
\begin{eqnarray}
{\cal L}_{2} & = & G_{2}(\phi,\,X)\,,\label{eq:L2}\\
{\cal L}_{3} & = & -G_{3}(\phi,\,X)\Box\phi\,,\label{eq:L3}\\
{\cal L}_{4} & = & G_{4}(\phi,\,X)R+G_{4X}(\phi,\,X)\left[\left(\Box\phi\right)^{2}-\phi_{;\mu\nu}\phi^{;\mu\nu}\right]\,,\label{eq:L4}\\
{\cal L}_{5} & = & G_{5}(\phi,\,X)G_{\mu\nu}\phi^{;\mu\nu}-\frac{1}{6}G_{5X}(\phi,\,X)\left[\left(\Box\phi\right)^{3}+2{\phi_{;\mu}}^{\nu}{\phi_{;\nu}}^{\alpha}{\phi_{;\alpha}}^{\mu}-3\phi_{;\mu\nu}\phi^{;\mu\nu}\Box\phi\right]\,,\label{eq:L5}
\end{eqnarray}
encode the dynamics of the Jordan-frame metric $g_{\mu\nu}$ and the scalar field $\phi$.
The theory is fully specified by four arbitrary \textit{Horndeski functions} $G_{i}(\phi,X)$ of the scalar field value $\phi$ and its canonical kinetic term, $2X\equiv-\partial_{\mu}\phi\partial^{\mu}\phi$;
we use subscripts $\phi,X$ to denote partial derivatives, e.g.~$G_{iX}=\frac{\partial G_{i}}{\partial X}$.%
\footnote{Note that viable scalar-tensor theories beyond Horndeski are known \cite{Zumalacarregui:2013pma,Gleyzes:2014dya,Langlois:2015cwa,BenAchour:2016fzp}, a class of which has already been implemented in a private branch of \hiclass \cite{Traykova:2019oyx}.}
Cosmological solutions can be obtained from this covariant action by assuming homogeneity and isotropy. 

The dynamics of background cosmology in any Horndeski model is then fully characterized by the (time-dependent) equation of state of the dark energy $w(\tau)$ and the energy density fraction today  $\Omega_\text{DE,0}\equiv\mathcal{E}(\tau_0)/H_0^2$ of the Horndeski scalar (or equivalently the dark energy density $\mathcal{E}(\tau))$). At the level of linear perturbations, we write the perturbed equations in terms of the scalar field velocity potential 

\begin{equation}\label{eq:VXdef}
V_X\equiv a\delta\phi/\phi^\prime\,.
\end{equation}
Then four additional time-dependent \textit{$\alpha$-functions} are sufficient necessary to fully describe the dynamics of linear perturbations. The basis of these functions is chosen to separate physical effects to the largest extent possible. They are defined in terms of the Lagrangians in appendix~\ref{sec:appendix_equations_background}:
\begin{itemize}
 \item \textit{Kineticity $\alpha_K(\tau)$} modulates the kinetic term for scalar field perturbations, i.e.\ how hard/easy it is to excite the scalar field. Larger values of $\aK$ lead to a smaller sound speed for the scalar and therefore the field responds to external sources only on smaller scales. It receives contributions from all the four $G_i$ functions.
 \item \textit{Braiding $\alpha_B(\tau)$} measures the kinetic mixing between perturbations of the metric and the scalar field, i.e.\ what combination of scalar field excitations and the gravitational potentials is a propagating degree of freedom. In the presence of braiding the effective Newton's constant for matter is modified. It receives contributions from $G_{3,4,5}$.
 \item \textit{Cosmological strength of gravity $M_*^2(\tau)$} determines the ratio of the space-time curvatures per unit mass between the cosmological background and the solar system. It receives contributions from $G_{4,5}$.
 Large-scale structure is sensitive only to its derivative, the \textit{Planck-mass run rate} $\alpha_M\equiv \frac{d\log(M_*^2)}{d\log(a)}$. The value of this quantity today and Solar-System measurements of the strength of gravity are not directly connected absent a particular model as a result of screening mechanisms in these theories and other small-scale physics.

\item \textit{Tensor Speed Excess $\alpha_T(\tau)$} captures the difference between GW speed and the speed of light $c_g^2 = c^2(1+\alpha_T)$. It receives contributions from $G_{4,5}$. The observation of electromagnetic and gravitational radiation from a neutron star merger \cite{Monitor:2017mdv} severely constraints $\alpha_T$ today \cite{Lombriser:2015sxa,Bettoni:2016mij,Ezquiaga:2017ekz,Baker:2017hug,Creminelli:2017sry,Sakstein:2017xjx,Kase:2018aps}.%
 \footnote{Note that GW constraints from ground-based detectors involve GW frequencies which are slightly beyond the limit of validity of Horndeski as a low energy classical description 
 (assuming a characteristic energy scale $\sim H_0$ consistent with DE physics) \cite{deRham:2018red}. A $c_g$ measurement at lower GW frequencies can be achieved with space detectors using galactic binaries \cite{Bettoni:2016mij} or cosmological standard sirens \cite{Belgacem:2019pkk}.}
 
\end{itemize}
Note that at least one of $\aM$ or $\aT$ must be non-zero in order to produce gravitational slip from perfectly-fluid matter sources. On the other hand, this causes a modification of the propagation of gravitational waves and the implied constraints from GW observations \cite{Saltas:2014dha}. For further details on the $\alpha$-functions and their cosmological effects see \cite{Bellini:2014fua,Zumalacarregui:2016pph}. Note that as a result of the definition for $V_X$ (\ref{eq:VXdef}), this efficient description does not allow one to follow an evolution through $\phi'=0$, e.g.\ a background scalar field oscillating around some value.

The relationship between $w$, the $\alpha$-functions and the Lagrangian is sufficiently non-trivial (see appendix~\ref{sec:appendix_equations_background}) that it leads to the expectation that any expansion history and any choice of time-dependence for the $\alpha$-functions can be achieved given sufficient tuning of the $G_i$ and the initial conditions for the background scalar field. This sort of structure with five unknown arbitrary functions of time defining the linear structure formation equations was also derived using the approach of effective theory for dark energy \cite{Gubitosi:2012hu,Gleyzes:2013ooa}, where one tries to derive the most general theory compatible with the symmetries of the cosmological background and the requirement that there is just one degree of freedom in addition to GR.

\begin{figure}
\center
\includegraphics{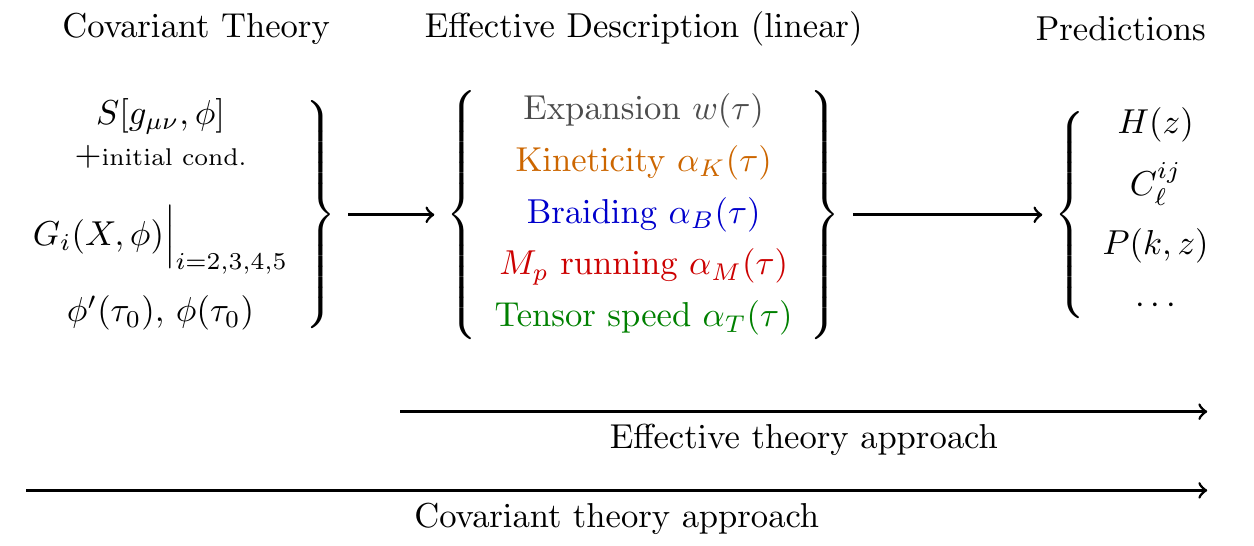}
\caption{Covariant action, effective-theory description and observables. 
In the covariant theory approach, a model is specified through the Horndeski functions $G_i(X,\phi)$ ($i=2,3,4,5$) in the action \ref{eq:action}, the free parameters they may contain, and the initial conditions of the scalar field $\phi(\tau_0),\phi^\prime(\tau_0)$. This choice fully fixes the expansion history (e.g.\ the equation of state $w$) and the alpha-functions that completely characterize the evolution of (linear) cosmological perturbations, and provide a set of cosmological observables.
In the effective theory approach one parametrizes deviations from GR, irrespective of any underlying covariant action.
\label{fig:test_approaches}}
\end{figure}

There are thus two ways in which \hiclass can be used to test gravity:
\begin{enumerate}
 \item The \textit{effective-theory approach} consists of providing a parametrization of the $\alpha$-functions and the equation of state, independent of any underlying theory, with the aim of testing for particular physical effects.
The implementation of the effective-theory approach was described in detail in the first \hiclass paper \cite{Zumalacarregui:2016pph}, and was, up to now, the only way of using the public version of the code.
 
  \item The \textit{covariant-theory approach} involves starting with a particular model defined in terms of the Horndeski functions $G_i$ (\ref{eq:L2}-\ref{eq:L5}). Upon specifying the initial conditions for the background field and the additional parameters of the theory in the $G_i$ functions, \hiclass automatically computes the cosmological background and the $\alpha$-functions are fully determined. We are introducing this method to the public version of \hiclass with this release.
\end{enumerate}
The two methods, summarized in Fig.~\ref{fig:test_approaches}, are complementary.%
\footnote{A third alternative is to study phenomenological modifications of GR by introducing \textit{effective gravitational constants}. This requires two generic functions of time and scale, multiplying the source term in the equations for each gravitational potential \cite{Kunz:2012aw}. Such a description may not be connected to a consistent covariant theory, but it can in principle elucidate that physics not allowed by known consistent theories are necessitated by the data. See Ref.~\cite[sec.~2.2]{Ezquiaga:2018btd} for a comparison of descriptions of gravity in cosmology.}
The effective-theory approach is simpler to implement and use because the $\alpha$-functions only depend on time and are independent of the properties of the cosmological background. Meanwhile the Horndeski functions depend on two variables ($X,\phi$) and all results depend on the field's initial conditions and, through the scalar's equation of motion, on the evolution of the other matter species in the universe.
In addition, the $\alpha$-functions represent generic features of deviations from GR present in theories other than Horndeski, and can be used for more agnostic tests.

An advantage of the covariant approach is that all properties of a model (background, perturbations...) are computed from the same parameters. This approach often leads to tighter constraints on the model and allows to address questions that involve the cosmic expansion history, such as the Hubble parameter tension \cite{Renk:2017rzu}.
More importantly, having a covariant theory means that information from other regimes (e.g.\ beyond linear perturbations, or the existence of screening mechanisms) or even other probes (e.g.\ gravitational waves, local tests of gravity), can be folded in into the constraints. This is especially relevant in the case where e.g.\ GW-speed constraints have reduced the functional freedom of the Horndeski action quite significantly, and with it the ability to construct reasonable theories with an arbitrary background expansion and completely independently choose the $\alpha$-functions. By starting from a full covariant action, one is able to discuss the level of fine-tuning required to explain both the background, perturbations and any other physics introduced by the model. 
In the effective approach, combining different regimes would necessitate introducing additional $\alpha$-functions, obscuring the connection to the linear regime
\cite{Bellini:2015wfa,Bellini:2015oua,Cusin:2017mzw}.
Owing to their relative advantages, both approaches are complementary and important for the optimal exploitation of cosmological and other data to test gravity and dark energy.

\vspace{0.5cm}
In this paper we report on three major upgrades of  the public version of the \hiclass code:
\begin{enumerate}
 \item[(i)] Dynamics of covariant Lagrangians: the evolution of the cosmological background, the scalar $\phi$ and the $\alpha$-functions can be solved self-consistently, given the functional form of a covariant action, the theory parameters and initial conditions for the scalar (Section \ref{sec:background_evolution}).
 \item[(ii)] Initial conditions: the perturbations in the scalar field, gravitational potentials and all matter species are solved starting from their attractor values deep in the radiation era, corrected for early dark energy and deviations from GR (Section \ref{sec:ICs}).
 \item[(iii)] Approximation schemes: the equations for cosmological perturbations are optimized using suitable approximations, speeding up the code --- the so-called \emph{Quasi-Static regime} --- but only in conditions where the scalar-field dynamics can be neglected without impact on accuracy (Section \ref{sec:QS}).
\end{enumerate}

\section{Background Evolution in Covariant Theories}
\label{sec:background_evolution}

Here we describe the implementation of theories based on covariant Lagrangians in \hiclass  (complementary to the effective-theory approach, which was described in the first \hiclass  paper \cite{Zumalacarregui:2016pph}). 
In section \ref{sec:background_difficulties} we discuss the difficulties of evolving the cosmological background in Horndeski theories. In section \ref{sec:background_solution} we present a solution inspired by numerical relativity methods and in \ref{sec:background_dynamics} we discuss some practical considerations such as adjusting the dark energy fraction today and discuss example models (Galileon gravity, nKGB, Quintesence and Brans-Dicke Theory).
These examples serve as a guideline for users to implement other covariant theories in \hiclassx.

\subsection{Cosmological Evolution in Generally Covariant Theories}\label{sec:background_difficulties}

The equations of generally covariant theories (such as GR and Horndeski) contain redundancies. 
The equations for the metric can be classified as: 
\textit{constraints}, with at most one time derivative of the metric
and \textit{dynamical equations}, with two time derivatives of the metric.
The evolution of the metric is fully determined by the dynamical equations given some initial conditions, which have to be compatible with the constraint equations. 
Exact solutions are guaranteed to satisfy the constraints, but approximate or numerical solutions induce constraint violations and may rapidly become unphysical. 
This is a known problem in numerical relativity which can be solved by incorporating terms to the dynamical equations that damp the constraint violations  \cite{Lindblom:2005qh}.

For a homogeneous and isotropic Friedman-Robertson-Walker (FRW) space-time with flat spatial sections
\begin{equation}
ds^2 = a(\tau)^2\left(d\tau^2 - d\vec x^2\right)\,, \label{eq:frw_metric} 
\end{equation}
the constraint and dynamical equations correspond to the Friedmann  and acceleration equation
\begin{eqnarray}
 H^2 &= \rho + \mathcal{E}\,,
 \hspace{2.3cm}  &\text{(constraint)}  \label{eq:frw_schematic_constraint} \\
 H^\prime &= - \frac{3}{2}a\big(\rho+p + \mathcal{E} + \mathcal{P} \big)\,,
 \qquad &\text{(dynamical)} \label{eq:frw_schematic_dynamical}
\end{eqnarray}
where $\mathcal{E}$, $\mathcal{P}$ are the effective energy and pressure density stemming from terms beyond GR (see Appendix \ref{sec:appendix_equations_background} for the definitions).%
\footnote{We are using CLASS conventions in which the units are fixed by the Friedmann equation (\ref{eq:frw_schematic_constraint}), primes denote derivatives with respect to conformal time and the Hubble rate is $H={a^\prime/a^2}$.}
The above equations correspond to the variation with respect to $g^{00}$ and $g^{ii}$, defining the time coordinate according to (\ref{eq:frw_metric}) and with the remaining equations fixed by the symmetries of the problem (homogeneity and isotropy).
The space-time is fully determined by solving for the scale factor $a(\tau)$, that fully specifies the metric, as well as the evolution of matter and scalar field (if present).

The standard approach in cosmological codes is to use the constraint equation (\ref{eq:frw_schematic_constraint}) to integrate $a(\tau)$, together with the matter conservation equations, which have analytic solutions for any matter fluid with a constant equation of state.
Using the constraint prevents the obtained solutions from becoming unphysical. 
However, one still needs to specify whether the universe is contracting or expanding at a given time by choosing the correct branch $H =  \pm \sqrt{\rho+\mathcal{E}}$. Although exotic in the standard cosmological scenario, $H(\tau)\to 0$ might occur during the evolution. The universe then would either bounce (from expansion to contraction or vice versa) or continue after coasting. From the point of view of the Friedman constraint, the difference amounts to whether or not one switches the branch.
One has to use the dynamical equation~\eqref{eq:frw_schematic_dynamical} to determine the sign of $H'$ and therefore the outcome.

\begin{figure}
\centering
  \includegraphics[width = 0.5\textwidth]{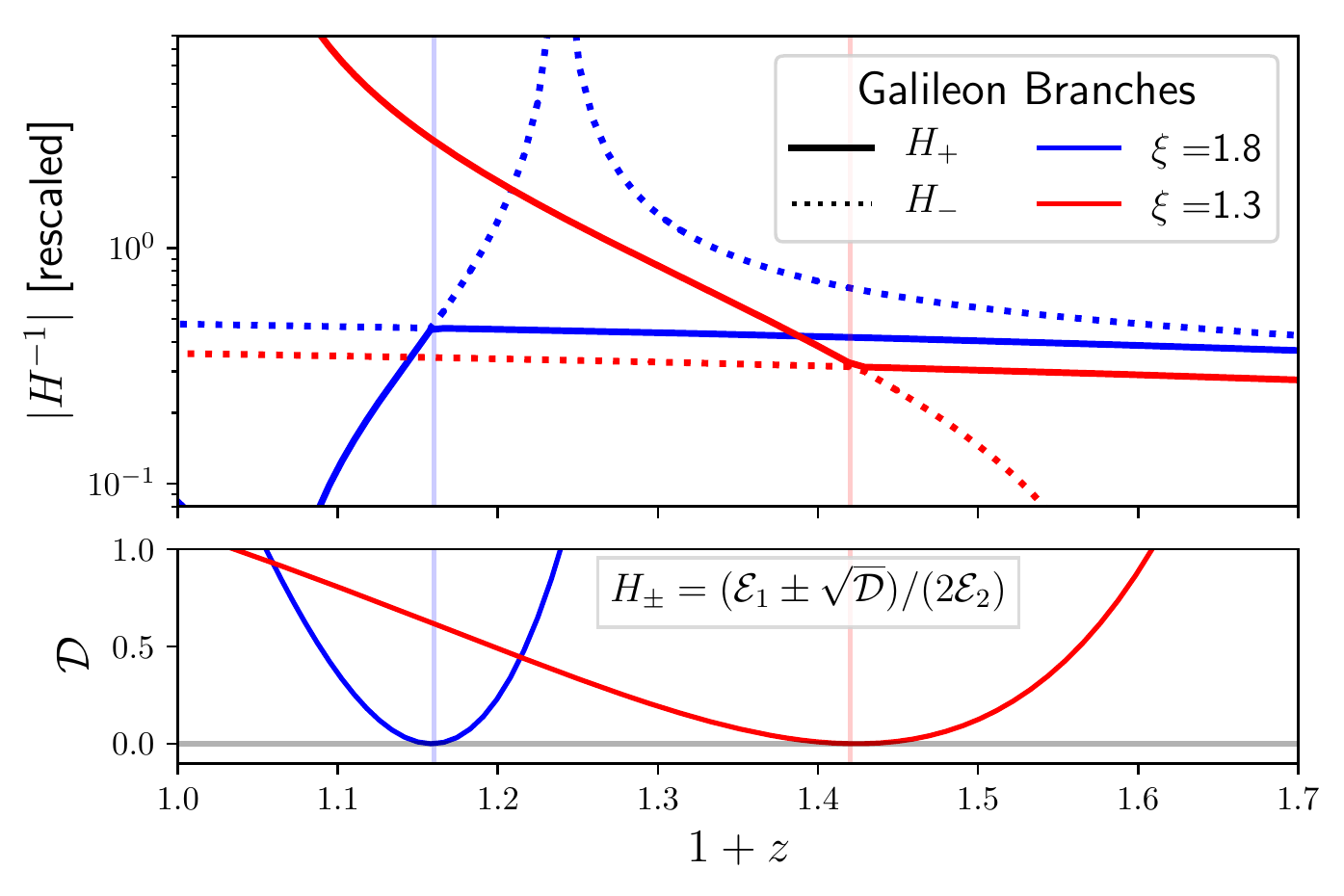}
 \caption{
 Difficulties solving the constraint equation in Horndeski theories. The dynamical evolution often ``jumps'' branches in the constraint equation (\ref{eq:hornd_friedman}). 
 Two quartic Galileon models (blue, red) with identical background expansion (here rescaled for display, see Sec. \ref{sec:galileon_example}) and ${\cal E}_3=0$, 
 jump from the $H_+$ branch (solid, tends to the expanding GR solution $H_+\to \sqrt{\rho}$) to the $H_-$ branch (dotted) at low $z$, producing a smooth solution. Both branches intersect when the discriminant becomes zero (lower panel).
 \label{fig:hubble_problem_constraint}}
\end{figure}

Using the constraint equation to integrate $a(\tau)$ is far more involved in theories beyond GR.
In Horndeski theories the constraint is, in general, a cubic polynomial in $H$
\begin{equation}\label{eq:hornd_friedman}
 {\cal C}(H)  \equiv  {\cal E}_3 H^3 - {\cal E}_2 H^2 + {\cal E}_1 H + {\cal E}_0 = 0\,,
\end{equation}
where we have arranged Eq.\ (\ref{eq:frw_schematic_constraint}), including $\mathcal{E}$, by powers of $H$. The coefficients ${\cal E}_i$ depend on $\phi,\dot\phi$ and the model parameters, the exact expressions are provided in Appendix \ref{sec:appendix_equations_background}.
Eq.~(\ref{eq:hornd_friedman}) has 1, 2 or 3 solutions depending on the values of the coefficients.%
\footnote{If ${\cal E}_3\neq0$ there can be up to 3 solutions, one of which diverges in the limit ${\cal E}_3H\to0$, including the GR limit (often satisfied in the early universe for dark energy models). 
Closed form solutions and numerical methods exist, but they are difficult to implement numerically when there is a large hierarchy in the ${\cal E}_i$ values.}
While it is possible to find all the solutions at any given time, the true solution (as given by the dynamical equation) often evolves between different branches, making any general scheme to integrate $a(\tau)$ based on Eq. (\ref{eq:hornd_friedman}) very difficult to implement for general theories.
And unlike in GR (where a branch switch can be readily interpreted as a bounce), there is little intuition for what these branches mean in Horndeski gravity: they are quite generic even for simple scenarios and are not associated to any sudden change in the dynamics of the cosmological expansion.
Figure \ref{fig:hubble_problem_constraint} shows this phenomenon for two examples of Galileon gravity with different model parameters but identical expansion history (see section \ref{sec:galileon_example} for details on the model).

\begin{figure}
\centering
 \includegraphics[width = 0.5\textwidth]{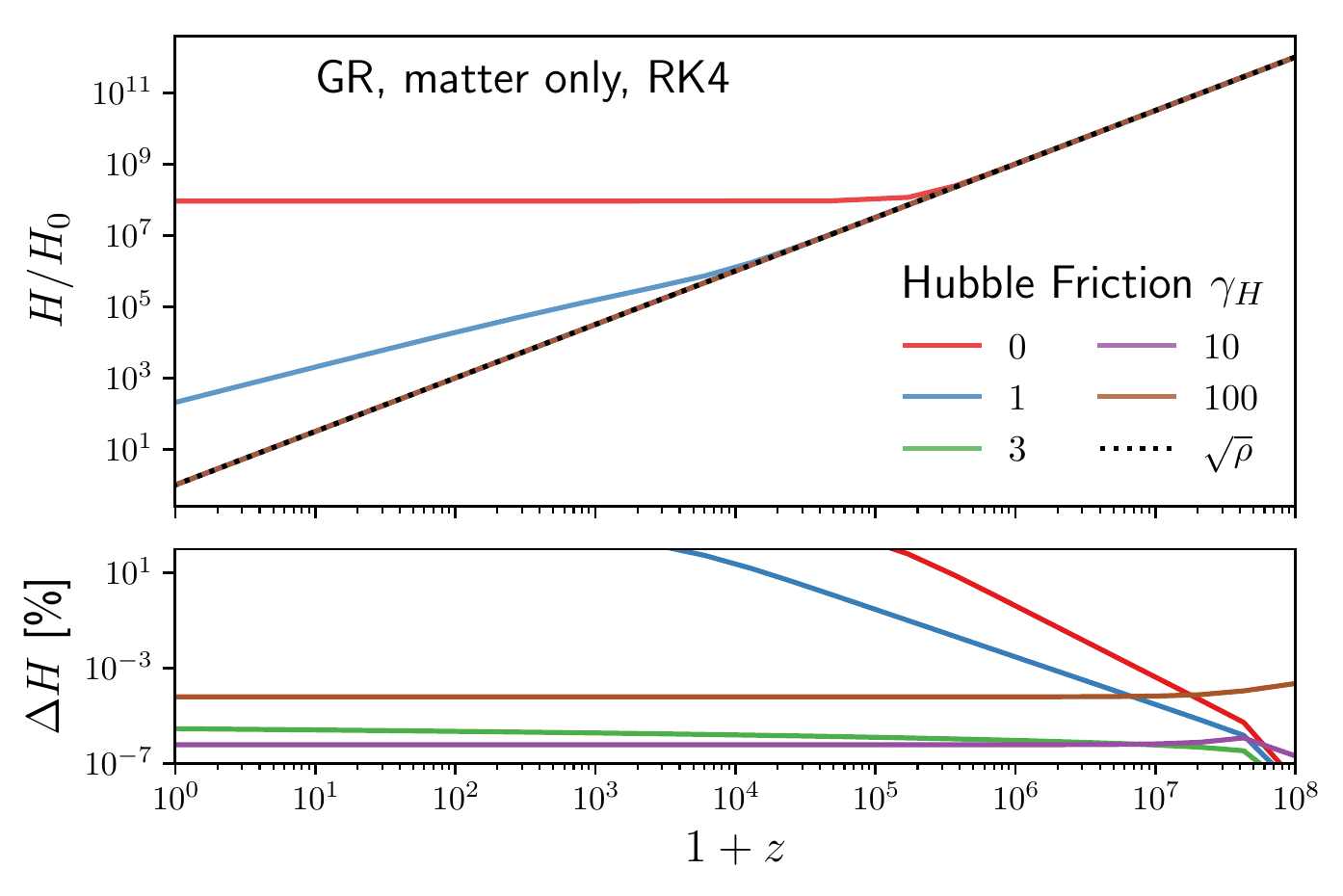}
 \caption{Difficulties integrating the dynamical equation for $H^\prime$. The panels show numerical solutions of the dynamical equation (\ref{eq:frw_schematic_dynamical}) and the relative deviation with respect to the solution obtained from the constraint (\ref{eq:frw_schematic_constraint}).
 The standard solution (red) fails to satisfy the constraint equation (black dotted) (\ref{eq:frw_schematic_constraint}) at a finite time, even with increased numerical precision.
 This issue is solved adding an artificial friction term $\propto \gamma_H\, \mathcal{C}(H)$ to the dynamical equation, eq. (\ref{eq:friedmann_friction_prescription}).
 Large values of the friction coefficient $\gamma_H \sim 10^2$ worsen the convergence, although deviations remain very small ($\lesssim 10^{-5}$).
 All curves correspond to a matter-only universe evolving under Einstein's GR, integrated using a fourth-order Runge-Kutta method.
 \label{fig:hubble_problem_dynamical}}
\end{figure}

\vspace{0.5cm}
An alternative method is to solve the dynamical equation (\ref{eq:frw_schematic_dynamical}) with initial conditions $\dot a(\tau_0)$ consistent with the constraint (\ref{eq:frw_schematic_constraint}).
However, directly solving this equation (e.g.\ with a standard Runge-Kutta integrator) rapidly leads to a violation of the constraint and an unphysical solution with constant $H$. 
This is true even for Einstein's GR, as shown in Fig. \ref{fig:hubble_problem_dynamical}.
The dynamical equation (\ref{eq:frw_schematic_dynamical}) is delicate under numerical evolution and is less robust than the constraint (\ref{eq:frw_schematic_constraint}):
\begin{itemize}

\item A small numerical error in the dynamical equation $\Delta H^\prime = \epsilon H^2$ (with $\epsilon$ dimensionless) is equivalent to a constraint violation  
\begin{equation}
 H^2 = \rho + \mathcal{E} + 2 \int d\tau \, \epsilon H^3 + \mathcal{O}(\epsilon^2)\,.
\end{equation}
The last term becomes approximately constant once $H$ drops below a threshold value, and dominates the evolution after $\rho$ dilutes sufficiently (c.f.\ Fig.~\ref{fig:hubble_problem_dynamical}).

 \item The constraint equation is numerically robust for any source. In constrast, a cosmological constant $\Lambda$ does not contribute to Eq. (\ref{eq:frw_schematic_dynamical}), since the source term vanishes $\rho_\Lambda+p_\Lambda = 0$. The difference between a universe with $\Lambda=0$ and $\Lambda\neq 0$ is encoded in the initial conditions, set by the constraint equation through a negligible term $\rho_\Lambda \ll \rho_m(\tau_{\rm ini})$. In order to capture $\Lambda$, the equations need to be solved with astonishing precision $\sim {\rho_\Lambda}/{\rho_m(\tau_{\rm ini})} \sim a_{\rm ini}^{3}$, beyond the capability of any numerical code.
\end{itemize}

A method to avoid unphysical solutions and incorporate information from the constraints can be obtained inspired by the methods of numerical relativity \cite{Lindblom:2005qh}. 
The key idea is to numerically integrate a combination of the dynamical and constraint equations. In the following we shall see that the correct branch of the constraint equation is always chosen.

\subsection{Dynamical Evolution with Constraint-violation Damping}
\label{sec:background_solution}

To solve the background dynamics numerically, \hiclass  integrates the dynamical equation (\ref{eq:frw_schematic_dynamical}), modified as
\begin{equation}\label{eq:friedmann_friction_prescription}
 H^\prime = - \frac{3}{2}a\big(\rho+p + \mathcal{E} + \mathcal{P} \big)  - a \gamma(H) \mathcal{C}(H)\,.
\end{equation}
The new term is proportional to the constraint, Eq. (\ref{eq:hornd_friedman}). 
It vanishes identically when the solutions are physical, $\mathcal{C}=0$, recovering the original dynamical equation. 
A suitable choice of the coefficient
\begin{equation}\label{eq:friedmann_friction_term}
\gamma(H) = \gamma_H \,\cdot \text{sign}\left(\frac{\partial{\cal C}}{\partial H}\right)\,,
\end{equation}
with $\gamma_H>0$ ensures that the last term in Eq. (\ref{eq:friedmann_friction_prescription}) acts as a friction term, driving the evolution towards ${\cal C} = 0$. 

Small violations of the constraint are exponentially damped by the new friction term. 
Let us express the numerical solution as $H_n = H + \delta H$, where $H$ is the physical solution satisfying both the constraint and acceleration equation (\ref{eq:frw_schematic_constraint}, \ref{eq:frw_schematic_dynamical}). 
Then  deviations are damped by the friction term in Eq.~(\ref{eq:friedmann_friction_prescription})
\begin{equation}
 \delta H^\prime = -a\gamma(H)\frac{\partial \mathcal{C}}{\partial H}\delta H + \mathcal{O}(\delta H^2)
 \quad \Rightarrow \quad 
 \delta H \propto \exp\Big(-\gamma_H\left|\frac{\partial \mathcal{C}}{\partial H}\right| t\Big) \,,
\end{equation}
and constraint violations vanish exponentially on a timescale $\sim \left(\gamma_H\left|\frac{\partial{\cal C}}{\partial H}\right|\right)^{-1}$. This prescription drives the evolution towards $\mathcal{C}\to 0$ and prevents unphysical results.
Values of $\gamma_H\gtrsim 2$ provide sufficient stability of the constraint under time evolution, for typical precision parameters, as shown in Fig.~\ref{fig:hubble_problem_dynamical}. 
Note that it is not necessary to solve for $H$ in the constraint (\ref{eq:hornd_friedman}) to integrate equation (\ref{eq:friedmann_friction_prescription}), it suffices to evaluate $\mathcal{C}(H)$.

The constraint equation (\ref{eq:hornd_friedman}) needs to be solved only when the initial conditions for the background are set.
This gives a value of $H$ that depends on the scale factor, the matter/radiation density and the initial configuration of the scalar field $\phi(\tau_{\rm ini}),\,\phi^\prime(\tau_{\rm ini})$. For models that approximate GR at early times, the usual solution $H=\sqrt{\rho_M}$ is good enough, but one has to be more careful for theories that modify gravity in the early universe (a simple example is Brans-Dicke, in which $M_*^2\propto \phi(\tau)$ can differ from unity and modify the Friedmann equation at early times). Even in those situations, one expects that the corrections will be small, and therefore an iterative correction to the GR solution (e.g. Newton's root finder) is enough. Note also that an initial violation of the constraint would also be damped by the prescription (\ref{eq:friedmann_friction_prescription}).

The method of constraint damping has been tested systematically in the simple setup shown in Fig. \ref{fig:hubble_problem_dynamical}. The results have also been validated against other codes for covariant theories (Brans-Dicke and Galileons), as described in the Einstein-Boltzmann code comparison beyond GR \cite{Bellini:2017avd}.
All the performed tests showed excellent agreement, with $\lesssim 0.01\%$ accuracy on the background expansion and sub-percent accuracy on CMB and matter power spectra.
Note that other codes did not rely on constraint damping because they are restricted to relatively simple models, with either no branch jumps (e.g.\ Brans-Dicke) or where solutions are fixed by a symmetry of the theory (e.g.~Galileons, c.f.~Section \ref{sec:galileon_example}).

\subsection{Initial and Final Conditions for the Background}
\label{sec:background_dynamics}

The main difference between the covariant and effective-theory approach is the need to solve the dynamical evolution of the background scalar field $\phi(\tau)$. 
This requires specifying the scalar field \textit{initial conditions, $\phi(\tau_{\rm ini}),\phi^\prime(\tau_{\rm ini})$ together with parameters entering the $G_i(X,\phi)$ functions.%
\footnote{Note that the choice of background initial conditions might be restricted in some theories to less than two arbitrary choices for the scalar and its velocity. This might be due to symmetry, the existence of attractor solutions or viability conditions (see examples in Sec.~\ref{sec:shift_symmetry}, \ref{sec:shift_symmetry_broken}).}}
In addition, model parameters need to be adjusted to satisfy the \textit{final condition} on the dark energy fraction today
\begin{equation}\label{eq:tuning_condition}
 H_0^2 \OmDE^{(0)} = \mathcal{E}(\phi(\tau_0),\phi^{\prime}(\tau_0),H_0) \,.
\end{equation}
Here the left-hand side involves cosmological parameters provided by the user and the right-hand side is a result computed by the code given the parameters of the Horndeski model, (i.e.\ evolving the field from initial conditions until today). 
The background evolution does not guarantee that the scalar field evolution is compatible with other input cosmological parameters, such as the expansion rate $H_0$ and the scalar field energy fraction $\OmDE^{(0)}$.%
\footnote{It is also possible to use all the Horndeski parameters (including field initial conditions) as input and treat $H_0$ and $\Omega_{\rm DE}^{(0)}$ as derived parameters when sampling the parameter space. Then the data would automatically prefer the right values, avoiding the need to adjust final conditions. 
However, understanding the subset of parameters leading to reasonable values of $H_0$ and $\OmDE^{(0)}$ is necessary for realistic convergence times, even in this type of approach.}
In some situations it might be desirable to adjust additional final conditions. 
An example is Brans-Dicke theory, in which the absence of a screening mechanism requires a tuning of scalar field value $\phi(\tau_0)$ to satisfy Solar System tests (see Section~\ref{sec:brans_dicke}).

The correct final conditions (\ref{eq:tuning_condition}) can be obtained by adjusting one or more of the model parameters (which may include the scalar field initial conditions).
While this choice is straightforward in simple dark energy models, it can be highly non-trivial in general cases, often depending on the values of multiple parameters at the same time.
When they exist, solutions to Eq. (\ref{eq:tuning_condition}) can be found numerically. Doing so efficiently requires a good guess of the solution and its dependence with the parameter being varied, which in turn requires a good understanding of the model's dynamics.
Finding the right initial and final conditions for a Horndeski model is in itself a craft and no simple or universal recipe exists.
While numerical root-finding is often unavoidable, there are several tricks to understand the role of model parameters in the field equations and improve the initial guess.

First, since it is only an unobservable variable, we can perform arbitrary field redefinitions on the scalar field. The Horndeski family of actions is closed under redefinitions which only depend on the scalar field value, i.e.\
\begin{equation}
\phi\to \tilde\phi = F(\phi)\,,
\end{equation}
yielding an equivalent theory with different Horndeski functions $\tilde G_i$ \cite{Flanagan:2004bz,Zumalacarregui:2012us,Bettoni:2013diz,Bettoni:2015wta,Ezquiaga:2017ner,Babichev:2019twf}.
A suitable choice may simplify one of the Horndeski functions in simple theories (for a quintessence example see Ref.~\cite{Garcia-Garcia:2018hlc}).
A constant rescaling $\phi \to \tilde\phi = f\cdot \phi$ can often be used to fix one of the model parameters (see \cite{Barreira:2014jha} and Sec. \ref{sec:galileon_example} for an example with Galileon models).%
\footnote{More complicated redefinitions, e.g.\ mixing the scalar and its derivatives, or even the scalar and the metric would also leave the observables unmodified. However, the action would generically no longer be of Horndeski form and even contain higher derivatives, which could be removed by new constraints. Such related actions have not been implemented in \hiclassx, which assumes minimal coupling of the matter.}

The modified Friedmann equation (\ref{eq:hornd_friedman}) can be used to understand the scale of the model parameters (cf. \ref{eq:tuning_condition}) and provide some useful relations. 
Since it only contains first and zeroth time derivatives of the metric and the scalar field, it provides a relatively simple relation between the expansion rate, the matter energy density and the scalar field configuration at any time.
Since the Friedmann equation is at most an order 3 polynomial in $H$ (order 2 if $G_{5,X}=0$), the solutions for $H$ can be obtained in closed form.
These expressions can be used to identify the contributions of each term to the energy density and obtain a guess for the model parameters.

Another important quantity is the equation of state for the scalar
 \begin{equation}
  w \equiv \frac{\mathcal{P}}{\mathcal{E}}\,.
 \end{equation}
This expression can be related to the evolution of the field by writing the scalar field equation as covariant energy-momentum conservation
\begin{equation}
 \frac{\mathcal{E}^\prime}{a} + 3H(1+w)\mathcal{E} = 0\,.
\end{equation}
Any bound on the value of $w$ limits how rapidly the field's energy density can evolve.
For instance, if $w>-1$ then the energy density $\mathcal E$ can only decrease as the universe expands (e.g.~in quintessence, Sec.~\ref{sec:quintessence_example}).

Further insight into the dynamics of the scalar field can be gained by writing the scalar equation of motion as
\begin{equation}
\frac{\mathcal{J}^\prime}{a} + 3 H \mathcal{J} = \mathcal{S}_{\phi}\,. \label{eq:field_shift-current}
\end{equation}
The left hand can be written as a covariant conservation law $\nabla_\mu\mathcal{J}^\mu$, for a Noether current $\mathcal{J}^\mu$ associated with a shift symmetry in the scalar field $\phi\to\phi+c$, while the right hand side includes contributions from terms violating the shift symmetry.
On a homogeneous and isotropic space-time the current $\mathcal{J}^\mu = (\mathcal{J},\vec 0)$ is fully characterized by the \textit{shift charge density} $\mathcal{J}$. The shift charge density is an order 3 polynomial in  $H$, whose coefficients depend on $\phi^\prime/a$ and $\phi$ through derivatives of the Horndeski functions (see Eq.~(\ref{eq:field_shift_charge}) for the full expression).
In general, Horndeski theories include terms $\mathcal{S}_{\phi}$ that violate shift symmetry (see Eq. (\ref{eq:field_shift_source}) for the full expression).
These terms act as a source for $\mathcal{J}$, which is no longer covariantly conserved if $\mathcal{S}_{\phi}\neq 0$.
Eq.~(\ref{eq:field_shift-current}) provides important insights into the dynamics of the theory, even when shift-symmetry is broken.%
\footnote{Note that it is sufficient that $\mathcal{S}_\phi=0$ for a theory to be shift-symmetric. The simplest method of realising this is setting all $G_{i,\phi}=0$, but it could well be a non-trivial condition. In principle, for a shift-symmetric theory that depends on $\phi$, one should be able to perform a field redefinition to obtain an equivalent theory which does does not depend on the scalar-field value, but, to the authors' knowledge, this has not been proven in the literature.}
It is important to remember that \hiclass integrates the full equations of motion, regardless of whether the dynamics can be simplified due to shift symmetry.

Below we provide the details of several theories with and without shift-symmetry. These include models already included in \hiclass that may provide the basis for further implementations. We emphasize how some of the above ideas can be used to understand the role of model parameters in fixing the initial and final conditions for the theory. 
Other techniques from dynamical systems applied to cosmology may provide further simplifications (see Ref.~\cite{Bahamonde:2017ize} for a review).

\subsubsection{Shift symmetry and current conservation: covariant Galileon \& nKGB} \label{sec:shift_symmetry}

In shift-symmetric theories, covariant conservation of the shift-current (\ref{eq:field_shift-current}) can be readily solved
\begin{equation}
 \frac{\mathcal{J}^\prime}{a} + 3H \mathcal{J} =  0 
\quad \Rightarrow \quad
 \mathcal{J} = \mathcal{J}_0 a^{-3} 
 \,. \label{eq:shift_symmetry_solutions}
\end{equation}
That is, the shift charge density $\mathcal{J}$ dilutes with the universe's volume, as expected of a conserved charge \cite{Deffayet:2010qz}. 
As a consequence, the vacuum state $\mathcal{J} = 0$ is a good approximation in the late universe unless the shift charge density is substantial at early times. Moreover, unless the co-dependence of $\OmDE$ (or any other late-time deviations from GR) on $\mathcal{J}$ is negligible,  the initial shift current needs to be very fine-tuned for its value today to play a role: note that the shift charge density decreased by a factor $\gtrsim 10^{28}$ since primordial nucleosynthesis, with this factor being much larger if initial conditions were set earlier (e.g.\ at the end of inflation). 

The condition $\mathcal{J}=0$ can have multiple solutions, which  are vacua of the theory with vanishing shift-charge density (see e.g.\ Ref.~\cite{Pujolas:2011he} for a discussion of the hydrodynamical interpretation of some such models). In an expanding universe, all these vacua are stable attractors of the system, as can be seen from Eq.~(\ref{eq:shift_symmetry_solutions}).
Which vacuum is reached depends on the initial conditions, and it may, in principle, never actually be reached if the Universe bounces during is evolution (e.g.\ \cite{Easson:2011zy}).

Apart from very particular choices for $G_i$, involving fractional or negative powers of $X$, shift-symmetric theories always allow for a trivial vacuum $\phi'=0$ on which (1) all $\alpha$-functions are zero (2) only a constant term is allowed in $\mathcal{E}$, so the cosmological predictions reproduce GR exactly (allowing at most a nonzero cosmological constant $\Lambda$). 
However, it is not necessarily true that this vacuum is stable, but the scalar could be a ghost, depending on the choice of the Horndeski functions.
Other, non-trivial, $\mathcal{J}=0$ solutions can exist depending on the choice of $G_i$. If they are stable, they are then the final states of evolution of the cosmology, at least for some basin of attraction. These non-trivial vacua are often known as \textit{trackers}, the exact properties of which depend on the other matter content in the universe, but which typically evolve toward a final accelerating state \cite{Deffayet:2010qz}.
In this kind of models it frequently happens that the scalar field is a ghost on flat spacetime, indicating that Minkowski spacetime is not a stable solution.

When the field reaches a tracker solution $\mathcal{J}\approx 0$ its dynamics are governed by an algebraic equation relating $H,\phi^\prime,\phi$ and the parameters of the theory, as given by Eq.~(\ref{eq:field_shift_charge}).
Below we will see how this dependence simplifies the treatment of two example theories, Covariant Galileon and Kinetic Gravity Braiding.

\paragraph{Covariant Galileon} \label{sec:galileon_example}

The covariant Galileon \cite{Nicolis:2008in,Deffayet:2009wt,DeFelice:2010pv} is a Horndeski theory defined by
\begin{equation}
 G_2=c_1\phi - c_2 X\,,\; G_3 = \frac{c_3}{\Lambda_3^3}X\,,\;
 G_4 = \frac{\Mpl^2}{2} - \frac{c_4}{\Lambda_3^6}X^2\,,\; G_5 = \frac{3c_5}{\Lambda_3^9}X^2\,, \label{eq:cov_gal_action}
\end{equation}
where $\Lambda_3^3 = H_0^2 \Mpl$ (note that it is possible to rescale the field and fix one of the galileon coefficients \cite[Eq.~22]{Barreira:2013jma}; see also the discussion in Appendix~\ref{sec:dimless}).
The theory is shift symmetric provided that the linear potential vanishes $c_1=0$. 
While Covariant Galileons are ruled out by a combination of gravitational wave and cosmological observations, they have interesting properties as cosmological models (see Refs.~\cite{Barreira:2014jha,Renk:2017rzu,Peirone:2017vcq,Ezquiaga:2017ekz} for updated constraints).

\begin{figure}
\centering
\includegraphics[width = 0.49\textwidth]{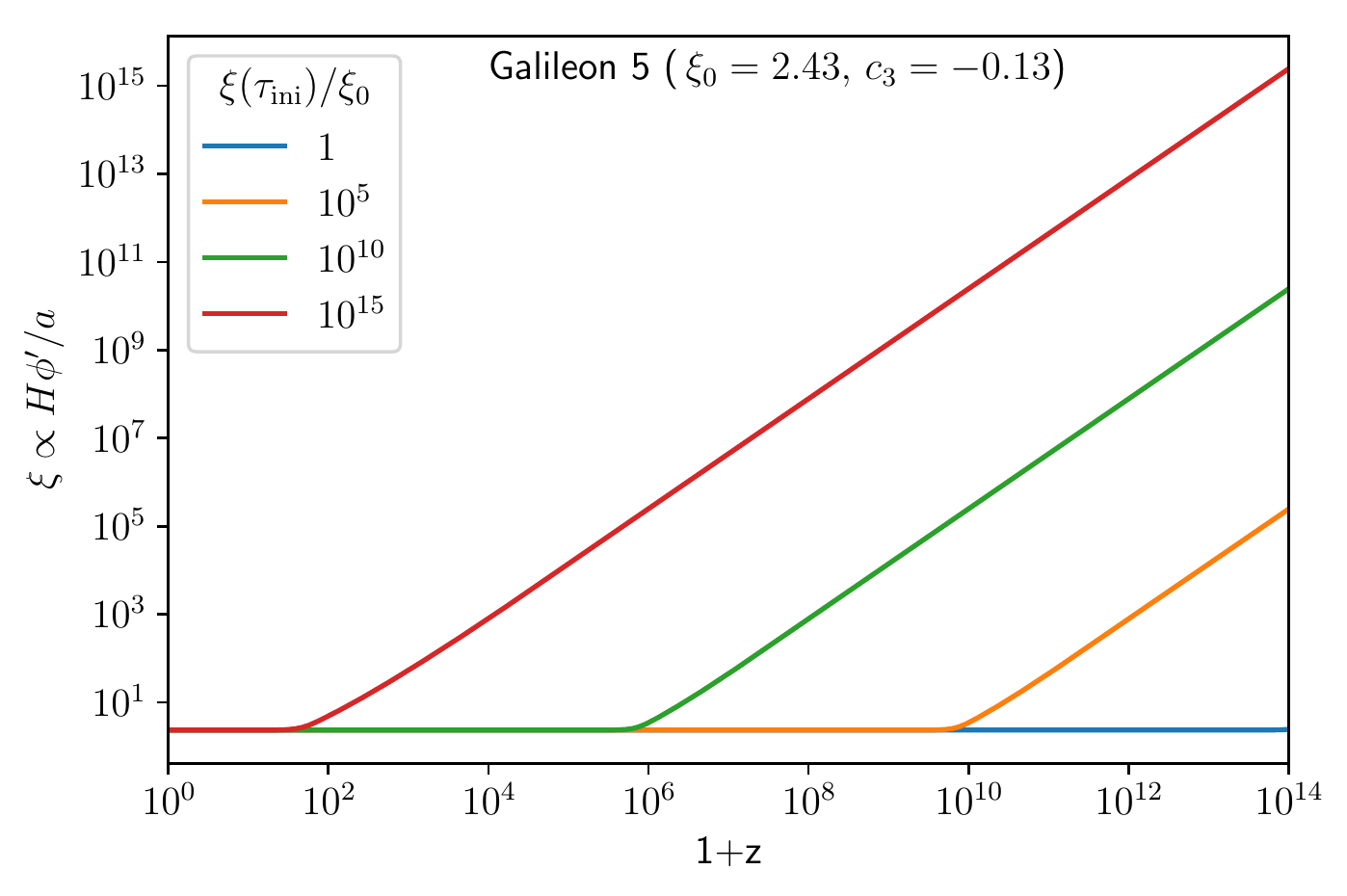}
  \includegraphics[width = 0.49\textwidth]{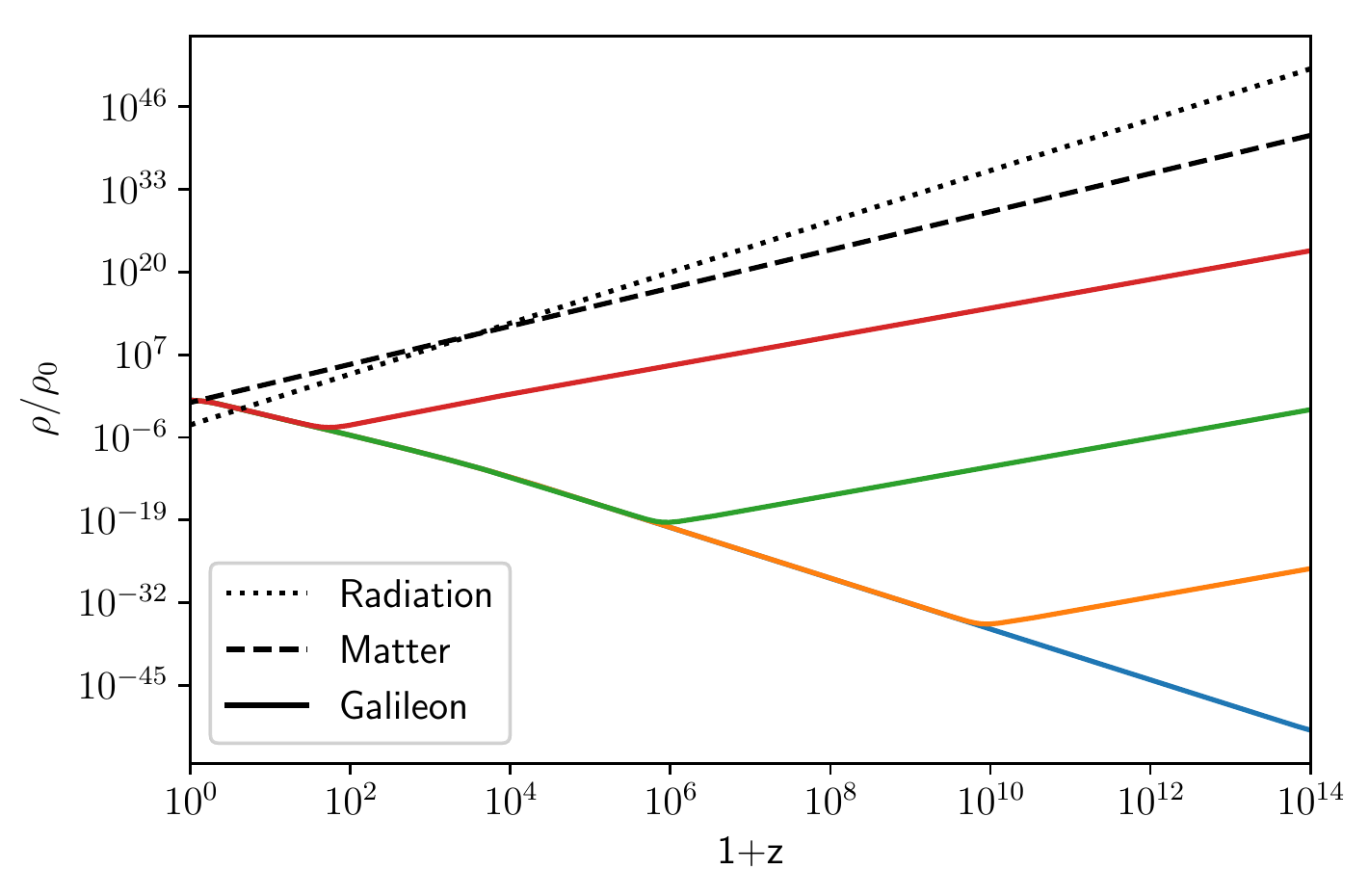}
 \caption{
 Covariant galileon dynamics: general initial velocities of the scalar field eventually converge to the tracker solution $\xi(z)\to\xi_0$ (left panel). The energy density of the galileon remains subdominant at all epochs (left panel). 
 The initial conditions do \textit{not} affect cosmological predictions, unless the tracker is reached in the late universe when DE is non-negligible.
All lines are for the same quintic model with $\xi_0 = 2.43,c_3 = -0.132$.
 \label{fig:galileon_tracker}}
\end{figure}

The shift charge density (\ref{eq:field_shift_charge}) only depends on the combination $H\phi'/a$.
Defining the dimensionless quantity
\begin{equation}\label{eq:gal_xi_def}
  \xi\equiv\frac{H\phi'}{aH_0^2\Mpl}\,,
\end{equation}
(see Refs.~\cite{Barreira:2014jha,Renk:2017rzu} for details), it can be expresed as
\begin{equation} \label{eq:galileon_track}
\frac{H}{\Mpl H_0^2}\mathcal{J} =  \xi\left(c_{2}-6c_{3}\xi+18c_{4}\xi^{2}+5c_{5}\xi^{3}\right) \to 0\,,
\end{equation}
Starting from initial conditions, the field velocity approaches the tracker $\xi\to \xi_0$, with $\mathcal{J}(\xi_0)=0$ in Eq.~(\ref{eq:galileon_track}), as dictated by the dilution of the shift charge density, Eqs~(\ref{eq:shift_symmetry_solutions}).
As the solution approaches the tracker $\xi\sim \xi_0$ is approximately constant and $\phi'\propto a/H$ as per Eq.~(\ref{eq:gal_xi_def}).
This behaviour is shown for different initial conditions in Fig.~\ref{fig:galileon_tracker}.
On the tracker, it is further possible to write the galileon contribution to the energy budget as
\begin{equation}\label{eq:galileon_Omega}
 \Omega_{\rm DE}^{(0)}=\frac{c_{2}}{6}\xi_0^{2}-2c_{3}\xi_0^{3}+c_{4}\frac{15}{2}\xi_0^{4}+c_{5}\frac{7}{3}\xi_0^{5}\,.
\end{equation}

The problem of the final condition reduces to solving Eqs~(\ref{eq:galileon_track},\ref{eq:galileon_Omega}). In practice it is convenient to treat $\xi_0$ as a free parameter of the theory, and use it to fix one of the coefficients $c_i$. It is conventional to define 3 models, ordered by increasing complexity:
\begin{itemize}
 \item Cubic Galileon: $\xi_0,c_3$ set by $\Omega_{\rm DE}^{(0)}$, $\mathcal{J}=0$ with $c_4,c_5=0$.
 \item Quartic Galileon: $\xi_0$ free, $c_3,c_4$ set by $\Omega_{\rm DE}^{(0)}$, $\mathcal{J}=0$ with $c_5=0$.
  \item Quintic Galileon:  $\xi_0,c_3$ free, $c_4,c_5$ set by $\Omega_{\rm DE}^{(0)}$, $\mathcal{J}=0$.
\end{itemize}
In all these models Eqs. (\ref{eq:galileon_track}, \ref{eq:galileon_Omega}) are used to fix two parameters and $c_2=-1$ is set by the normalization of the field --- cubic models require $c_2<0$ in order to have $\Omega_{\rm DE}^{(0)}>0$,  quartic \& quintic models allow $c_2 >0$ but are disfavoured by observations \cite{Barreira:2013jma}.
The initial conditions for $\phi^\prime$ are fixed by setting the field in the tracker $\mathcal{J}=0$ initially --- non-tracker solutions are possible, cf.~\cite{Leloup:2019fas}, but fine tuned as discussed after Eq. (\ref{eq:shift_symmetry_solutions}). 
Note that all shift-symmetric galileons contain a trivial tracker, leading to $\xi\to0$ for which $\mathcal{E}\to0$, $\alpha_i\to 0$ as long as the initial conditions are sufficiently near $\xi=0$. 

One may in principle consider different initial conditions for the scalar field, parameterized by $\xi(\tau_{\rm ini})$. However, for most values of $\xi(\tau_{\rm ini})$ the tracker solution will be reached while the galileon is subdominant (see Fig.~\ref{fig:galileon_tracker}), leading to identical cosmological predictions. For the sake of simplicity it is conventional to set $\xi(\tau_{\rm ini})=\xi_0$. 
The initial conditions may change the model predictions only if the tracker solution is reached at low redshift, when the galileon energy density is not negligible. Ref.~\cite{Barreira:2013jma} found that this possibility was disfavoured by CMB data, placing a bound on $\xi(\tau_{\rm ini})$ (or equivalent $\rho_{\rm DE}(\tau_{\rm ini})$, cf. right panel of Fig.~\ref{fig:galileon_tracker}).

\paragraph{nKGB} \label{sec:nKGB_example}
nKGB is a two-parameter family of models in the class of Kinetic Gravity Braiding \cite{Deffayet:2010qz,Kobayashi:2010cm,Kimura:2010di}, which generalizes the cubic galileon. In this theory, the propagation of gravitational waves is not modified $\aT=0$. The Lagrangian is given by the choice
\begin{equation}
	G_2 = -X\,,\qquad G_3 = g^{(2n-1)/2} \Lambda \left(\frac{X}{\Lambda^4}\right)^n \,, \qquad G_4= \frac{\Mpl^2}{2}\,,\qquad G_5 =0 \,.
\end{equation}
with $g$ a dimensionless constant, of order one if one chooses $\Lambda^{4n-1}=H_0^{2n}\Mpl^{2n-1}$. Minkowski spacetime is unstable in this model and the final state of the Universe is a self-accelerating vacuum with equation of state $w=-1$, which is approached from \emph{below}. The shift-symmetry of the action leads to a non-trivial vacuum characterised by the dimensionless quantity
\begin{equation}
\xi \equiv \frac{\phi' H^{1/(2n-1)}} {a H_0^{2n/(2n-1)}\Mpl} \,, \label{eq:nKGBtrac}
\end{equation}
with $\xi\rightarrow \text{const}$ as the vacuum $\mathcal{J}=0$ is approached.
In this vacuum state equation of state for DE is tracking the external matter, $w$,
\begin{equation}
	1+w = -\frac{1}{2n-1}(1+w_\text{m})\,,\quad n\neq\frac{1}{2}\,, \qquad (\text{vacuum}, \mathcal{J}=0)
\end{equation}
where $w_\text{m}$ is the total equation of state for the standard relativistic and non-relativistic matter content, weighted by density fractions. The density fraction in dark energy today is
\begin{equation}
	\Omega_{\text{DE,0}}^{-1} = g (2\cdot 3^{2n})^{1/(2n-1)} = 2^\frac{2n+1}{2n-1}3^\frac{2n+2}{2n-1}\xi^{-2}\,.
\end{equation}
Larger values of $n$ lead to an equation of state closer to the consmological constant, and $w=-1$ is the final state reached in any case in the asymptotic future for all parameter values. The choice of $n$ and $\Omega_\text{DE,0}$ fixes the parameter $\xi$ (or $g$) and through the tracker equation  \eqref{eq:nKGBtrac}, sets the initial condition for $\dot\phi$. The limit $n\rightarrow \frac{1}{2}$ is singular: $w\rightarrow -\infty$ and the value of $\phi'$ at initialisation rapidly decreases; for sufficiently small $n$, $G_{3X}$ and $G_{3XX}$ exceed machine precision and the evolution cannot be followed properly.

Away from the vacuum, when the energy density is dominated by the shift-current $\mathcal{J}$, the scalar field also tracks the matter. The shift charge dilutes as $\mathcal{J}\propto a^{-3}$, while the energy density evolves with the effective equation of state 
\begin{equation}
	w = \frac{1}{4n}(1-w_\text{m}) \,, \qquad (\mathcal{J}\gg a^{-1}\phi')
\end{equation}
i.e.\ for $n>\frac{1}{2}$ redshifts more slowly that the radiation. If the initial shift-charge density is too large, a period of shift-charge domination would occur prior to matter-radiation equality modifying the cosmic microwave background significantly.
\subsubsection{Broken Shift Symmetry: Quintessence \& Brans-Dicke} \label{sec:shift_symmetry_broken}

General DE and modified gravity models break shift symmetry.
This leads to richer dynamics and makes it necessary to adjust the model parameters numerically to obtain the desired $\OmDE$. 
Even in that case, understanding the dynamics as a sourced shift current can simplify the problem of the initial and final conditions by providing a good guess of reasonable model parameters.
Below we will describe the dynamics of quintessence and Brans Dicke gravity, focusing on the initial and final conditions for the field.

\paragraph{Quintessence}\label{sec:quintessence_example}

Quintessence is a class of models defined by
\begin{equation} \label{eq:quint_action}
 G_2 = X - V(\phi)\,,\quad  G_4 = \frac{\Mpl}{2}\,, \quad G_3 = G_5 = 0\,,
\end{equation}
i.e.~a canonical kinetic term for the scalar field and a potential $V(\phi)$ that depends only on the scalar field value. 
Quintessence only modifies the expansion of the universe and the growth of perturbations on horizon scales due to the clustering of the scalar field (since $c_s^2 = 1,\alpha_B=\alpha_M=\alpha_T=0$), and is therefore not considered as a modified gravity model.
See Refs~\cite{Copeland:2006wr,Tsujikawa:2013fta} for comprehensive reviews of quintessence models and their dynamics.

Quintessence can exhibit very rich dynamical behavior, despite the model's apparent simplicity. 
Nonetheless, some straightforward limits may be readily understood from the expression for the equation of state
\begin{equation}\label{eq:quint_eos}
 w_{\rm DE}\equiv\frac{\mathcal{P}}{\mathcal{E}} = \frac{X-V(\phi)}{X+V(\phi)}\quad \longrightarrow \quad
 \left\{ 
 \begin{array}{l l}
 -1 & \quad (V \gg X) \\
 1 & \quad (X\gg V)
 \end{array}
 \right.
 \,.
\end{equation}
It is clear from this expression that $-1\le w \le 1$, and the limiting cases appear when either the kinetic ($X$) or potential ($V$) energy dominate.
The limit in which the potenial dominates corresponds to a cosmological constant (i.e.~$V(\phi)$ becomes a cosmological constant if $\phi$ is constant).
The limit of negligible potential corresponds to a very rapidly decaying energy density%
\begin{equation}\label{eq:quint_kination}
\mathcal{E} \propto a^{-3(1+w)} \sim a^{-6} \qquad  \text{(if $V\ll X$)}\,.
\end{equation}
This regime is usually known as \textit{kination}.%
\footnote{This limit can be understood in terms of shift-symmetry alone. If $V_{\phi}\to 0$, shift symmetry is restoredand the tracker condition (\ref{eq:field_shift_charge}), 
$ n \propto \phi^\prime/a = 0$, implies that the only vacuum is the trivial solution. This solution is approached as $\phi^\prime/a \propto a^{-3}$, leading to a very rapidly vanishing kinetic energy $ 
 X\equiv \frac{(\phi^\prime)^2}{2a^2} \propto a^{-6}$.}
Another important limit is when the scalar field oscillates around a minimum of the quintessence potential.
If $V(\phi)= m_\phi^2 \phi^2$, then $w_{DE}$ oscillates between $-1$ and $1$, with average $\langle w_{\rm DE}\rangle = 0$. The scalar field can act as dark matter if $m_\phi$ is sufficiently large (see Ref.~\cite{Marsh:2015xka} for a review). 

\begin{figure}
\centering
  \includegraphics[width = 0.49\textwidth]{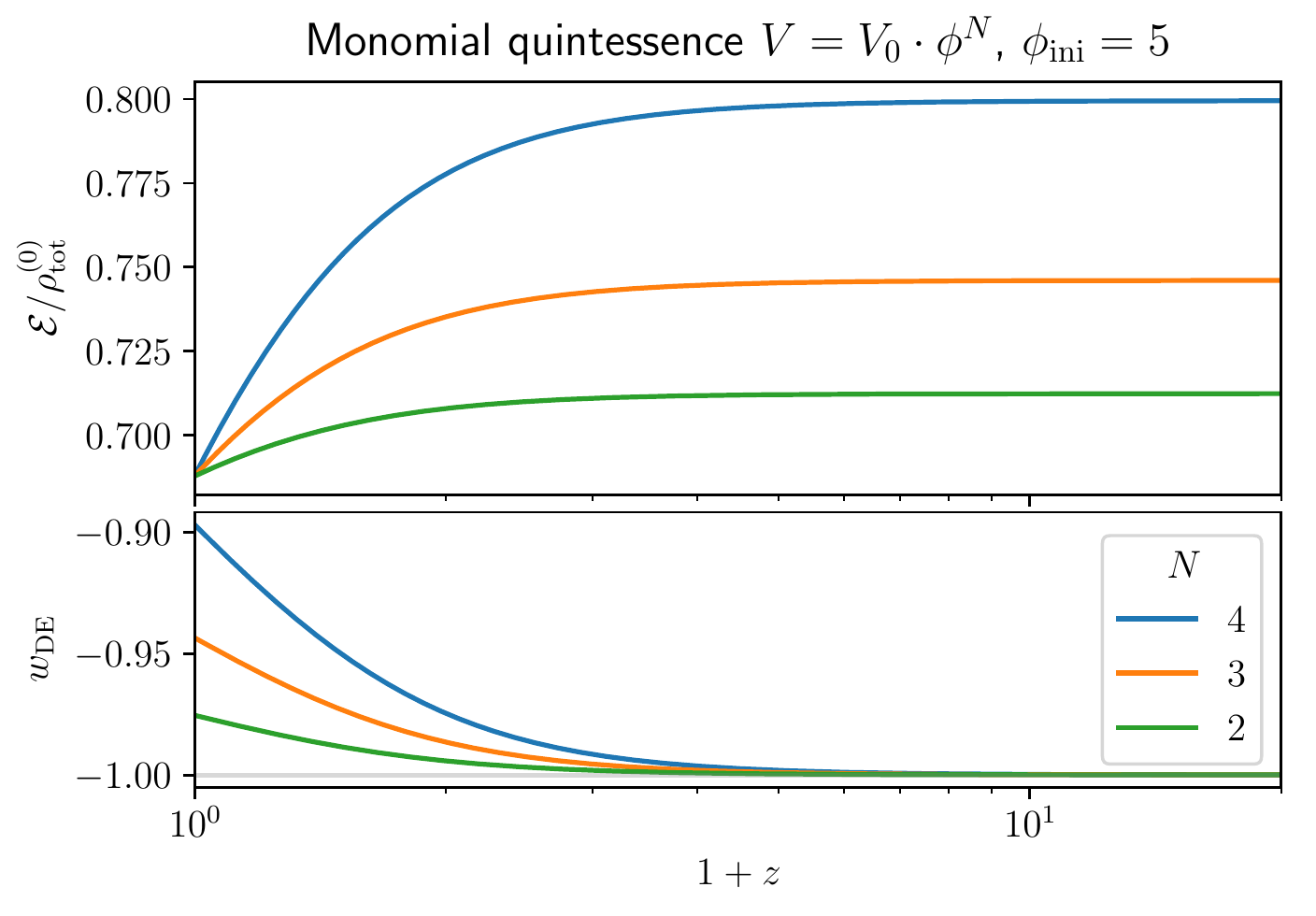}
  \includegraphics[width = 0.49\textwidth]{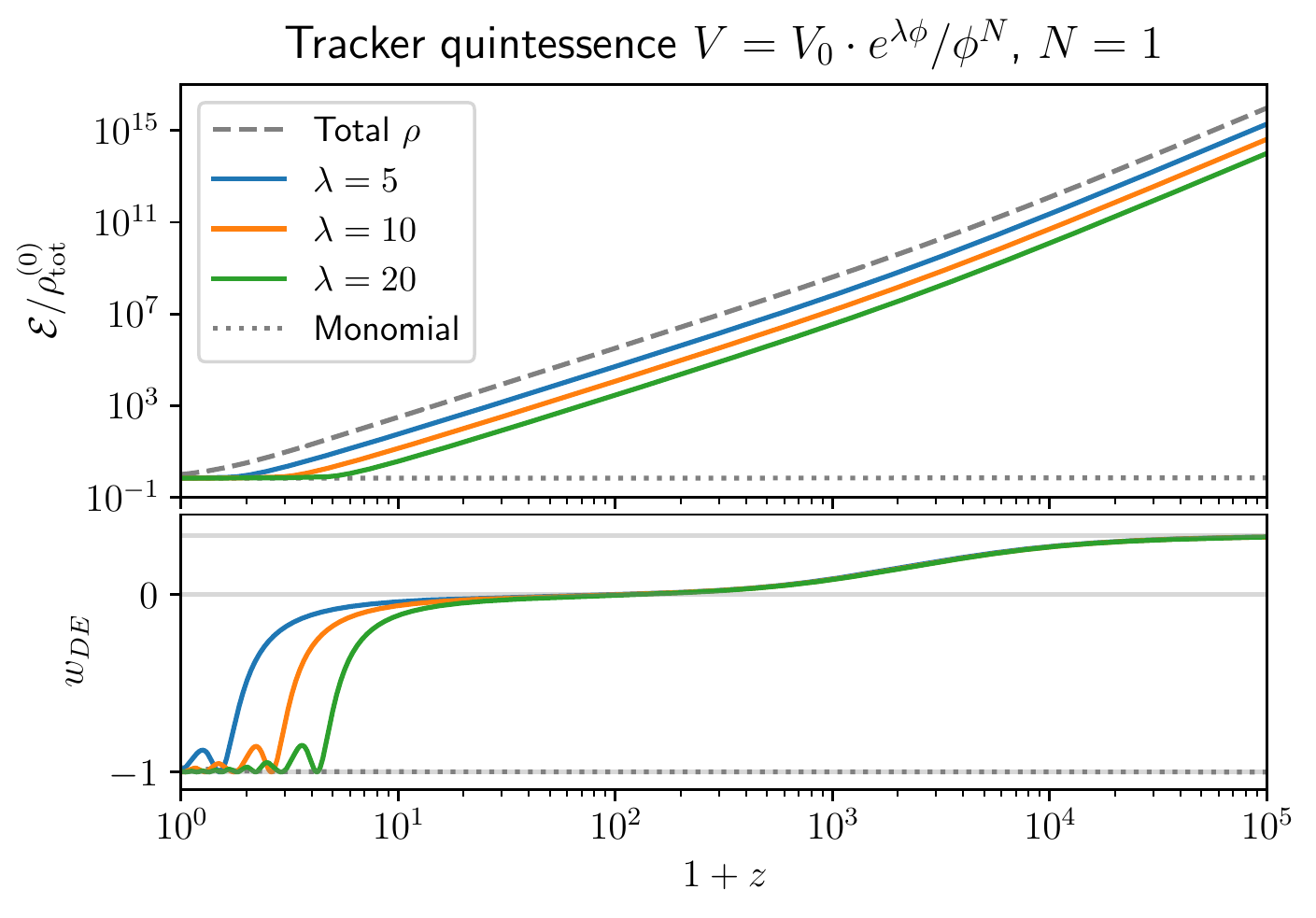}
 \caption{
Example quintessence models, showing the energy density (top) and equation of state (bottom) for \textit{thawing} and \textit{freezing} dynamics (left/right panel respectively). 
\textbf{Left panel:} Monomial quintessence behaves like a cosmological constant initially ($X\ll V\Leftrightarrow w_{\rm DE}\approx -1$), but becomes dynamical when DE becomes non-negligible.
\textbf{Right panel:} Tracker quintessence follows the dominant matter component initially until the field reaches a minimum of the potential. The larger the exponent coefficient $\Lambda$, the lower the amount of early DE and the earlier the minimum of the potential is reached. 
Note that the tracker behaviour requires an initially sizeable $\mathcal{E}_{\rm ini}$: a very small value (e.g. $\sim H_0^2$) would lead to a thawing model.
 \label{fig:quintessence_examples}}
\end{figure}

Quintessence models can be divided into two main categories \cite{Caldwell:2005tm}, depending on whether they evolve from/into an effective cosmological constant 
\begin{itemize}
 \item \textit{Thawing} models start with $X\ll V$, then the scalar starts evolving when DE becomes non negligible. An example of this is given by \textit{monomial quintessence}, with
 \begin{equation}\label{eq:monomial_V}
  V(\phi) = V_0\cdot \phi^N\,.
 \end{equation}
 The models behaves very close to a cosmological constant initially, deviating at late times (left panel of Fig. \ref{fig:quintessence_examples}). 
 \item \textit{Freezing} models are intially dynamical (e.g.~$X\sim V$), until the field slows down, dominates and accelerates the expansion. An example is given by \textit{tracker quintessence}, with 
 \begin{equation}\label{eq:tracker_V}
  V(\phi) = V_0\cdot \frac{e^{\lambda \phi}}{\phi^N}\,.
 \end{equation}
 Initially, the field's energy density tracks the dominant matter component (hence the name, see Fig. \ref{fig:quintessence_examples}, right panel), producing non-zero \textit{early dark energy}.
 This solution is described in \cite[Sec V]{Copeland:2006wr}.  
 Once the field reaches the minimum of the potential $\phi_{\rm min}=N/\lambda$, it starts oscillating around $\phi_{\rm min}$ until the Hubble friction damps the oscillations into an effective cosmological constant with $\mathcal{E}\approx V(\phi_{\rm min})\gg X$.
\end{itemize}
While the classification between thawing and freezing is rather general, the details are tied to the specific model (for other examples see \cite{Marsh:2014xoa}). 
These details are often critical to understand the initial and final conditions of the model.

Two parameters play a very important role: the initial value of the field, $\phi(\tau_{\rm ini})$, and the energy scale of the potential $V_0$ (in Eqs. (\ref{eq:monomial_V}, \ref{eq:tracker_V}).
In most cases, the final condition for $\Omega_{\rm DE}^{(0)}$ can be obtained by varying  $V_0$, starting with a guess value $V_0\sim \Omega_{\rm DE}^{(0)}H_0^2$ (monomial) or $V_0\sim \Omega_{\rm DE}^{(0)}V(\phi_{\rm min})$ (tracker).
However, whether this is possible depends in general on the rest of the parameters: for instance, if $\phi(\tau_{\rm ini})$ is sufficiently close to a minimum and $V(\phi_{\rm min})=0$ the only way to obtain a sizeable amount of DE is to choose a different initial field value (this is the case in monomial quintessence \ref{eq:monomial_V}).%
\footnote{More generally, a model for which the abundance crosses $\Omega_{\rm DE}^{(0)}$ at \emph{any} $z>0$ will not be able to satisfy the final conditions. This is because $w>-1$, and hence the energy density of quintessence never grows $\mathcal{E}^{\prime}\leq 0$.
Note that a change of $\phi(\tau_{\rm ini})$ can often be absorbed by redefining other model parameters, such as $V_0$ in the above examples.}

The initial field velocity, $\phi^\prime(\tau_{\rm ini})$, is typically irrelevant for quintessence. 
In freezing models like tracker quintessence (\ref{eq:tracker_V}), $\phi^\prime(\tau_{\rm ini})$ is set by early time fixed-point solution in which the scalar mimics the dominant matter component. Larger values lead to rapid kination until the tracker solution is reached, while smaller values lead to dynamics that remain frozen until $V\sim \rho_m$, after which the scalar starts behaving like a tracker. If the field's potential energy is frozen to a value $V\sim \Omega_{\rm DE}^{(0)}H_0^2$, then the thawing behaviour is recovered. 
In thawing models a large initial kinetic energy will be loss very rapidly, until dynamics are restored when $V\sim \rho_m$. 
Any sizeable residual kinetic energy requires that quintessence dominates in the early universe, cf. Eq. (\ref{eq:quint_kination}), dramatically impacting recombination and big-bang nucleosynthesis, and violating bounds on the abundance of additional species in those epochs.

\paragraph{Brans-Dicke}\label{sec:brans_dicke}

Brans-Dicke theory \cite{Brans:1961sx} is characterized by the following action
\begin{equation}
G_2 = \frac{\omega_{\rm BD}}{\phi}X - \Lambda\,,\quad
G_4 = \phi \frac{\Mpl}{2}\,,\quad
G_3 = G_5 = 0\,.
\end{equation}
The modifications of gravity stem from the  coupling between the scalar field and the Ricci scalar, which is characteristic of a broad range of gravitational theories, including chameleon, symmetron and $f(R)$ models \cite{Burrage:2017qrf}.%
\footnote{$f(R)$ gravity is Brans-Dicke with $\omega=0$, but with a potential function $\Lambda\to V(\phi)$ related to $f$ \cite{Chiba:2003ir}. Brans-Dicke theories and some generalizations can be recast as GR plus an explicit conformal coupling between matter and the scalar field \cite{Flanagan:2004bz}, while such a relation is not possible in more general Horndeski theories (\cite{Zumalacarregui:2012us,Bettoni:2013diz}).}
This coupling is modulated by the Brans-Dicke parameter $\omega_{\rm BD}$. The GR limit is recovered for $\omega_{\rm BD}\to\infty$: in that case the kinetic term is large and the scalar field becomes effectively ``frozen'' (i.e.~it becomes increasingly difficult ot excite the scalar field).
The canonical Brans-Dicke theory has no potential term. A cosmological constant $\Lambda$ has been added to act as dark energy. The final condition is set by varying $\Lambda$ to obtain the desired $\Omega_{\rm DE}^{(0)}$.

The initial field velocity can be set to $\phi^\prime(\tau_{\rm ini})\approx 0$ via an approximate recovery of shift-symmetry in the radiation era. Brans-Dicke violates shift-current via two terms. The first term is the coupling between $\phi$ and the Ricci scalar $\bar R = 
6\left(2H^2+H^\prime/a\right)\propto \rho_m-3p_m +\mathcal{E}-3\mathcal{P}$, which vanishes for ultra-relativistic matter (i.e. if $p=\rho/3$). 
The second term involves $G_{2,\phi}\propto X/\phi^2$ and is proportional to the kinetic energy, hence subdominant unless $X$ is very large initially.
This implies that in the radiation era the scalar field approximately recovers shift symmetry, up to the sub-dominant contributions from non-relativistic matter ($\rho_m \ll \rho_r$) and the field's kinetic energy. 
Thus, $\phi^\prime(\tau_{\rm ini})\approx 0$ is a good approximation for non fine-tuned initial conditions, similarly to many quintessence models, c.f.\ Eq.~(\ref{eq:quint_kination}).

Brans-Dicke theories have been used as a benchmark model to test gravity in different systems.
The coupling to gravity $G_4\propto \phi$ produces two effects. First, the background value of $\phi$ modulates the gravitational force, i.e.\ how much curvature is produced by unit mass. 
This can be seen via the expression for cosmological strength of gravity
\begin{equation}\label{eq:brans_dicke_cosmo_strength}
\frac{G_{\rm FRW}}{G} = \frac{1}{M_*^{2}(\phi)} = \frac{1}{2G_4(\phi)} = \frac{1}{\phi}\,,
\end{equation}
as it can be read off the modified Friedmann equation.
Second, scalar-field excitations mediate an additional, attractive interaction. The small-scale gravitational force experienced by \textit{non-relativistic} test particles is given by \cite{Brans:1961sx,Clifton:2011jh}
\begin{equation}\label{eq:brans_dicke_local_force}
\frac{G_{\rm NR}}{G} = \frac{1}{M_*^2(\phi)} \frac{4+2\omega_{\rm BD}}{3+2\omega_{\rm BD}}\,,
\end{equation}
were the first factor accounts for the strength of gravity and the second factor accounts for the scalar force. On the other hand, ultra-relativistic species (e.g.\ light) do not experience the scalar force, with $G_{\rm UR}/G=M_*^{-2}=G_\text{FRW}/G$. This difference implies that Brans-Dicke gravity violates the weak equivalence principle and is therefore constrained by tests of gravity.
In fact, since the canonical Brans-Dicke theory lacks any screening mechanism, the strongest tests come from astrophysical systems  \cite{Berti:2015itd}. Current cosmological bounds set $\omega_{\rm BD}\gtrsim 700$ \cite{Avilez:2013dxa} (or $\gtrsim 300$, depending on dataset and priors \cite{Ballardini:2016cvy}), but future surveys will reach a level comparable to astrophysical tests $\omega_{\rm BD}\gtrsim 1.7\cdot 10^4$ \cite{Alonso:2016suf}.

The above expression motivates several possible choices of initial or final conditions for the scalar field:
\begin{enumerate}
 \item $\phi(\tau_{\rm ini}) = 1$ fixes the cosmological strength of gravity at early times, leading to a standard recombination history and CMB predictions. 
 \item $\phi(\tau_0) = 1$ fixes the cosmological strength of gravity today, leading to a standard $\Lambda$CDM behaviour at late times.
 \item $\phi(\tau_0)=\frac{4+2\omega_{\rm BD}}{3+2\omega_{\rm BD}}$ ensures the correct value of Newton's constant, as measured by laboratory experiments. 
\end{enumerate}
Choice (1) sets the initial condition for the field $\phi(\tau_{\rm ini})$ directly.
Choices (2,3) require adjusting the initial field value $\phi(\tau_{\rm ini})$.
All choices are in addition to fixing $\Omega_{\rm DE}^{(0)}$ by varying $\Lambda$.

\section{Initial Conditions for Cosmological Perturbations\label{sec:ICs}}

The new feature of \hiclass compared to CLASS is the presence of an extra degree of freedom in the perturbation equations. The initial conditions for this field's perturbations must thus be appropriately set. Depending on the  values of the $\alpha$-functions and the energy density in dark energy in the \emph{early} universe (i.e.\ early dark energy, e.g.\ \cite{Pettorino:2013ia}), these initial conditions may also affect the configuration of the standard species at early times. In this release of \hiclassx, we introduce the proper initial conditions, allowing for a study of consistent early modified gravity models, as well as a dynamical elimination of models of modified gravity which exhibit instabilities in the early universe.

Einstein-Boltzmann codes such as \hiclass begin solving the linear perturbation equations mode by mode at a large redshift, some initial conformal time $\tau_*$, significantly before recombination, but also significantly after reheating, with some appropriate initial conditions for all the species. Observations of the CMB require that the dominant mode is the adiabatic one ($<2\%$ contribution from isocurvature modes to the CMB power spectrum \cite{Akrami:2018odb}). For this mode, the curvature perturbation $\zeta$ is conserved on superhorizon scales. Thus the value of $\zeta$ at $\tau_*$ --- for the modes which are still superhorizon at that time --- is the same as when it was imprinted upon horizon exit during the initial phase of the evolution of the Universe, e.g.\ inflation. It is only this fundamental assumption that allows us to relate the observations of the CMB to the properties of this primordial power spectrum. As we modify the theory of gravity, we must bear in mind that this connection must be maintained in order to interpret the output of the code in the standard manner.

In this section, we show that standard $\Lambda$CDM physics in the radiation era remains approximately unchanged if two conditions are satisfied: 
\begin{enumerate}
	\item[(i)] the evolution of the scalar field fluctuations and the gravitational field is driven by the standard matter species collapsing on superhorizon scales
	\item[(ii)] the modification of gravity is such that no relevant new timescales are introduced through the dynamics of the scalar and thus scaling attractor solutions can actually be found.
\end{enumerate}

The first condition is violated whenever the dark-energy isocurvature modes grow faster than the adiabatic mode: in such a case the configuration is dominated by the initial conditions for the scalar field set at some primordial juncture. This is a typical outcome when the scalar is tachyonic in the radiation era (cf.\ Ref.~\cite[section 2.3]{Zumalacarregui:2016pph}). The most important result of this section is to find conditions under which this happens so that such models can be discarded. It turns out that models with such pathological features can easily be constructed, even starting from a covariant action (e.g.\ even a covariant galileon with some particular choices of parameters $c_i$, see section \ref{sec:galileon_example}). We discard such models since not only do they need to be augmented with some sort of physics which would set the appropriate primordial initial conditions, but also because these isocurvature modes can get sourced from numerical noise and in any case end up dominating the solutions at late times, leading to a loss of predictivity.

The second is essentially the statement that $(aH)^{-1}$ is the only relevant timescale in superhorizon dynamics and therefore all the species settle to universal solutions (or alternatively, other timescales are much longer or shorter and thus irrelevant for setting initial conditions for the mode). This requirement is on some level already violated by the presence of the timescale of matter-radiation equality, which in the end leads to the existence of (matter) isocurvature modes. For a simple well-behaved universal attractor, one needs to initialise the modes deep in the radiation era.

A completely arbitrary Horndeski model will in general exhibit a variation of the $\alpha$-functions with its own time-scale even during radiation domination and therefore a universal attractor for the scalar field might not even exist. If the field is subdominant gravitationally, then it is just a technical problem for finding the right initial conditions for the scalar field. If it is not --- this time variation will impact all the other species and therefore it would in general be impossible to find universal initial conditions for any of them.

We thus concentrate on two quite general classes of models in which such a universal attractor exists: the limit in which deviations from GR at early times are small and do not introduce corrections to the dynamics of the standard species (Sec. \ref{sec:EFIC}), and then when deviations from GR are significant but time-independent (Sec.~\ref{sec:GAIC}). These conditions are specific, but should be at least approximately satisfied in any reasonably model which is not very tuned.

For the models with a dominant adiabatic mode, we find that the correct superhorizon configurations are in any case approached within a decade of initialisation in the scale factor. This allows us to delay the time from which the modes are evolved without sacrificing accuracy, although this does not significantly improve the run time. The proper identification of dark-energy isocurvature-mode dominated models is thus the main result of this section.

\subsection{Review of Perturbation Initial Conditions in General Relativity}
Ma \& Bertschinger \cite{Ma:1995ey} showed how to derive the appropriate adiabatic initial conditions deep in the radiation-domination era: the Universe is dominated by a photon-baryon fluid and neutrinos redshifting with a constant ratio of energy densities $R_\nu \equiv \rho_\nu/(\rho_\gamma+\rho_\nu)$. Pressure support for radiation disappears on scales outside the sound horizon, $k/aH\ll 1$, and any radiation overdensities collapse under the force of gravity.
\paragraph{The adiabatic mode:}
In synchronous gauge, the radiation-gravity system can be written down as a fourth-order differential equation for the metric potential $h$ (for each mode in Fourier space) with power-law solutions,
\begin{equation}
	h = A + B (k\tau)^{-2} + C(k\tau)^2 + D(k\tau) \label{eq:MBICs}\,.
\end{equation}
It can be shown that the constant ($A$) and decaying ($B$) solutions are gauge modes, resulting from the fact that the gauge is not completely fixed in synchronous gauge comoving with dark matter. The $C$ and $D$ solutions are the physical solutions. The $D$ solution actually represents a decaying mode for curvature $\eta$, so it is neglected. The $h=C(k\tau)^2$, with $C$ a dimensionless constant, is the one of relevance to us since it is the dominant mode at late times, it corresponds to conserved curvature superhorizon and is the adiabatic mode for perturbations. Its normalisation is chosen to as to give a fixed value of $\eta=1$ to the curvature perturbation superhorizon, i.e.\ $C=1/2$.

Given this growing solution for the potential $h$, the configuration of the individual matter species can be obtained by solving the individual Euler and continuity equations, assuming this solution for $h$ and expanding in $k\tau$ as an order parameter. This leads to the standard adiabatic initial conditions for the species, once the choice is made to neglect the homogeneous part of the solutions, leaving only the particular solution:
\begin{align}
	\delta_\text{c,b}&=\frac{3}{4}\delta_{\gamma,\nu} = -\frac{h}{2} = -\frac {1}{4}k^2\tau^2\,,\label{eq:adiabICs}\\
	v_\gamma&= v_\text{b}= -\frac{1}{36}k^2\tau^3 = \frac{15+4R_\nu}{23+4R_\nu}v_\nu \,,\notag\\
	\sigma_\nu &= \frac{2}{3(15+4R_\nu)}k^2\tau^2 \notag\,,
\end{align}
for the energy-density perturbation, $\delta_i$, the velocity potential $v_i$ and the neutrino anisotropic stress $\sigma_\nu$, with the subscript $i\in \{\text{c},\text{b},\gamma,\nu\}$ denoting the cold dark matter, baryons, radiation and neutrinos respectively.%
\footnote{$v_c=0$ defines the synchronous gauge comoving with dark matter.} %
These are the superhorizon attractor solutions for the species during radiation domination which follow the gravitational field created by the dominant collapsing radiation. Finally, the other scalar potential in the synchronous gauge, $\eta$, can be shown to remain constant at leading order on superhorizon scales,
\begin{equation}
	\eta=1 - \frac{5+4R_\nu}{12(15+4R_\nu)}k^2 \tau^2\,.
\end{equation}
At superhorizon scales, $\eta$ is equal to the curvature perturbation, the amplitude of which is the major prediction of inflationary theory. Its normalisation is the amplitude of the primordial fluctuations, so for the transfer functions being calculated here it is taken to be unity.

\paragraph{Isocurvature modes}

In addition to the adiabatic mode above, there are also four isocurvature modes, which in the limit $k\tau\rightarrow0$ have $\eta=0$, but one of the other  variables is non-zero: $\delta_\text{c}$ for CDI (CDM density isocurvature), $\delta_\text{b}$ for BI (baryon density isocurvature), $\delta_\nu$ for NID (neutrino density isocurvature) and $v_\nu$ for NIV (neutrino velocity isocurvature). These modes can be generated by new physics but there is no evidence that any are present in the CMB anisotropies, and they are limited to approximately a maximum 2\% contribution to the CMB power spectrum \cite{Akrami:2018odb}.

The method to derive their superhorizon attractor solutions is a power-law expansion in conformal time in units of time of matter-radiation equality $\tau/\tau_\text{eq}$ and was comprehensively derived in Ref.~\cite{Bucher:2000kb} and we refer the reader for detail there. For our purposes, what is relevant is that all of these isocurvature modes have the effect of producing an evolution of the metric potential $h$, which, to the relevant precision, can be expanded as a sum of integer power laws in $\tau$.
	
Let us now list, following Ref.~\cite{Bucher:2000kb} improved by some higher-order corrections, the evolution of $h$ for all the five types of modes in $\Lambda$CDM:
	\begin{align}
	&\text{Adiabatic:} 	&& h = \frac{(k\tau)^2}{2}-\frac{1}{10}\omega k^2\tau^3\,, \label{eq:allsols} \\
	&\text{CDI:}		&& h = R_\text{c} \omega \tau - \frac{8}{3}R_\text{c} \omega^2 \tau^2\,,\notag\\
	&\text{BI:}			&& h = R_\text{b} \omega \tau - \frac{8}{3}R_\text{b} \omega^2 \tau^2\,,\notag\\
	&\text{NID:}		&& h = \frac{R_\nu R_\text{b}}{40 R_\gamma} \omega k^2\tau^3 -\frac{R_\nu k^4\tau^4}{36(15+4R_\nu)} - \frac{R_\nu R_\text{b}(R_\nu+R_\text{b})\omega^2k^2\tau^4} {128R_\gamma^2} \,,\notag\\
	&\text{NIV:}		&& h = \frac{9R_\nu R_\text{b}}{32R_\gamma} \omega k\tau^2 - \frac{3R_\nu R_\text{b}(3R_\text{b}+5R_\gamma)}{160R_\gamma^2}\omega^2 k\tau^3 - \frac{4R_\nu}{15(5+4R_\nu)}k^3\tau^3\,,\notag
	\end{align}
	where the $R_{\nu,\gamma}\equiv \rho_{\nu,\gamma}/(\rho_\nu+\rho_\gamma)$ are density fractions for radiative species, while $R_\text{c,b}\equiv \rho_\text{c,b}/(\rho_\text{b}+\rho_\text{c})$ are the equivalent for the non-relativistic species. The time-scale $\omega\equiv a(\rho_\text{c}+\rho_\text{b})/\sqrt{\rho_\gamma +\rho_\nu}$ is a dimension-one constant depending only on cosmological parameters, determining how close to matter-radiation equality $\tau_*$ is.%
\footnote{Note the CLASS definition of density here, $H^2 = \rho$.} %

This power-series expansion is only valid provided that both $k\tau\ll1$ and $\omega\tau\ll 1$. For the neutrino isocurvature modes, the hierarchy is not necessarily obvious since the relative size of $k$ and $\omega$ depends on the mode and the time of initialisation of it by \hiclassx, so the higher-order terms must be included for precision.

\vspace{0.5cm}

Introducing the Horndeski scalar field does not change these, provided that radiation remains dominant, i.e.\ is approximately the only source of the gravitational field. The appropriate solution for the scalar-field perturbation $V_X$ can be obtained by solving the equation of motion assuming the solutions \eqref{eq:adiabICs} as an external source. We will call this the \emph{external-field attractor initial conditions}. On the other hand, if the Horndeski scalar contributes significantly (i.e.\ it modifies gravity in the early universe), then an appropriate self-consistent solution, involving both the radiation and the Horndeski scalar, must be developed. We will call this setup the \emph{gravitating attractor initial conditions} and show that this modifies the solutions \eqref{eq:adiabICs}. 

\subsection{External-Field Attractor Initial Conditions\label{sec:EFIC}}

The \emph{external-field attractor} initial conditions are relevant when the scalar field does not source the gravitational potential during radiation domination, around $\tau_*$ and the potential $h$ is driven in the standard manner for the mode under consideration by the matter species collapsing on superhorizon scales. This requires that, at that time, $\OmDE, \alpha_i\ll 1$. We assume that this evolution can be described as a (sum of) power law(s) $h\propto\tau^{n_h}$. We obtain the solution for $h = \mathcal{A}_{n_h} \tau^{n_h}$ (e.g.\ $n_h=2$ for the standard adiabatic mode, giving $\mathcal{A}_2=k^2/2$) by assuming that there is no contribution of $V_X$ in the Einstein equations.
 
On the other hand, the equation of motion for $V_X$ mixes with time derivatives of $h$ and $\eta$. We eliminate $\eta$ for $h$ and therefore obtain an equation purely for $V_X$ and $h$. We use the previously obtained power-law solutions for $h$ to reduce the scalar equation of motion to a second-order differential equation for $V_X$ with an explicit function of time for the source on the right-hand side:
\begin{equation}
	\tau^2 V_X'' + B_1 \tau V_X' + B_2 V_X = -B_3 \mathcal{A}_{n_h} \tau^{n_h+1} \,.
\end{equation}
The $B_i$ coefficients are functions of $\OmDE$ and the $\alpha_i$ and their derivatives, with only $B_3$ also depending on $n_h$  (the full expressions are very involved and can be found in the code, below we present a simple example).

If $B_i$ are constant, then we have simple power-law in $\tau$ solutions for the $V_X$. Time variation in the $B_i$ coefficients implies that there is also a time-scale other than $\mathcal{H}^{-1}$ in the problem and the attractor is at best approximate.

On the assumption that the $B_i$ be constant, the solution we are interested in is the particular solution
\begin{equation}
	V_X = A_{n_h} \tau^{n_h+1}\,,\qquad A_{n_h}\equiv -\frac{\mathcal{A}_{n_h} B_3}{n_h^2 + n_h(B_1+1)+B_2}\,. \label{eq:vxsol}
\end{equation}
with $n_h=2, \mathcal{A}_2=k^2/2$ for the dominant adiabatic mode. If there are multiple power laws in the source coming from the expansion, then the total solution is just a sum of individual ones, since the equation is linear.

\vspace{0.5cm}

Additional solutions exists, satisfying the homogeneous equation, left hand side of Eq.~(\ref{eq:vxsol})
\begin{equation}
	V_X \propto \tau^{n_\pm+1}\,,\qquad n_\pm = -\frac{1+B_1}{2}\pm \sqrt{\frac{(1-B_1)^2}{4}-B_2} \,.
\end{equation}
These are the dark-energy isocurvature modes which have their initial condition set independently of the standard modes of interest. If $\mathfrak{Re}(n_+)>2$, then they grow more quickly than the adiabatic solution and eventually dominate. Quite generically such situations occur when $B_2<0$ which implies that the scalar $V_X$ is tachyonic and therefore are really a sign that there is an instability at large scales: the background of the scalar implied by the choice of $H, \OmDE$ and the $\alpha$'s is not a good one over the timescales of the evolution of the universe. On the other hand, this is just a Jeans instability, arrested at small scales by our stability requirement that the sound speed be positive, so one may argue that it is not a fundamental problem and let observations decide.

Nonetheless, for reasons of predictivity, the default choice is to exclude models with dark-energy isocurvature modes growing relatively to the adiabatic solution during the radiation era, i.e.\ such models where $\mathfrak{Re}(n_+)>2+\epsilon$, with some user-specified small parameter $\epsilon$. Otherwise, if even a small admixture of the DE isocurvature mode is present initially, it will completely dominate the evolution of the universe when the scalar field begins to gravitate, unless another mechanism is introduced to erase it at small scales. The prediction of the magnitude of this effect would not be universal but rather would depend on $\tau_*$, the initial time at which \hiclass begins to evolve the mode.
 The appropriate solution for $V_X$ is given by a sum over with each term in \eqref{eq:allsols} giving a source for \eqref{eq:vxsol}.

\begin{figure}[ht]\begin{center}
	\includegraphics[width=0.5\textwidth]{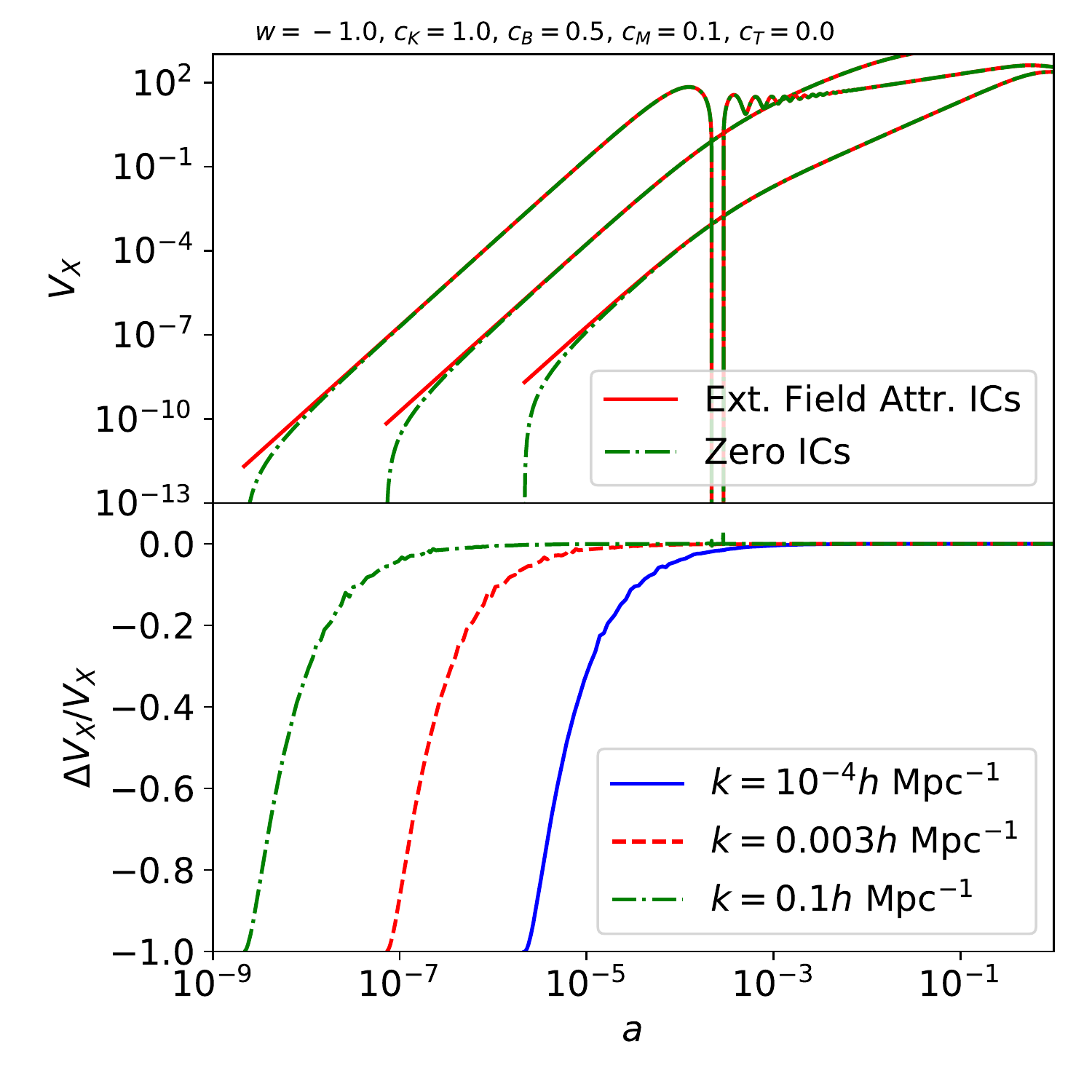}
	\caption{Comparison of evolution of $V_X$ with two choices of initial conditions for the scalar field for three $k$-modes ($k=0.1,0.03,0.0001 h$~Mpc$^{-1}$) for the adiabatic mode in a model with $w=-1$ and $\alpha_i = c_i a$ with non-zero $c_K=1.0, c_B=0.5, c_M=0.1$ proportional to the scale-factor. The newly implemented external-field attractor ICs put the field $V_X$ directly on the attractor solution, while allowing the mode to start with $V_X=0$ leads to a convergence toward the attractor lasting around one decade in $a$. In the bottom panel we show the relative difference between $V_X$ set on the external-field-attractor solution and zero ICs for different $k$-modes. \label{fig:extfldICs}}
\end{center}\end{figure}

Since, by assumption, the scalar field does not gravitate, none of the coefficients in Eqs~\eqref{eq:allsols} contain any corrections from $\Omega_{\rm DE}$ or $\alpha_i$. On the other hand, the $B_i$ coefficients only depend on the model of gravity and $n_h$. The full expressions for a general Horndeski model are very lengthy and can be found in the code. But here we show an explicit example in  the simplified case $\aT=\aM=0$ and $M_*=1$:
\begin{align}
	B_1 &=	\frac{\tau \aK'}{\aK} + \OmDE(1-3w)	\,,\\
	B_2 &\approx -2+18\frac{(1+w)\OmDE-\aB}{\aK} +\frac{\tau(\aK'+6 \aB')}{\aK}\,, \notag\\
	B_3 &\approx -n_h\frac{ \left((3(1+w) \OmDE +(n_h-1)\aB)+ \tau\aB'\right)}{2\aK}\,.		\notag
\end{align}
where we present only the leading order terms. 
We can see from this that the conditions for the approximate constancy of the $B_i$ and therefore for the existence of the attractor are that ratios of those $\alpha$'s and $\OmDE$ that are dominant are approximately constant and that the $\alpha$'s evolve as power laws of conformal time $\tau$. This is the case when parameterisations such as $\alpha_i \propto \OmDE$ or $\alpha_i\propto a^p$, with $p$ some constant power are used. This is also the limit in full models derived from an action such as the covariant galileon. In principle, only the slowest-varying parameters need to have a common time-dependence since the others will be negligible sufficiently early, at least if they evolve monotonically. In figure~\ref{fig:extfldICs}, we present a comparison of the evolution of $V_X$ for the external-field attractor initial conditions for the adiabatic mode comparing it with initialising with $V_X=0$, showing that we place the field on its attractor correctly, avoiding a relatively slow approach of the incorrect solution.

We note here that one can recover the results obtained in Ref.~\cite{Ballesteros:2010ks} for the case of a perfect-fluid dark energy with constant parameters $w$ and $c_\text{s}^2$ by setting $\aK=3\OmDE(1+w)/c_\text{s}^2$, $\aB=\aT=\aM=0$ and $M_*^2=1$.

The implementation in \hiclass of these external-field attractor ICs is quite generic: it assumes the $B_i$ are constant but uses the full expressions, evaluating the derivatives of the $\alpha$'s numerically. Thus if the model being considered happens to achieve the approximate constancy of the $B_i$ in some non-trivial fashion (e.g.\ a special cancellation), the implemented code should still give the approximately correct initial conditions. 

Using these initial conditions when the variation of the $B_i$ is fast, or some of the $\alpha$'s or $\OmDE$ are of order 1 will not give the correct result. For example, initialising the evolution during a transient feature in the behaviour of the $\alpha$-functions can lead to an order-one misprediction of the initial value of $V_X$. Forcing the mode initialisation to occur at an earlier time is simplest way to properly evolve through such a feature. Moreover, such features can source dark-energy isocurvature modes, but these decay back to the universal external-field-attractor initial conditions in models where the isocurvatue modes grow less quickly than the adiabatic, which is the default requirement we impose on the models.

We also stress that it is key that for isocurvature modes, all the modes are initialised sufficiently before matter-radiation equality. The series expansion fails there and if initialised too close the final observables can vary significantly from the correct solution.

We also note here that $B_2$ is the effective mass of the scalar field fluctuation, in the superhorizon limit during radiation domination. When it is large, $B_2\gg1$, the initial condition for $V_X\rightarrow0$ (at least for some appropriate $B_3$) and therefore any modification of gravity is switched off at early times. This is the case, for example, in models such as $f(R)$ gravity \cite{Song:2006ej}. Despite this simple initial condition, the scalar perturbation evolves by oscillating rapidly during radiation domination as a result of the large mass, resulting in a very slow integration by \hiclassx.  For such cases, we have implemented the quasi-static approximation which removes the dynamics from the evolution of the scalar field at judiciously chosen time intervals to accelerate the computation without affecting the final results, see section~\ref{sec:QS}. For the purposes of this section, this quasi-static approximation can be thought of as an alternative IC scheme for such models. 

\subsection{Gravitating Attractor Initial Conditions: Adiabatic Mode} \label{sec:GAIC}

The initial conditions derived in section~\ref{sec:EFIC} apply when gravity is not modified: i.e.\ they describe the behaviour of the scalar field fluctuation $V_X$ during radiation domination when the gravitational potential is being driven by relativistic matter collapsing superhorizon in the standard manner. The scalar is assumed not to backreact onto the rest of the matter content and therefore all the other initial conditions are standard. It is essentially the test-field approximation.

In this section, we describe the contrary situation, when the gravitational backreaction of the scalar field is significant, i.e.\ gravity is modified by the scalar field already during radiation domination. We compute the self-consistent initial conditions for all the matter species, including the scalar field $V_X$. This is not the usual situation, but it applies in the case of early modified gravity, i.e.\ it allows one to test the extent to which gravity is allowed to be modified e.g.\ during recombination (see Refs.~\cite{Pettorino:2013ia,Brax:2013fda,Lin:2018nxe} for some examples). We call this the \emph{gravitating attractor initial conditions}. At this time, we have only implemented the gravitating-attractor initial conditions for the adiabatic mode. Isocurvature gravitating-attractor ICs are not yet available, the code will return an error if this combination is requested. The rest of the discussion in this section will assume that the mode is adiabatic.

The technical difference with respect to the \emph{external-field attractor} is that, in this case, at least some of the $\{\OmDE,\alpha_i\}$ are not negligible during radiation domination. Since in the equations of motion the kinetic terms for the gravitational potential $h$ and the scalar field fluctuation $V_X$ mix with coefficients involving $\alpha_i$, it is now inappropriate to solve the system in the standard manner, assuming that the solution for $h$ does not depend on $V_X$. Instead, the $(V_X,h)$ system must be solved simultaneously. Since, in addition, the coefficients of the $(V_X,h)$ system are generically not homogeneous in the $\{\OmDE,\alpha_i\}$,  the only generic possibility of obtaining an attractor similar to the standard case (i.e.\ one with a single timescale $(aH)^{-1}$) is when those $\alpha$'s which are non-negligible are constant. In our solution, we thus assume that all of $\{\OmDE, \alpha_i\}$ are constant, which, for example, implies that the scalar during radiation domination tracks the radiation energy density, $w=1/3$. 

We again present only the simple subcase of kinetic gravity braiding with constant $\alpha$ parameters, \begin{align}
	&\OmDE,\aK,\aB=\text{const}\,, \label{eq:KGB}\\
	&\aM=\aT=0\,,\notag\\
	&M_*^2=1\,,\notag
\end{align}
for readability. The \hiclass code contains the full general expressions for Horndeski. We will comment where appropriate to discuss the additional complications in this general case.

For KGB, eq.~\eqref{eq:KGB}, the $(V_X,h)$ system during radiation domination becomes
\begin{align}
	2\aK \tau^2 &V_X'' - \aB \tau^3 h'' - 4\OmDE \tau^2 h' + 4(12\OmDE -\aK -9 \aB)V_X=0 \,,\label{eq:kgbRDeqs} \\
	2D \tau^4 &h''' -2\aK(12\OmDE+\aK-3\aB )\tau^2 V_X'' + (6\aK + \aK\aB+6\aB^2+12\OmDE\aB)\tau^3 h'' + \notag\\ &(-6\aK+\aK\aB+4\OmDE(2\aK+3\aB))\tau^2 h' - 2(\aK-6\aB)(\aK+9\aB-12\OmDE) \tau V_X' \notag\\
	&+ \notag 2(\aK-6\aB)(\aK+9\aB-12\OmDE)V_X =0\,,\notag 
\end{align} 
where $D\equiv\aK+\frac{3}{2}\aB^2>0$ is the normalisation of the kinetic term for the scalar degree of freedom. Note that we have used a slightly different method here to arrive at the equations than the solution \eqref{eq:MBICs} in Ref.~\cite{Ma:1995ey}: we eliminate all $\eta$ and density perturbations $\delta$, leaving the system containing only the metric potential $h$ and the velocity potentials for the scalar $V_X$ and for the matter species. We then take the $k\rightarrow0$ limit, removing the radiation/neutrino velocity potential terms $v_{\gamma,\nu}$. In the GR limit, this system is lower by one order than the one considered by \cite{Ma:1995ey}, with three of the four solutions in Eq.~\eqref{eq:MBICs}: the adiabatic $C$, and the gauge $A,B$, but not the decaying $D$ solution. This is enough for our purposes, since it recovers the adiabatic mode and also the DE isocurvature modes.%
\footnote{Note that the exponents $n_\pm$ are in general not integers and therefore these modes do not appear in the standard approach e.g.~Ref.~\cite{Bucher:2000kb} where a power-law expansion is used. A resummation method generalising this approach, such as presented in \cite{Kopp:2016mhm} would have to used to see these modes in such an expansion.}%

The two DE isocurvature solutions of eqs~\eqref{eq:kgbRDeqs} are
\begin{align}
		h &\propto \tau^{n_\pm}\,,\quad V_X \propto \tau^{n_\pm+1}\,, \\ 
		&n_\pm = -\frac{1}{2}\pm \frac{\sqrt{D-8(1-\OmDE)(12\OmDE-\aK-9\aB)}} {2\sqrt{D}}\,. \notag 
\end{align}
Again, models exist in which $\mathfrak{Re}(n_+)>2$ and the scalar field isocurvature grows faster than the adiabatic mode, as given by the values of $\{\OmDE, \alpha_i\}$ during radiation domination.

Typically this is a result of some sort of tachyonic instability in the scalar field for this choice. In inflation, if the scalar is light, it will obtain its own isocurvature initial conditions, and this mode will evolve to eventually dominate the gravitational field over the standard adiabatic perturbations. 

Thus, to retain the standard connection of cosmological observables to the properties of the primordial power spectrum the default choice is to disallow such parameters enforcing $\mathfrak{Re}(n_+)<2$ to keep the DE isocurvature modes subdominant with respect to the adiabatic mode.
\hiclass performs this test, in addition to absence of ghost and gradient instabilities ($D>0$ and $Dc_\text{s}^2=8\OmDE-\aB-\frac{\aB^2}{2}>0)$.
These constraints can be satisfied simultaneously in various disjoint ranges of parameters which can be easily obtained from the above conditions, but are not particularly illustrative and we will not quote.%
\footnote{Note that positive energy density, $\OmDE>0$, is \emph{not} a requirement: stable configurations with $\OmDE<0$ are possible when the DE energy-momentum tensor is not of perfect-fluid form, as happens whenever any of $\aB,\aT,\aM\neq0$. One should bear this in mind when setting priors for the early Universe \cite{Deffayet:2010qz}.} %

Models with parameter choices which do pass these tests have an adiabatic solution very similar to the standard one, but the coefficients are somewhat modified. In particular, we have, again for the KGB sub-case,

\begin{align}
	&h= \frac{C_1}{2C_2}(k\tau)^2\,,\quad V_X = \frac{(4\OmDE + \aB)}{4C_2}k^2\tau^3\,, \\
	&\delta_\text{c}=\delta_\text{b}=\frac{3}{4}\delta_\gamma=\frac{3}{4}\delta_\nu = -\frac{h}{2}\,,\notag\\
	&v_\gamma = v_\text{b} = -\frac{C_1}{36C_2} k^2 \tau^3\,,  \notag\\
	&v_\nu = -\frac{1}{36}\left(\frac{8}{15+4R_\nu(1-\OmDE)}+\frac{C_1}{C_2}\right)k^2\tau^3\,, \notag \\
	&\sigma_\nu = \frac{2}{3(15+4R_\nu(1-\OmDE))}k^2\tau^2\,,\notag\\
	&\eta = 1 + \frac{1}{12}\left( \frac{10}{15+4R_\nu(1-\OmDE)} - \frac{C_1}{C_2}  \right)k^2\tau^2 \,, \notag \label{eq:GAIC_matter_ics}
\end{align}

where to aid clarity we have defined two combinations of the $\alpha$-functions:
\begin{align}
	C_1 &= 12\OmDE + 2\aK-9\aB\,, \\
	C_2 &= 3D+(C_1-3\aK)(1-\OmDE) \,. \notag
\end{align}
It can be seen that the amplitudes for the matter species are all reduced in order to keep the curvature perturbation $\eta$ at a fixed value superhorizon. This is a result of the fact that the scalar field $V_X$ now gravitates and modifies the curvature perturbation produced by the other species. The velocity potential for the species which have anisotropic stress (e.g.\ neutrinos) is further modified compared to the standard GR case. However, whenever terms quadratic in $\{\alpha, \OmDE\}$ are negligible, $C_1/C_2\rightarrow1$ and the solutions for the matter species and metric potentials return to the standard GR results \eqref{eq:adiabICs}.

Note that  the results above naively imply that when $C_1=0$, there is some sort of cancellation. In this limit, the standard power-law solution is corrected by a logarithm term, $h\sim \tau^2 \log \tau$. This is a slowly evolving correction, resulting from an accidental cancellation, and not important for the appropriate initialisation of the system. 

Provided that the cosmological strength of gravity $M_*^2$ does not run, $\aM=0$, the adiabatic mode always has the metric potential evolving in the standard manner, $h\propto \tau^2$. However, the moment that $M_*^2$ evolves, it is no longer true that the solutions to Eqs~\eqref{eq:kgbRDeqs} are power laws, since the coefficients now contain both $M_*$ and $\aM$. Nonetheless, provided that $\aM$ is sufficiently small, the solution is close to a power law and the exponent slowly evolves with a correction linear in $\aM$. 

\begin{figure}[t]
	\begin{center}
		\includegraphics[width=0.5\textwidth]{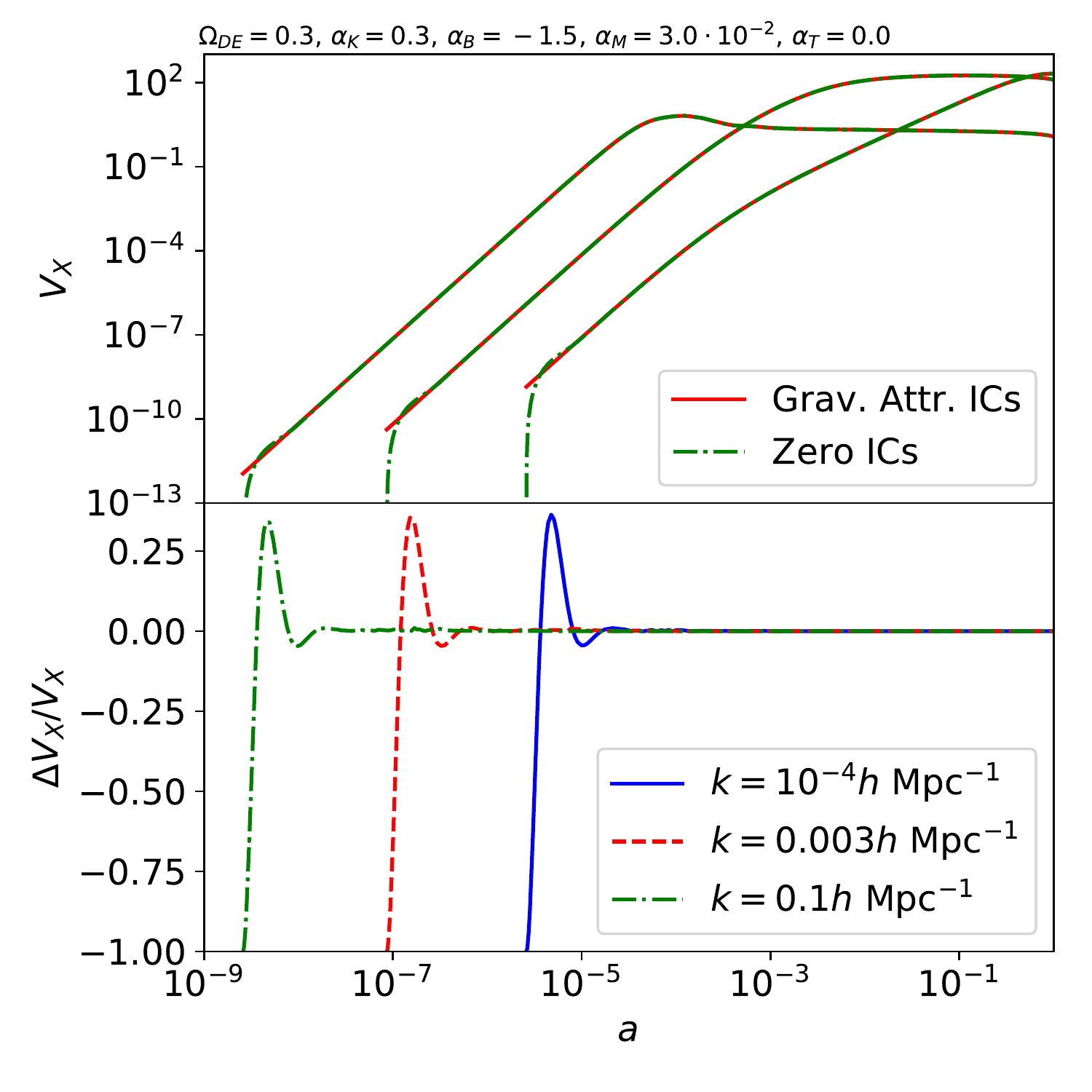}
		\caption{Comparison of evolution of $V_X$ for three choices of initial conditions for three modes ($k=(0.1,0.03,0.0001) h$~Mpc$^{-1}$) in a model with an early dark energy $\OmDE=0.3$ and non-zero $\aK=0.3, \aB=-1.5, \aM=0.03$, proportional to the dark-energy density fraction. The newly implemented gravitating attractor ICs put the field $V_X$ directly on the attractor solution. Both the external-field attractor and the naive zero ICs oscillate around the real solution for approximately a decade in the scale factor $a$. \label{fig:gravattrVx}\\ \textbf{Bottom Panel:} Relative difference between the evolution of $V_X$ on the gravitating attractor initial conditions and starting at zero. }
	\end{center}
\end{figure}

Technically, we obtain an approximate solution with $\alpha_M\neq0$, by combining the two equations~\eqref{eq:kgbRDeqs} into a single higher-order ODE, producing two versions: one for $h$ and one for $V_X$. If the solution were exact power laws, they would be the same for both. Since they are not, the approximate exponents differ slightly and we use the average of these differences to set the initial conditions in the code. We find that for small $\aM\ll1$ the system rapidly relaxes to the correct solution. We note here that despite the deviation of $h$ from the standard evolution when $M_*^2$ evolves, $\eta$ is still conserved superhorizon in the standard manner. We present the difference between the initial conditions in a case with early dark energy in figure~\ref{fig:gravattrVx}. 

\vspace{0.5cm}

We have thus presented two classes of approximate initial conditions: (i) the external-field attractor, in which the modified gravity parameters are small but can be evolving in time in the early universe and (ii) where the modified gravity parameters are constant in the early universe, but can be large. These initial conditions can be used in many other scenarios, since in a situation with non-homogeneous evolution of the parameters, only those which evolve most slowly will be relevant in the early universe and one of our two scenarios might still be applicable. \hiclass  cannot automatically set correct initial conditions if the modified gravity parameters were large and rapidly varying deep in the past, but such a specialised model is likely to require a severe rethinking of the connection between the primordial initial conditions and the late universe and an implementation on a case-by-case basis. We have also implemented isocurvature modes for the case of the external-field attractor only.

Note that, in principle, it is possible that the choice of the modified-gravity model parameters is such that in the limit $k\tau\rightarrow 0$, terms higher-order in $k\tau$ do not become subdominant compared to lower-order ones (most obviously, this happens in the case when the sound speed grows very rapidly to the past during radiation domination, but a large mass for the scalar in the past is technically similar). In such a case, a different method must be used to search for initial conditions: the mode cannot be initialised outside the scalar's mass horizon. Instead, we can use a quasi-static approximation scheme as a suitable initial condition. This has been implemented and is discussed in Section~\ref{sec:QS}.

\section{Quasi-Static Approximation Scheme \label{sec:QS}}

The key difference between the original CLASS code and \hiclass is the introduction of a new scalar degree of freedom (d.o.f.) that acts as a source for DE and/or modifies the laws of gravity at cosmological scales. In general, its full linear dynamics has to be taken into account at all times and scales of interest in order to provide accurate results. Then, it seems a bit contradictory 
implement a Quasi-Static (QS) approximation in the code. However, QS sometimes is the only method to properly evolve such a system, either because \hiclass would be too slow as a result of a highly oscillatory nature of the solution for the scalar field, or it may not be able to carry out the computation at all. 

To appreciate the usefulness of the QS approximation, it is instructive to look at the equation governing the dynamics of the new d.o.f., which in synchronous gauge schematically reads 
\begin{align}\label{eq:dynamical}
 &V_X^{\prime\prime} + aH f\left(\tau\right) V_X^{\prime} + a^2H^2 \mu\left(\tau,\,k\right)^2 V_X = S\left(\tau,\,k,\,\eta,\,\delta\rho,\,\delta p\right)\,.
\end{align}
This equation has been obtained using the Einstein equations to diagonalize the scalar field equation. $V_X\equiv a\delta\phi/\phi^\prime$
is the extra d.o.f., $f\left(\tau\right)$ a dimensionless friction term, $\mu\left(\tau,\,k\right)$ a scale-dependent mass term and $S$ is a source containing matter perturbations and the residual metric perturbation $\eta$. The full system of equations that \hiclass solves is given by the matter equations, Eq.~(\ref{eq:dynamical}) and the Einstein equations (which depend only on $V_X$ and $V_X^\prime$ and not on $V_X^{\prime\prime}$). The homogeneous part of Eq.~(\ref{eq:dynamical}) has oscillatory solutions with decaying amplitude due to the friction term and frequency determined by the mass term $\mu$. Models with large mass have very high oscillation frequencies, and consequently Eq.~(\ref{eq:dynamical}) has to be integrated with very small step sizes. This can cause the code to slow down and potentially to crash when the integration step size becomes smaller than  machine precision. On the other hand, as we shall see in the next section, these models have a natural and safe QS limit which allows us to simplify Eq.~(\ref{eq:dynamical}) and obtain accurate results.

Before moving forward to discuss the regime of validity of the QS approximation and our implementation in the code, it is important to stress here that the QS approximation in \hiclass is considered model-by-model, at specific scales and only during the necessary range of time. This is to say that, when running a particular model, QS can be switched on/off as many times as needed in order to be conservative and use it only when strictly necessary.\footnote{Strictly speaking, \hiclass can switch the QS approximation 6 times for each scale. This is enough for most realistic models, but it can be easily generalized if necessary. In the default implementation, if a model switches more than 6 times, the fully dynamical evolution is considered after the $6^{\rm th}$ time.}
This is in contrast with the other approximations used in CLASS, which only need to be switched on/off once per scale, see Ref.~\cite{Blas:2011rf} for details.

\subsection{Validity of the Quasi-Static approximation}\label{sec:validityQS}

A simplistic, but effective, definition of the QS approximation could be: space derivatives are dominant w.r.t.\ time derivatives on sub-horizon scales. This allows one to neglect time derivatives of the perturbations and convert a system of differential equations into a system of algebraic equations.\footnote{Clearly here we are talking about time derivatives of the additional scalar field and of constraints, such as the metric perturbations. Neglecting time derivatives of other d.o.f., such as matter perturbation, would kill completely the dynamics of our universe.} Usually, QS is considered on sub-horizon scales (i.e.\ when $k^2\gg a^2 H^2$) and assuming that the only time scale of the universe is the Hubble rate (i.e.\ $\Phi^{\prime\prime}\sim a^2 H^2 \Phi$, for any perturbation $\Phi$). We showed previously that in the idealised case of dust plus scalar field, the QS approximation is actually valid only inside the \emph{sound} horizon of the scalar d.o.f.\ ($c_\text{s}^2 k^2\gg a^2 H^2$) \cite{Sawicki:2015zya}. In order to take a step further and achieve a more rigorous definition of QS useful for the algorithm in \hiclassx, it is useful to generalise this discussion.

The idea of the QS approximation is that there is a degree of freedom which can react on a timescale much faster than any other in the system. Thus its dynamics can be neglected and its equation of motion can be approximated as a constraint (effective infinite speed of propagation). Determining which degree of freedom this is would require that one solves all the constraints, diagonalises the system of perturbation equations into normal modes and then determining the relative natural timescales. Such a statement would be gauge invariant. However, this is impossible not least since the universe is evolving the system of perturbation equations is changing as a function of time: the normal modes at one time are not the same as they are later.

Thus to remain practical, we make a particular choice of what we define as the QS approximation, making it in the synchronous gauge in which \hiclass is formulated. As we will show, in the range of scales where the QS limit is valid, we recover the true dynamics of the full system, and therefore also the usual Newtonian gauge QS approximation whenever it is valid:	
\begin{itemize}
 \item We define QS as neglecting the terms $V_X^{\prime\prime}$ and $V_X^\prime$ in Eq.~(\ref{eq:dynamical}), when they can be considered much smaller than the term proportional to $V_X$, because of the magnitude of $\mu^2$. On the other hand, there is no need (and it would be incorrect) to neglect terms proportional to $V_X^\prime$ (or time derivatives of other metric perturbations) in the Einstein equations, since such terms are typically not suppressed by a small coefficient. Even in the QS regime, $V_X$ is time dependent, and clearly its time derivative is non-zero; 
 \item QS has to do with the dynamics of a d.o.f. being approximated, not generically with time derivatives. The QS limit is reached when the amplitude of the homogeneous solutions of Eq.~(\ref{eq:dynamical}) is much smaller than the amplitude of the particular solution driven by the source term;
 \item QS is a mathematical statement about the solutions of a differential equation. This is to say that QS has to be discussed in the particular gauge used. Eq.~(\ref{eq:dynamical}) would be different in a different gauge, and so the physical meaning of the perturbations changes. \hiclass uses the synchronous gauge and here we discuss the synchronous gauge QS;
 \item If $S$, $\mu$ and $f$ are evolving smoothly, there is not a large hierarchy (i.e.\ there are not rapid oscillations in the background and in the perturbations due to radiation pressure) then the only time-scale of the universe is the Hubble rate, which means that $d/d\tau \sim \mathcal{O}(aH)$, e.g.\ $V_X^{\prime\prime} \sim \mathcal{O}(a^2 H^2 V_X)$. Thus, for $k^2\gg a^2 H^2$ the QS approximation reduces to a sub-horizon approximation.
\end{itemize}
Given these considerations, we are now able to show the system of equations that \hiclass is going to solve in the QS regime. On top of the usual matter and Einstein equations (considered without any approximation), we replace Eq.~(\ref{eq:dynamical}) with
\begin{align}
 &V_X = \frac{S\left(\tau,\,k,\,\eta,\,\delta\rho,\,\delta p\right)}{a^2H^2 \mu\left(\tau,\,k\right)^2}\,,\label{eq:qs_vx}\\
 &V_X^\prime = \frac{d}{d\tau}\left[\frac{S\left(\tau,\,k,\,\eta,\,\delta\rho,\,\delta p\right)}{a^2H^2 \mu\left(\tau,\,k\right)^2}\right]\,.\label{eq:qs_vxp}
\end{align}
As already discussed, Eq.~(\ref{eq:qs_vxp}) is necessary to feed the $V_X^\prime$ terms into the Einstein equations, and it is obtained by taking the analytic time derivative of Eq.~(\ref{eq:qs_vx}).

Now that we have the system of equations required to be solved in the QS regime, we are ready to discuss four situations where the QS approximation potentially fails:
\begin{enumerate}[(i)]
 \item when the coefficients of the perturbations ($\OmDE, \alpha_i$) are varying rapidly. In this regime there are clearly additional time-scales in our universe on top of $H$. It is then important to keep time derivatives in order to follow the proper dynamics of the universe;
 \item during the radiation era when radiation pressure can make the source $S$ oscillate. As in the previous point, when $k$-modes are entering the sound horizon of radiation, new time-scales are introduced, and we have to properly follow their dynamics;
 \item when $\mu(\tau,k)^2<0$, the solution to the homogeneous equation is tachyonic and does not decay away. In fact, we must ensure that the source $S$ grows relative to the solution being neglected, which is a slightly more complex condition \cite{Sawicki:2015zya}. As an example, it is instructive to think about standard DM perturbations. DM has a small tachyonic mass term (caused by zero sound speed), and clearly it is not possible to apply QS to this fluid;
 	
 \item when $\mu\left(\tau,\,k\right)^2\lesssim 1$. If the mass term is not sufficiently large, it can not be considered the leading contribution on the left hand-side of Eq.~(\ref{eq:dynamical}). 
\end{enumerate}
Note that we have not diagonalised Eq.~\eqref{eq:dynamical}: the right-hand side contains $\eta$. Solving for $\eta$ using Einstein equations bring a dependence of $h'$ which cannot be eliminated without taking derivatives of Einstein equations, which cannot be sensibly implemented within the structure of \hiclassx. Moreover, as discussed earlier, $V_X$ is in any case only a part of the actual combination of variables forming the normal mode.

We thus in principle make an error in e.g.\ the definition of the mass, but given sufficiently conservative criteria for the thresholds over which the QS approximation can be applied, this is not going to be a relevant source of error.

Thus our QS approximation refers purely to the solution for the value of the scalar-field perturbation $V_X$. We should stress here that even if this solution were grossly incorrect, it does not necessarily mean that observables are affected. For that to happen, the scalar field must also gravitate, so that any error is translated to the gravitational field and therefore the matter species. The effect on the gravitational field is mostly communicated through coefficients involving the $\alpha$-functions, so only when those are non-negligible it is necessary to worry about the correct solution. This is similar to the situation for setting the correct initial conditions, described in Section~\ref{sec:ICs}, where, provided that a model is well-behaved, an error in the very early Universe does not significantly change the final observables. Thus, in many cases it may not be material to worry about the correct solution  for $V_X$ at high redshifts. Nonetheless, it definitely is important to obtain the correct approximate solution in the late universe, and it may be important to obtain at early times in the case of early modified gravity.

\subsection{Conditions for Quasi-Staticity}

In CLASS, and consequently in \hiclassx, the approximation schemes are evaluated for each mode independently prior to evolving the perturbations. The strategy then is to formulate and implement conditions to use an approximation based only on the particular $k$-mode, combinations of background functions and on triggers that are user defined, but not on perturbation variables or actual solutions for them. The goal of this Section is to illustrate how we implemented the conditions discussed in Section~\ref{sec:validityQS} (for a step-by-step description of the alghorithm used see Section~\ref{sec:appendix_triggers}). We define the following precision parameters:

\begin{itemize}

\item \emph{Onset of forced dynamics ($z_{\rm FD}$):} Observables are most sensitive to deviations from GR at late times, at least for typical Dark Energy models.
$z_{\rm FD}$ allows the user to prevent the use of the QS approximation at late times $z<z_{\rm FD}$, in order not to miss any interesting dynamics associated to DE.

\item \emph{Mass trigger ($T_\mu$):} In order to ensure that the effective mass $\mu$ of the scalar field is sufficiently large we impose that
\begin{align}\label{eq:cond_mass}
 &\mu^2>T_\mu^2\,\gg 1\,,
\end{align}
where $T_\mu$ is a user defined trigger (we will discuss typical values for all the triggers in the next section). This allows for a hierarchy to be introduced in the equation of motion \eqref{eq:dynamical}.

\item \emph{Radiation trigger ($T_r$):}
Inside the radiation sound horizon, radiation pressure supports the relativistic species against collapse, leading to oscillations and damping of the gravitational potential. These sound waves gravitate and interact with the Horndeski scalar field through the source term $S$ in Eq.~\eqref{eq:dynamical}. We thus require that
\begin{align}\label{eq:cond_rad}
 &\mu_r^2\equiv \frac{\mu^2}{\Omega_r^2}\times\left(\frac{aH}{c_{r}k}\right)^2>T_r^2\,,
\end{align}
where $T_r$ is the radiation trigger and $c_{r}=1/\sqrt{3}$ is the radiation sound speed. This condition ensures that radiation is either gravitationally subdominant ($\Omega_r\ll1$), or, that the mass $\mu$ is large enough to compensate for the oscillations inside the radiation sound horizon. Note that $\mu^2\propto k^2$ at small scales, so the left hand side approaches scale independence for large enough $k$.

\item \emph{Decay factor ($\epsilon_{\rm decay}$):}
After the matter and radiation triggers are satisfied, we wait long enough for the full solution to approach the QS one, so their relative difference is at most a small factor $\epsilon_{\rm decay}$.%
\footnote{Eqs.~(\ref{eq:cond_mass}) and (\ref{eq:cond_rad}) form a set of necessary but not sufficient conditions for the validity of QS. To understand why, suppose that in our system a mode $k_e$ enters the mass horizon at some time $\tau_e$, i.e.\ $\mu\left(\tau>\tau_e,\,k_e\right)^2> 1$. This mode at $\tau>\tau_e$ oscillates with a -- at least in principle -- decreasing amplitude, towards the QS solution ($V_X^{QS}$). Ideally the code should switch to the QS approximation only once the amplitude of the oscillations is much smaller than the amplitude of the QS solution. Unfortunately, this moment can not be predicted in full generality without solving for the perturbations. We can nonetheless estimate when this happens with a few assumptions.}
The decay rate of the full, oscillating solution will depend on the friction term $f$ and the time variation of the effective mass $\mu^\prime$. On the assumption that $f$ be constant and $\mu$ be described by a power law, Eq.~\eqref{eq:dynamical} is a Bessel equation, with the amplitude of the deviation from the QS solution
described by
\begin{align}
 &\frac{V_X-V_X^{QS}}{V_X^{QS}}\propto a^{-s}\,,
\end{align}
where $s$ is the slope of the oscillations given by
\begin{align}
 &s = -\frac{1}{4}\left[1 - 2f - \frac{\left(\mu^2\right)^\prime}{\mathcal{H}\mu^2} + 3\frac{p + \mathcal{P}}{\rho + \mathcal{E}}\right]\,.
\end{align}
Note that oscillations decay toward the QS solution if $s>0$. Then, if Eqs.~(\ref{eq:cond_mass}) and (\ref{eq:cond_rad}) are first satisfied at time $\tau_{\rm ini}$, we check for an additional condition before switching on QS, namely
\begin{align}\label{eq:decay}
 &\left.\frac{V_X-V_X^{QS}}{V_X^{QS}}\right|_{\tau_{\rm fin}} < \left.\frac{V_X-V_X^{QS}}{V_X^{QS}}\right|_{\tau_{\rm ini}} \epsilon_\text{decay}\,,
\end{align}
where $\tau_{\rm fin}$ is the time at which the amplitude of the oscillations has decayed by a factor $\epsilon_\text{decay}$ (specified by the user). Eq.~(\ref{eq:decay}) can be rewritten as
\begin{align}\label{eq:cond_slope}
 &a_{\rm fin} > \epsilon_\text{decay}^{-1/s} a_{\rm ini}\,.
\end{align}
This is not an exact solution in general, but for sufficiently slowly varying $f$ and $\mu'/\mathcal{H}\mu$, it is usually good enough. The most tricky point of this approach is that we can estimate how much the oscillations decayed after a certain amount of time, but we cannot \emph{a piori} know what the amplitude of the oscillations was when the mode $k_e$ entered the mass horizon. In our investigations, we found that $\left(V_X-V_X^{QS}\right)/V_X^{QS}\simeq \mathcal{O}(1)$ at the time of mass-horizon crossing.

\end{itemize}

\begin{figure}
	\centering
	\includegraphics[width = 0.55\textwidth]{{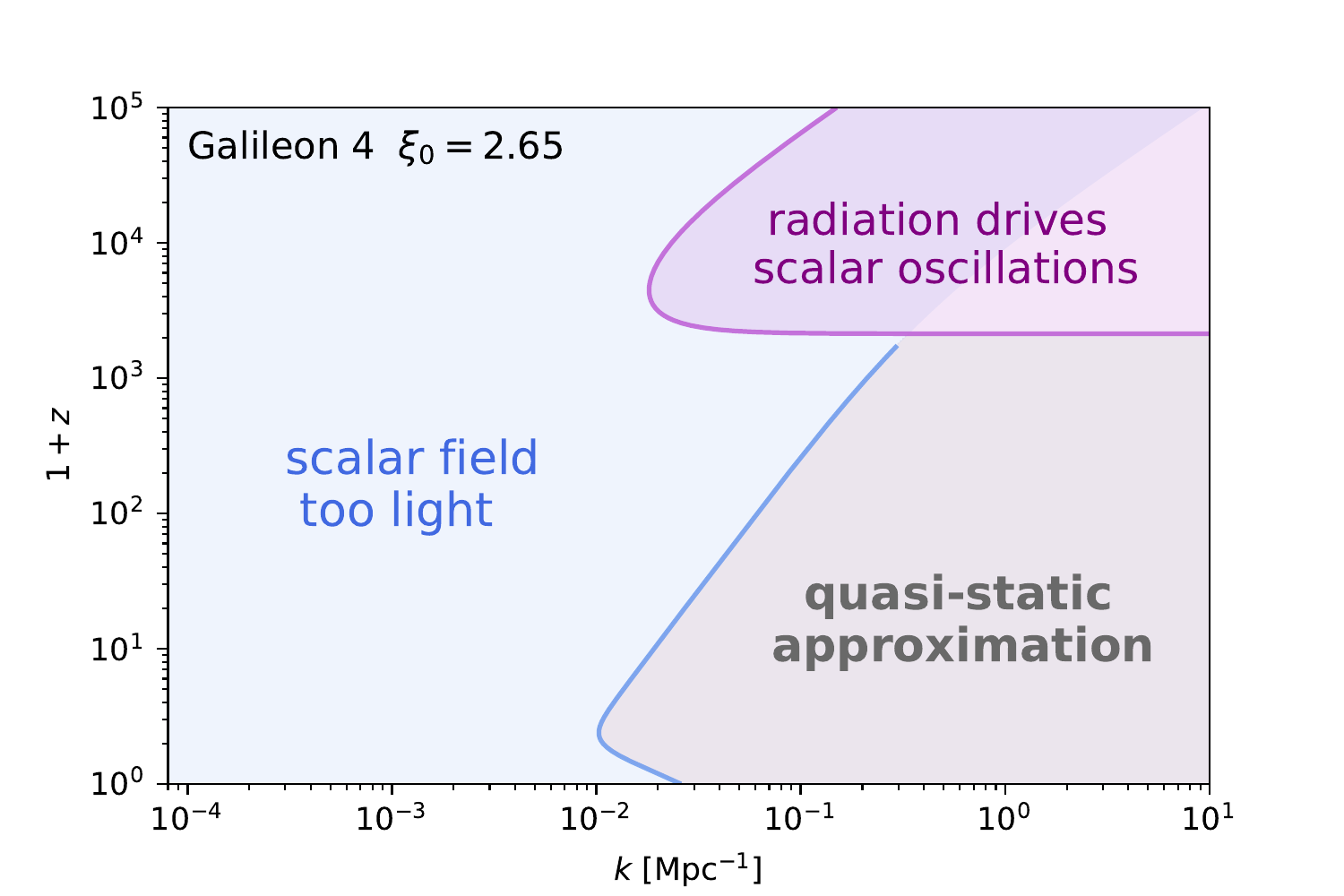}}
	\includegraphics[width = 0.49\textwidth]{{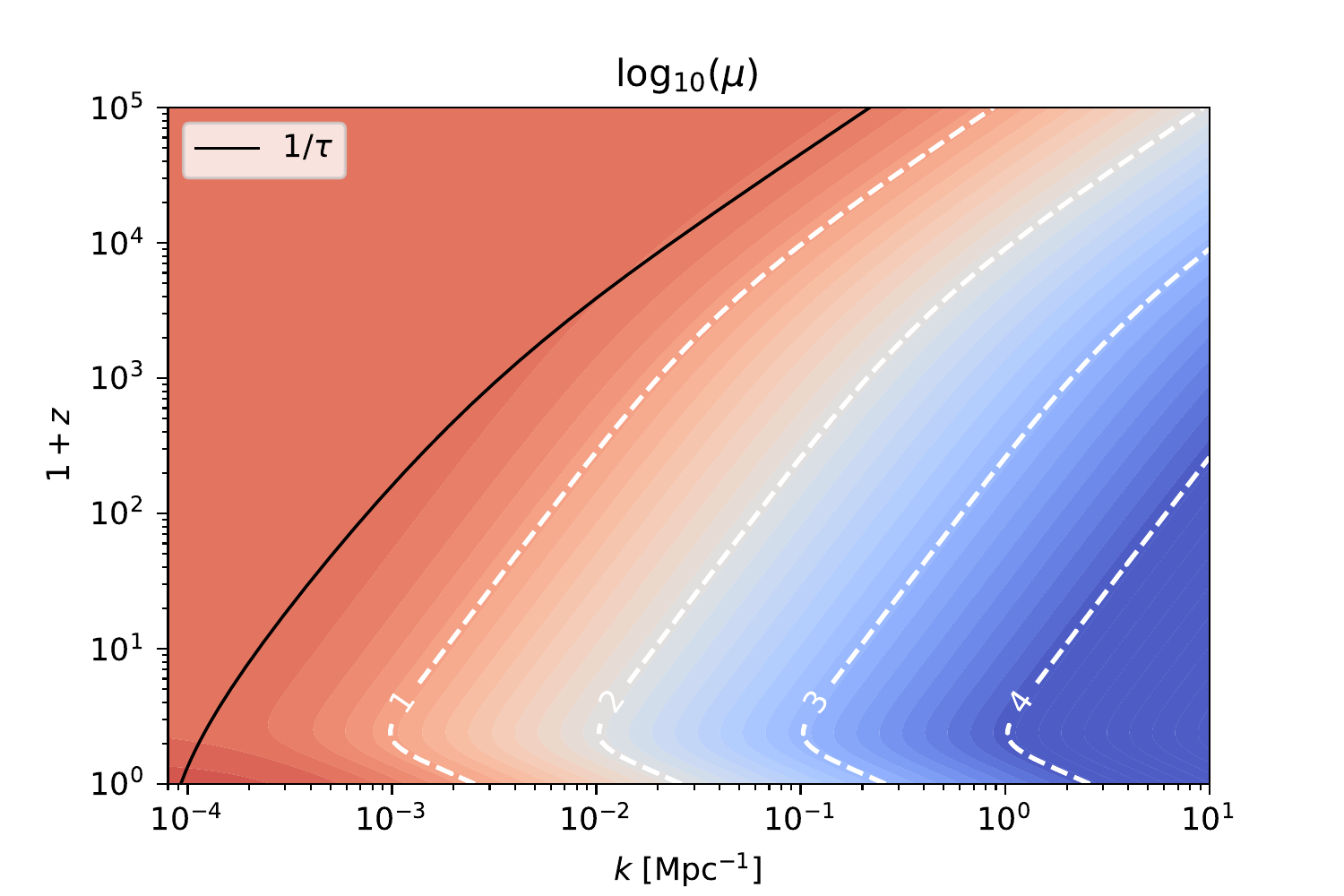}}
	\includegraphics[width = 0.49\textwidth]{{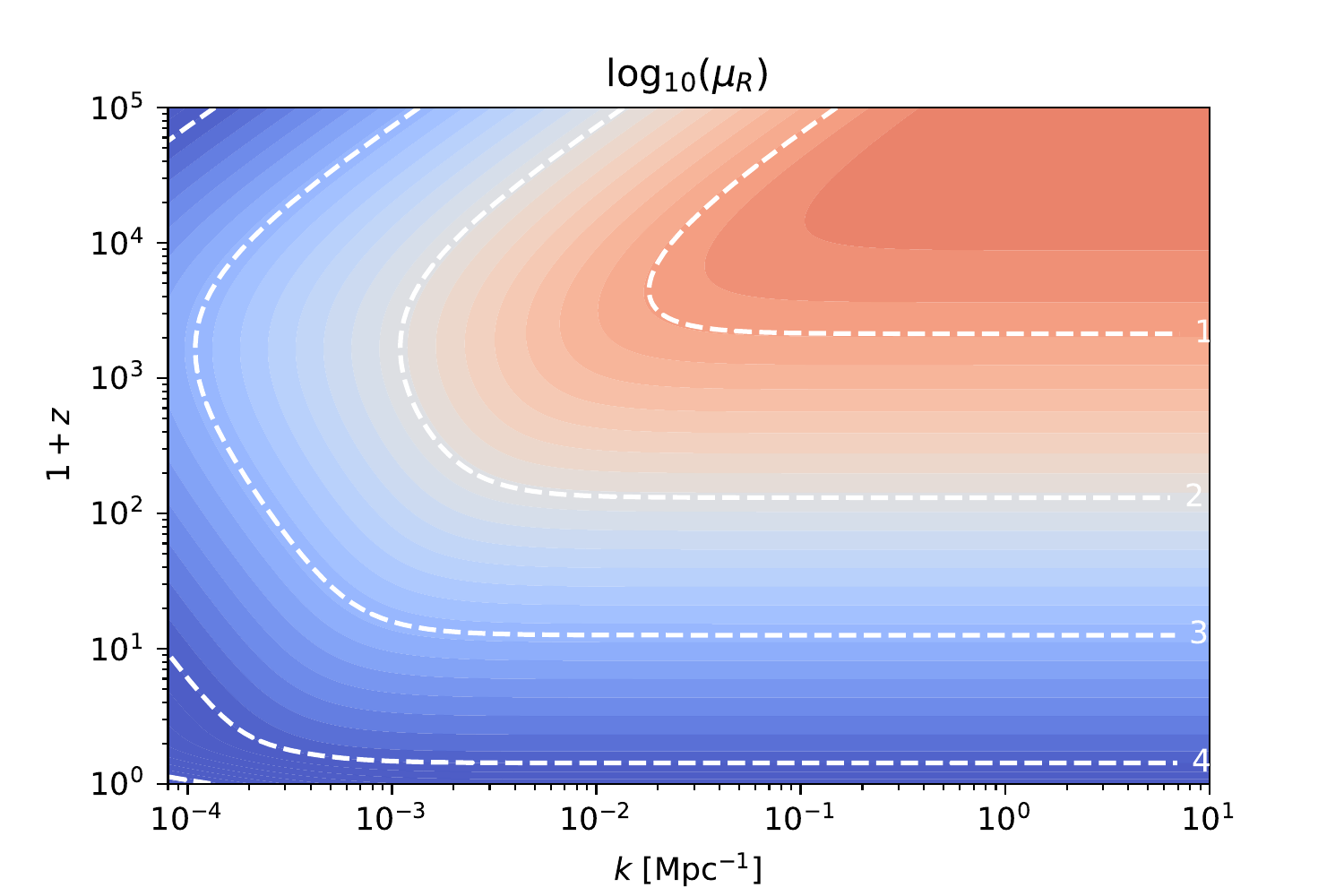}}
	\caption{Conditions for the QS approximation (top panel), as a function of the physical mode $k$ and redshift $1+z$ for an example Galileon model (cf. Sec. \ref{sec:galileon_example}).
	The regions are defined by comparing the triggering functions for mass (bottom left panel) and radiation (bottom right panel)
	to the choice of precision parameters, in this case $T_\mu = 100,\, T_r = 10$. Black line represents the cosmological horizon.
	Note that we are not considering the decay factor $\epsilon_{\rm decay}$ or onset of forced dynamics $z_{\rm FD}$, which would further reduce the region in which the quasi-static evolution is considered valid.
	The performance of the quasi-static approximation for this model is shown in Fig. \ref{fig:qs_gal_example}. 
		\label{fig:qs_trigger_example}}
\end{figure}

In Figure~\ref{fig:qs_trigger_example} we show epochs and scales on which the different conditions apply. These regions follow from the time and scale dependence of the mass, Eq.~(\ref{eq:cond_mass}), and radiation, Eq.~(\ref{eq:cond_rad}), triggering functions for a particular model. As naively assumed, the effective mass $\mu$ is large at small scales, in principle admitting a QS approximation. 
One should note that all modes at early times are deeply superhorizon and therefore the QS approximation at early times typically is poor. This nonetheless may not translate to a meaningful error, since DE is subdominant and does not significantly affect the gravitational potentials. On the other hand, radiation oscillations have no effect on large scales, but subhorizon, especially at early times, they drive the oscillations of the gravitational potential invalidating the QS approximation also at small scales and high redshift. This behaviour is typical but not universal: e.g.\ models with low kineticity lead to large values of the mass $\mu$ on all scales at prior to acceleration era.

\subsection{Implementation and Performance of the Quasi-static Approximation}\label{sec:appendix_triggers}

For each run of \hiclassx, the user can choose between three different methods to solve for the scalar field:
\begin{itemize}
 \item Fully Dynamic (\texttt{FFD}): force the fully dynamic evolution of the perturbations at all times;
 \item Quasi Static (\texttt{FQS}): force the quasi-static evolution of the perturbations at all times;
 \item Automatic (\texttt{AUTO}): use the approximation scheme implemented (see following description).
\end{itemize}
In \texttt{AUTO} mode, the QS approximation scheme performs the following steps for each $k$-mode:
\begin{enumerate}
 \item All the relevant quantities are sampled in time: time itself $\tau$, $\mu^2$ for Eq.~(\ref{eq:cond_mass}), $\mu_r^2$ for Eq.~(\ref{eq:cond_rad}) and $s$ for Eq.~(\ref{eq:cond_slope}). The time step is chosen as
  \begin{equation}
   \delta\tau=\left|\frac{2\mu^2}{\left(\mu^2\right)^\prime\sqrt{N_{min} N_{max}}}\right|\,,
  \end{equation}
  in order to be smaller when the mass is varying rapidly and larger when it is a smooth function. The time steps are also regulated by two parameters, $N_{min}$ and $N_{max}$. They set a minimum and maximum number of steps that the sampler has to take to avoid extreme cases;
 \item For each time step, if $z<z_{\rm FD}$ we flag it as \texttt{FD}, otherwise the triggers $T_\mu$ and $T_r$ are used to evaluate Eqs.~(\ref{eq:cond_mass}) and (\ref{eq:cond_rad}). If both are satisfied that time step is flagged as \texttt{QS}, otherwise \texttt{FD};
 \item The arrays are shortened in order to get just intervals where the approximation scheme is constant. The array for $\tau$ will represent the list of times at which the code switches approximation (from \texttt{FD} to \texttt{QS} or from \texttt{QS} to \texttt{FD}). The array for $s$ is calculated with the weighted average of the slopes in all the time intervals where the approximation scheme is constant.
 \item At this point we use Eq.~(\ref{eq:cond_slope}) to calculate the delay we have to apply to the switches from \texttt{FD} to \texttt{QS} in order to have the oscillation amplitude decayed enough (note that from \texttt{QS} to \texttt{FD} there is no delay). This is regulated by the trigger $s$. Suppose that at some time $\tau_i$ the system wants to switch from \texttt{FD} to \texttt{QS} and at $\tau_{i+1}$ back from \texttt{QS} to \texttt{FD}. Suppose also that according to Eq.~(\ref{eq:cond_slope}) $\tau_f$ is the final time when the oscillations have decayed enough. If $\tau_f<\tau_{i+1}$, $\tau_i$ is just replaced by $\tau_f$. If $\tau_f>\tau_{i+1}$, the whole interval between $\tau_i$ and $\tau_{i+1}$ is flagged as \texttt{FD};
 \item If the approximation scheme calculated in this way predicts more switches than implemented, i.e.~6, the first 6 switches are considered, and the rest is considered as \texttt{FD}. This is to be conservative, since we prefer to have \texttt{QS} applied at early times when DE does not gravitate, than at late times.
\end{enumerate}

There are a number of parameters that the user can modify within the QS approximation. First of all, the method has to be explicitely set to \texttt{AUTO}, otherwise the \texttt{FFD} equations are evaluated. Then, we chose default values of the parameters that should allow the users to run most of the safe models (with no pathological instabilities) but without pushing them to extreme values that would optimize \hiclass in terms of speed. The reason is that those extreme values are very model dependent, and we prefer accuracy over speed. It is the user's responsibility to tune these parameters in order to have an optimal balance between speed and accuracy for the particular model they are studying. 

It is possible that a mode is approximated as \texttt{QS} at the initial time, if the effective mass $\mu$ evolved from a very large value in the past. In theories where the actual mass of the scalar field becomes large in the past, this is not a problem, since any fifth forces are suppressed. However, a choice of parameters leading to a parametrically large sound speed for the scalar field in the past can allow for a modification of gravity even at large superhorizon scales and therefore a resulting non-conservation of the curvature perturbation. The connection to the primordial power spectrum is lost in such models.
If a mode is to be initialised in a \texttt{QS} regime, \hiclass performs a test to ensure that the configuration implied by the \texttt{QS} solution does not significantly modify the gravitational potentials and that they are still dominated by the radiation fluid. If this test is passed, the mode is evolved, otherwise the code returns an error. Moreover, if the user chooses to force \texttt{FQS} at all times, the standard stability tests for the isocurvature modes in the initial conditions are still performed, unless disabled (i.e.\ the code uses the tests related to the gravitating attractor and external field attractor initial conditions, see Section~\ref{sec:ICs}). This is to avoid the situation where forcing the QS approximation at initial time may hide the fact that the real dynamics at early times would lead to a rapidly growing solution of the homogeneous part of equation~\eqref{eq:dynamical} and therefore that nothing like the QS solution would ever be reached in reality.

\vspace{0.5cm}

\begin{figure}[t!]
	\centering
	\includegraphics[width = 0.49\textwidth]{{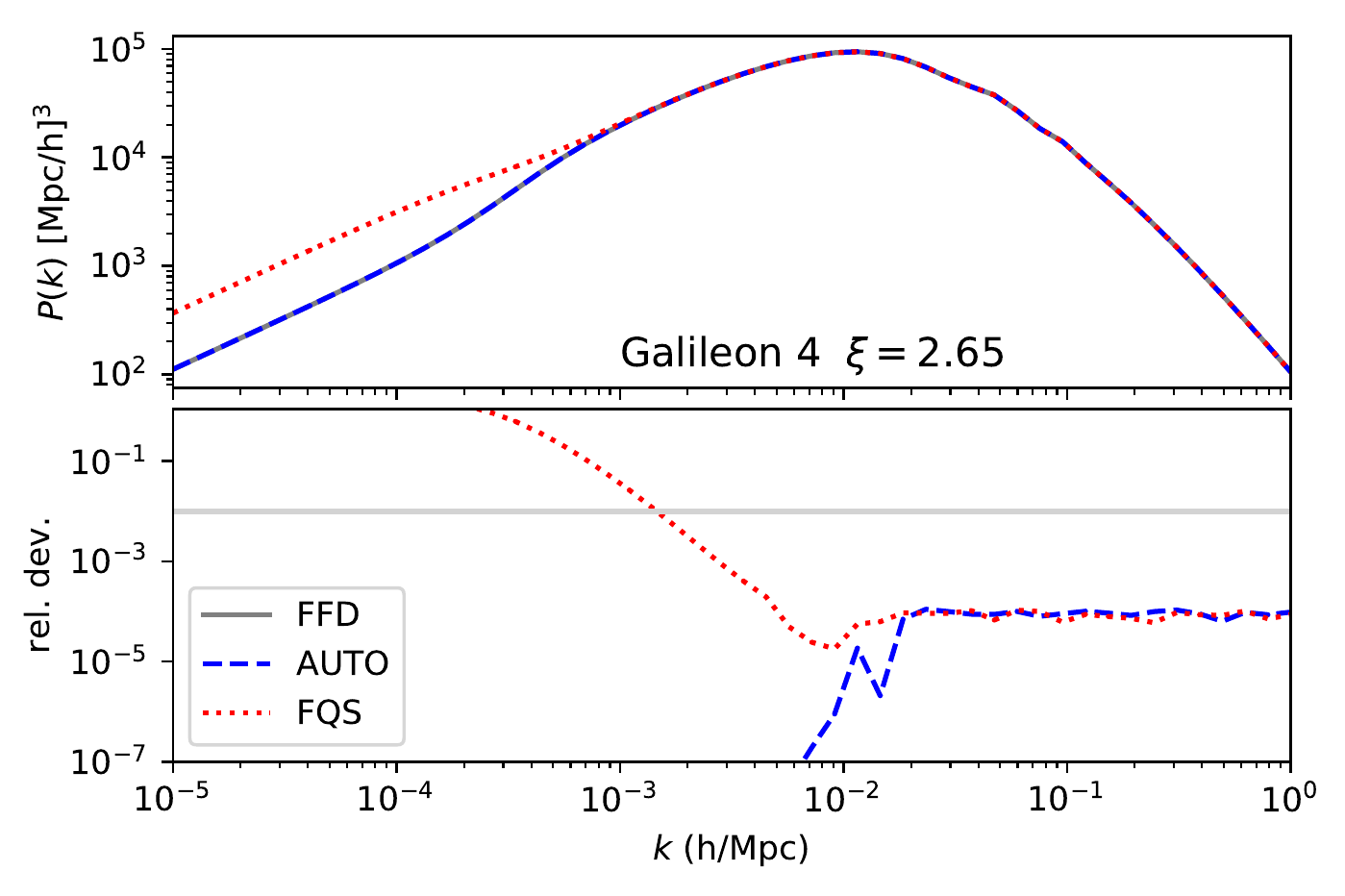}}
	\includegraphics[width = 0.49\textwidth]{{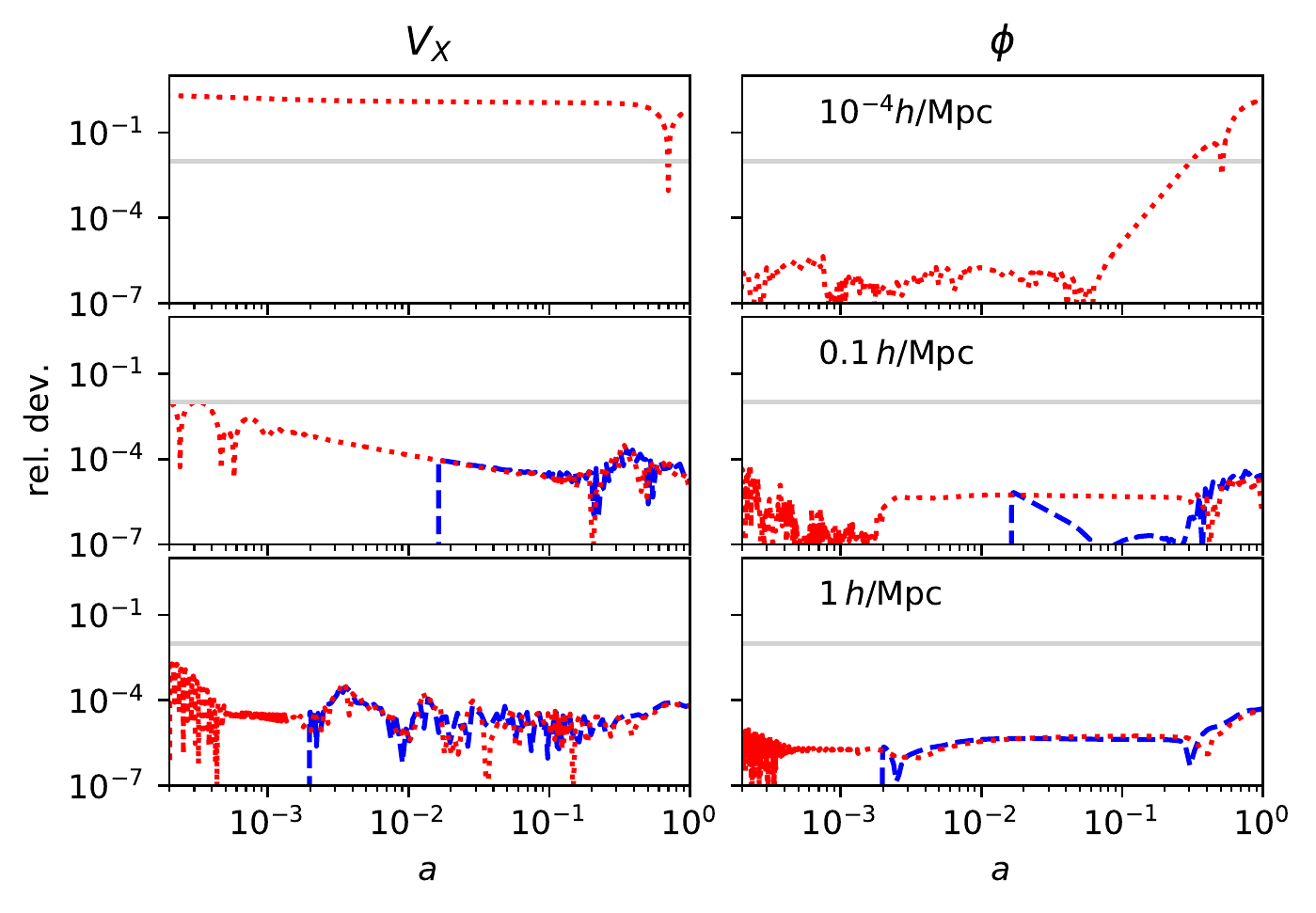}}
	\caption{QS performance for the quartic galileon model used in  Fig.~\ref{fig:qs_trigger_example} on the matter power spectrum (left panel) and the perturbation evolution (right panel). 
	Lines show the fully dynamic computation (black), full quasi-static (dotted red) and automatic (dashed blue) with $T_\mu= 100$, $T_{r}=10$, $\epsilon_\text{decay}= 0.01$, $z_{\rm FD}= 0$.
	Gray horizontal lines correspond to 1\% accuracy.}\label{fig:qs_gal_example} 
\end{figure}

We present the performance of the QS approximation and our implementation in Figure~\ref{fig:qs_gal_example}, for the Galileon model and precision parameters shown in Figure~\ref{fig:qs_trigger_example}.
Here, we plot the matter power spectrum (left panel) and the evolution of the perturbations (right panel) for a quartic Galileon model, which has a scalar field sound speed of the order of the speed of light. The results are quantitatively similar across the parameter space of viable quartic galileons.
The \texttt{AUTO} implementation yields a factor $\approx 2.6$ improvement in execution time with respect to the \texttt{FFD}, with differences in results well below the sub-percent level. The forced \texttt{FQS} computation improves the performance by a factor $\approx 6.5$, but leads to considerable deviations again on scales $k\lesssim 10^{-3}h/$Mpc and which arises from the closeness of the cosmological horizon for these modes. 
This scale can be considered as the boundary, beyond which QS approximation results cannot be trusted, at least for DE/MG models designed to drive the late-times accelerated expansion of the universe. This is in agreement with the results found in \cite{Peirone:2017ywi}.

\begin{figure}[t!]
	\centering
	\includegraphics[width = 0.49\textwidth]{{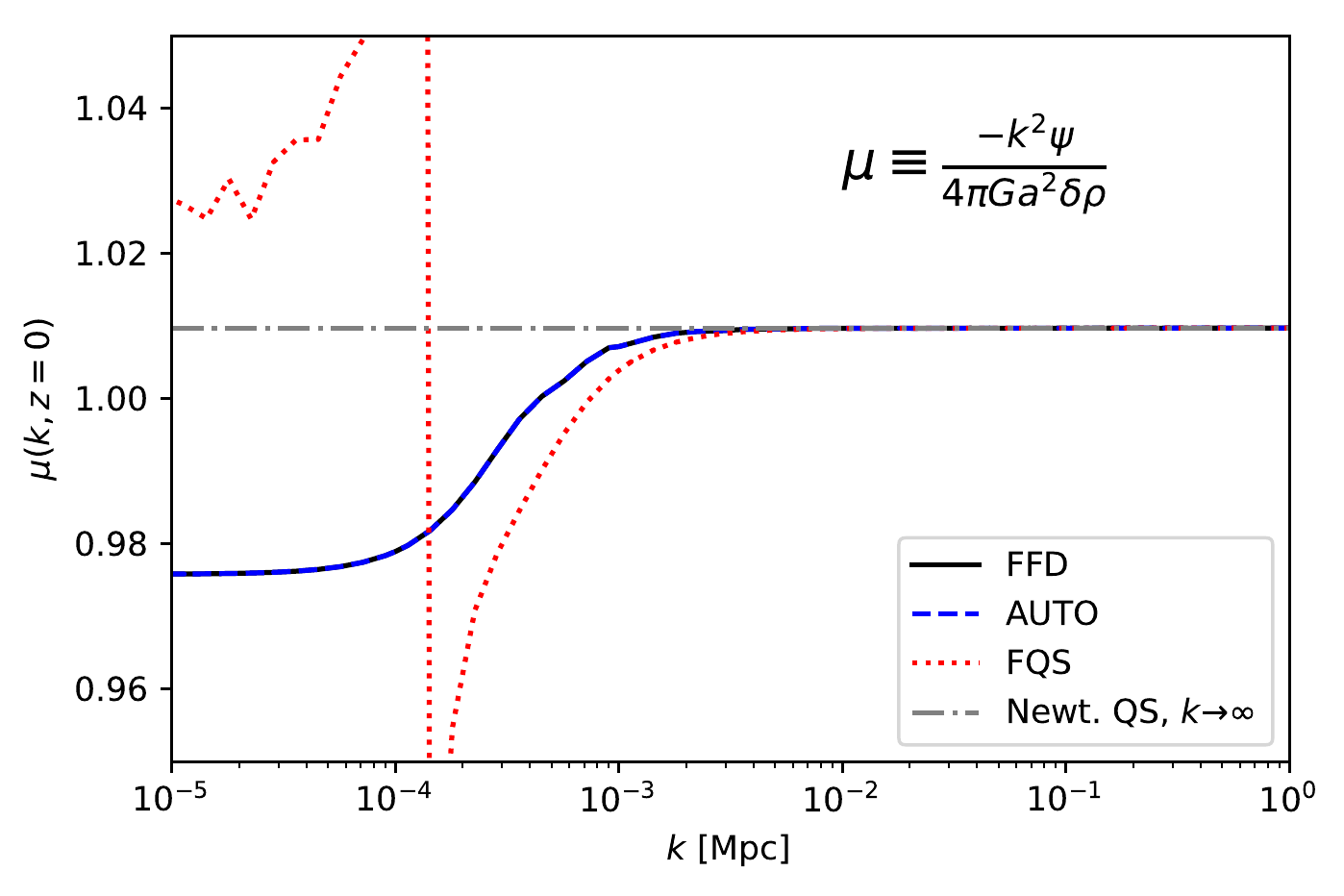}}
	\includegraphics[width = 0.49\textwidth]{{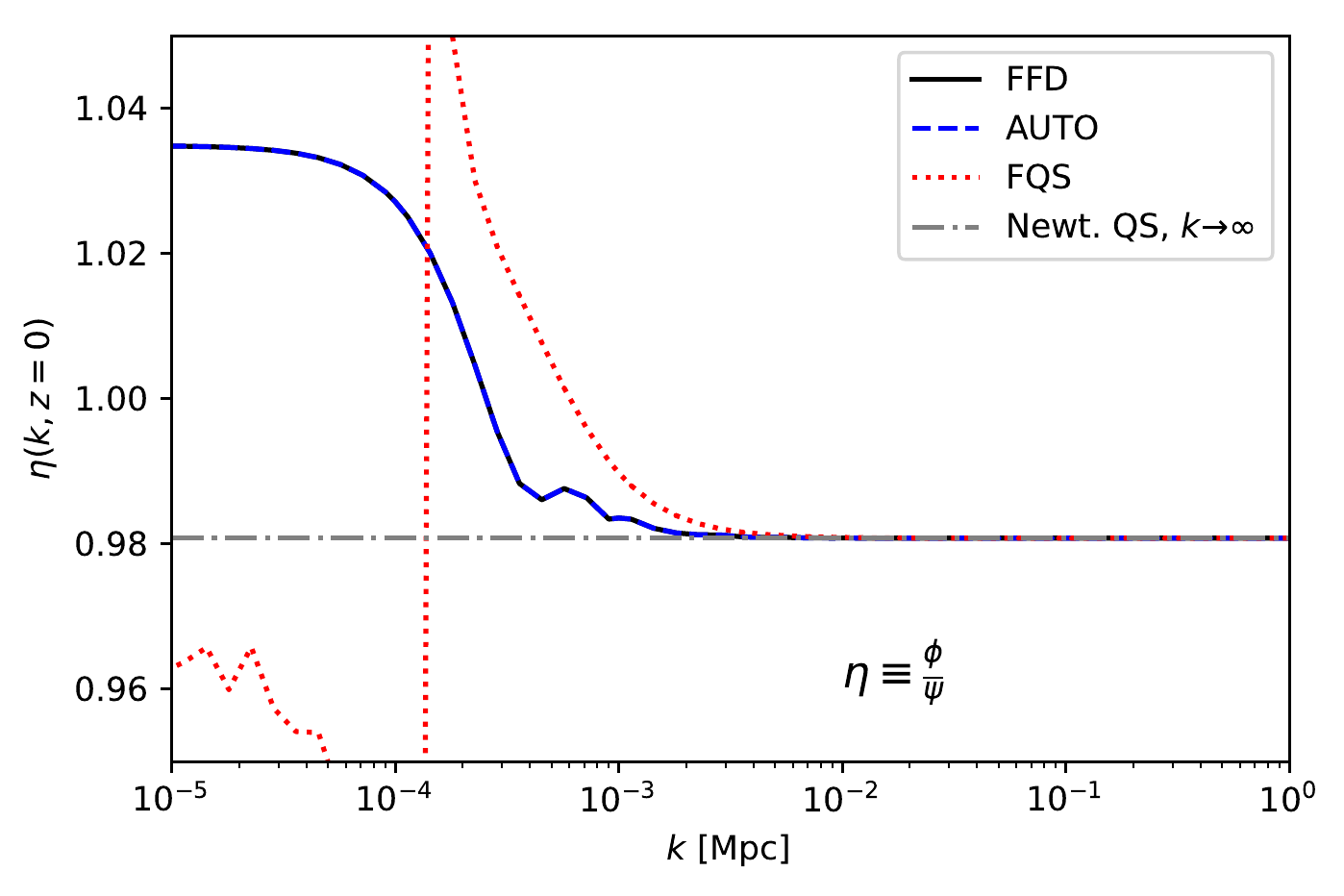}}
	\caption{QS performance for Brans-Dicke theory with $\omega_{\rm BD}=50$, $M_{*,0}^2=1$. The left and right panel show the effective Newton's constant and ratio of gravitational potentials (note the CLASS conventions on the Poisson equation) for different quasi-static approximation schemes. The fully dynamic (black) and automatic (dashed blue) agree on all scales while the quasi-static calculation (red dotted) agrees only on sufficiently small scales.
	All schemes reproduce the analytic quasi-static result computed in the conformal Newtonian gauge (gray dot-dashed) on sufficiently small scales.
	}\label{fig:qs_bd_example} 
\end{figure}

The results for Brans-Dicke theory under different approximation schemes are shown in Figure \ref{fig:qs_bd_example}. The model includes substantial deviations from GR, with $\omega_{\rm BD}=50$ and $M_{*,0}^2=1$ and scalar speed of sound $c_s^2=1$ at all epochs. The plot shows how the effective gravitational constant $\mu(k,z=0)$ and the ratio of gravitational potentials $\eta(k,z=0)$ agree with the fully-dynamic implementation on all scales for the automatic implementation ($T_\mu = T_r = 100,\, \epsilon_\text{decay}= 0.1,\, z_{\rm FD} = 0$), while the forced quasi-static approximation is a correct description for $k\gtrsim 10^{-3}\text{Mpc}^{-1}$, roughly where the scale dependence of growth sets in. We note that this model has $\mu^2<0$ around horizon scales in the matter era, hence the forced quasi-static approximation is implemented setting $T_\mu = 0,\, \epsilon_\text{decay}= 1,\,  T_r = 100,\, z_{\rm FD} = 0$. 
For this model the \texttt{AUTO} and \texttt{FQS} schemes improve the computation speed by factors $2$ and $1.25$ relative to the \texttt{FFD} calculation.
On small scales all calculations agree with the $k\to\infty$ quasi-static limit derived in the conformal Newtonian gauge \cite{Clifton:2011jh}, as expected from the gauge-independence of perturbations in this limit.

\begin{figure}
	\centering
	\includegraphics[width = 0.8\textwidth]{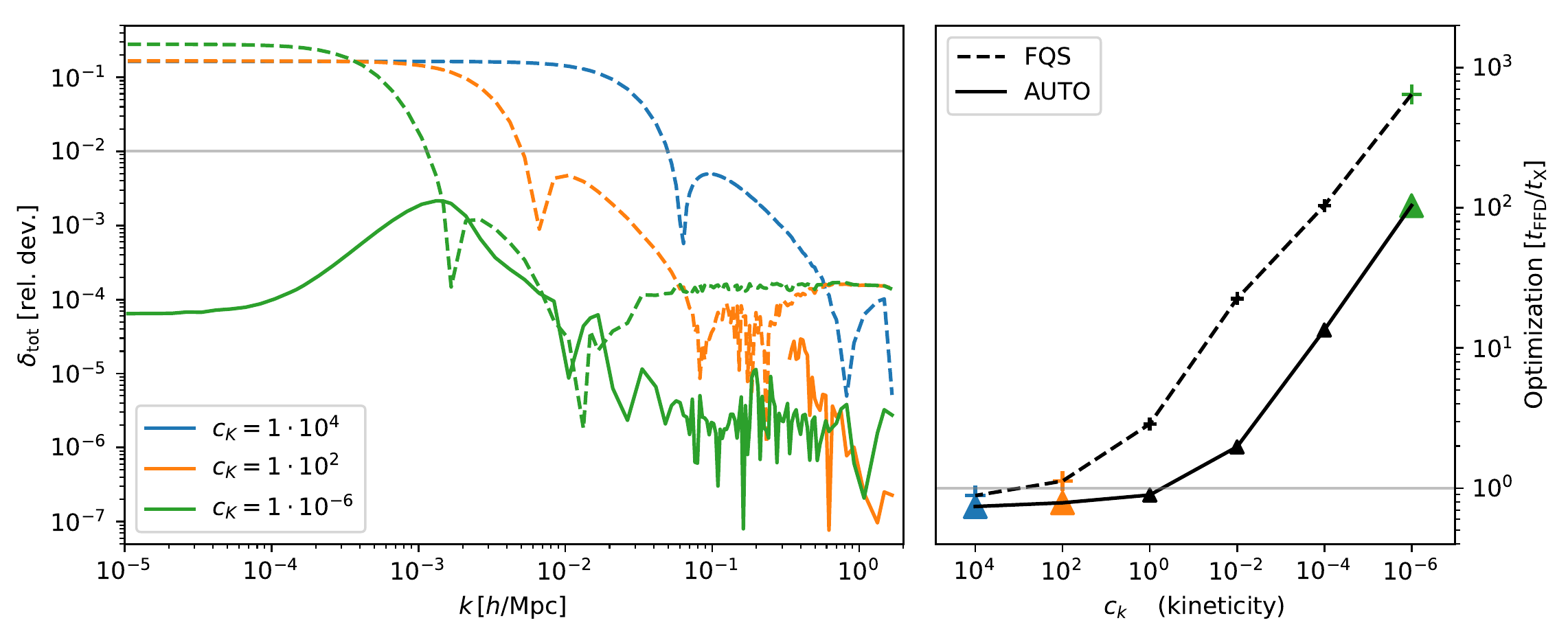}
	\caption{QS approximation performance and limit $\aK\to0$ for a model with $\alpha_i = c_i \times \Omega_{\rm DE}$, $c_B=1.5$, $c_M=c_T=0$ and a constant equation of state $w_0=-0.9$.  
	\textbf{Left panel:} Relative deviation of the matter density perturbation with respect to fully dynamic (\texttt{FFD}) computation at $z=0$. Solid lines correspond to \texttt{AUTO} ($T_\mu= T_r = 100$, $\epsilon_\text{decay}=0.01$, $z_{\rm FD} = 10$), dotted lines correspond to forcing the QS approximation at all times (\texttt{FQS}).
	Models with intermediate values of $c_K$ have relative deviations similar to $c_K=10^{-6}$ (green).
	\textbf{Right Panel:} Execution time ratio of the \texttt{FFD} computation relative to \texttt{AUTO} (solid) and  \texttt{FQS} (dashed) approximations. 
	Gray lines correspond to 1\% accuracy and equal execution time.
	\label{fig:qs_propto_performance}} 
\end{figure}

The performance of the QS approximation scheme on a substantially different model is shown in Figure~\ref{fig:qs_propto_performance}, allowing the kineticity (and hence the scalar's sound speed) to vary by several orders of magnitude. 
On one hand, models which have large sound speeds or scalar-field masses (leading to rapid oscillations) benefit from a very significant speed up --- by orders of magnitude for small kineticity --- when evolved in QS approximation. Without such an approximation these models cannot be sensibly studied. While this approximation is excellent at small scales, modes with $k<10^{-3}$~$h/$Mpc are inaccurate at more than 1\% level in the \texttt{FQS} runs, even in the case of very superluminal sound speeds. The size of the cosmological horizon sets a hard limit to the potential accuracy of the QS approximation. $k\sim10^{-3}$~$h/$Mpc should be thought of as the largest scale at which QS can be used to model modified gravity in the best-case scenario.
	
On the other hand, as the sound speed of the scalar-field fluctuations decreases, performance benefits of the QS approximation disappear. Simultaneously, the scale at which the error in the amplitude of the spectra breaches $1\%$ shrinks, to the extent that for sound speeds $c_s^2\lesssim\mathcal{O}(10^{-4})$ there is \emph{no} scale \emph{at all} where \emph{both} linear perturbation theory is valid and the prediction of the QS approximation is accurate to $<1\%$.

The lines of Figure~\ref{fig:qs_propto_performance} are obtained by varying a single parameter within one model, thus are very likely to all be explored in single MCMC. This demonstrates that the code must make a choice dynamically between applying the approximation during a part of the evolution and evolving the full system dynamically at other times. Indeed, we show that our \texttt{AUTO} implementation still benefits from the improved performance when it is appropriate, but predicts accurate spectra at all linear scales.
The above examples demonstrate why the \texttt{AUTO} algorithm must be sufficiently complex to be applicable to a variety of models.

\section{Conclusions}

\hiclassx's new features bring the code to the level of detail usually available to standard cosmology, while extending its functionality to a wide class of gravitational theories.
The covariant theory approach allows the user to work from a model's Lagrangian directly, probing the cosmological expansion and the growth of structures \emph{and their codependence} simultaneously.
Identification of the initial conditions of the field perturbations sheds light on the connection between the primordial universe and the radiation era, allowing the study of deviations from GR in the early universe.
A flexible approximation scheme improves the code performance without sacrificing accuracy, by using quasi-static evolution on scales and epochs in which it correctly describes the scalar field dynamics.

The main advances of the code and conclusions can be summarized as follows:
\begin{enumerate}
 \item We have developed a method to integrate the background evolution in covariant Horndeski theories, allowing \hiclass applications beyond the effective-theory approach. 
 The equations are stable against violations of the Hamiltonian constraint resulting from numerical errors.
 \item Analysis of covariant Horndeski theories requires understanding the initial and final conditions of the scalar field. While these choices are highly model-specific, we have presented methods to analyze the dynamics of several well-known models, which can serve as a template for further implementations. 

\item We have derived consistent intial conditions for the scalar-field perturbations in two limiting cases: \textit{external-field attractor}, in which Einstein's GR is recovered at early times, and \textit{gravitating attractor}, when gravity is significantly modified at early times but in a time-independent manner.
\item We have identified the conditions under which Horndeski isocurvature modes can grow faster than the (standard) adiabatic mode. Models that violate this condition require a separate analysis and new physical principles to connect primordial power spectrum to the late-time cosmological observables.
\item The initial conditions are correctly set for adiabatic mode in all cases including early modified gravity, and for the standard matter isocurvature modes in the case of the external-field attractor.

\item We have implemented flexible \textit{quasi-static} approximation schemes that allow \hiclass to speed-up calculations by neglecting the scalar field evolution in certain epochs/scales. The validity of this approximation relies on several physical triggers, whose thresholds can be set by the user.
\item Approximating with conservative triggers can lead to $\mathcal{O}(1)$ performance boost retaining sub-percent accuracy on all scales. Aggressive triggers can improve performance further, but lose accuracy on large scales. 
\item Approximation schemes allow computations for models with low/vanishing kineticity, such as those based on $f(R)$ gravity, in which the performance improves by orders of magnitude.
\end{enumerate}

These improvements of the code are available to the community, and can be readily used to test dark energy and constrain gravity with cosmological datasets. 
\hiclass users can now easily implement and test covariant models beyond those described in section \ref{sec:background_dynamics}, exploring the rich dynamics of cosmological background expansion beyond GR (e.g.\ bouncing models, coasting, future singularities...). 
The use of covariant models facilitates the joint study of the expansion history and the evolution of perturbations beyond $\Lambda$CDM, a program that might provide new insights into solving cosmological tensions.
The careful study of initial conditions beyond GR allows new questions to be addressed regarding the connection of dark energy and the very early universe.
Finally, the inclusion of flexible approximation schemes not only increases the applicability of \hiclass to a larger set of models, but also allows users to save valuable time and computational resources when exploring the viability of any model.
We encourage the scientific community not only to use \hiclass to obtain scientific results, but also to join the effort and contribute improving and extending this publicly available software.

Future observations of the universe will provide great opportunities to test the standard cosmological model, either pointing out the need for new physics, or strengthening it by ruling out alternatives beyond $\Lambda$CDM. 
Through current and future improvements, \hiclass will become an even more versatile tool to narrow down posible explanations of cosmic acceleration and the properties of gravity on the largest scales available to observations.
It is the interface between cutting-edge theoretical ideas, flexible computational tools and extensive datasets that will lead to the great discoveries of fundamental physics in the $21^{\rm st}$ Century.

\paragraph*{Acknowledgements: }
We are very grateful to
Pedro Ferreira, Carlos Garcia-Garcia, Michael Kopp, Julien Lesgourges, Eric Linder, Johannes Noller, Louis Perenon, Leo Stein, Filippo Vernizzi for useful discussions and comments on the manuscript.
We thank Johannes Dombrowski, Carlos Garcia-Garcia, Janina Renk, Dina Traykova for testing the development version.
We are grateful to Thomas Tram for patiently answering many questions pertaining to the CLASS code.

E.B.\ is supported by ERC H2020 693024 GravityLS project, the Beecroft Trust and the Science and Technology Facilities Council (STFC). 
I.S.\ is supported by the European Structural and Investment Fund and the Czech Ministry of Education, Youth
and Sports (MŠMT) (Project CoGraDS — CZ.02.1.01/0.0/0.0/ 15\_003/0000437).
M.Z.\ is supported by the Marie Skłodowska-Curie Global Fellowship Project “NLO-CO”. This project has received funding from the European Research Council (ERC) under the European Union’s Horizon 2020 research and innovation programme (grant agreement No 693024).

\appendix

\section{Background equations} \label{sec:appendix_equations_background}

In this section we present the background equations for Horndeski models. We show three equivalent formulations for the benefit of the reader. The first reflects what is implemented in \hiclassx. The second one defines an effective density and pressure of the additional d.o.f., while the third uses the shift-current description.

\subsection{Dynamical system}

In \hiclass\, we implemented the two Friedmann equations and the scalar field equation. As described in Section \ref{sec:background_solution} we use the Friedmann constraint to regularize the solution of the equation for the evolution of $H^\prime$. The complete system used to solve for the expansion of the universe and the scalar field evolution is
\begin{align}
 &-C\equiv -E_{0}-E_{1}H+E_{2}H^{2}-E_{3}H^{3}=0\,,\\
 & \phi^{\prime\prime}=\frac{AP-FR}{BF-AM}\,,\\
  & H^{\prime}=-\frac{R+B\phi^{\prime\prime}}{A} + a\gamma(H) C\,.
\end{align}
Here $\gamma(H)$ is defined in Eq.~(\ref{eq:friedmann_friction_term}) and $E_i$, $A$, $B$, $F$, $M$, $P$, $R$ are time dependent functions defined as
\begin{align}
E_{0}= & \frac{1}{3}\left[3\rho_{\textrm{m}}-G_{2}+2X\left(G_{2X}-G_{3\phi}\right)\right]\,,\\
E_{1}= & \frac{2\phi^{\prime}}{a}\left[-G_{4\phi}+X\left(G_{3X}-2G_{4\phi X}\right)\right]\,,\\
E_{2}= & 2\left[G_{4}-X\left(4G_{4X}-3G_{5\phi}+2X\left(2G_{4XX}-G_{5\phi X}\right)\right)\right]\,,\\
E_{3}= & \frac{2\phi^{\prime}X}{3a}\left(5G_{5X}+2XG_{5XX}\right)\,,\\
A= & -\frac{2}{9}M_{*}^{2}\,,\\
B= & \frac{2H^{2}X}{9a}\left(3G_{5X}+2XG_{5XX}\right)+\frac{2}{9a}\left[-G_{4\phi}+X\left(G_{3X}-2G_{4\phi X}\right)\right]\\
 & \qquad+\frac{4H\phi^{\prime}}{9a^{2}}\left[G_{4X}-G_{5\phi}+X\left(2G_{4XX}-G_{5\phi X}\right)\right]\,,\nonumber \\
F= & \frac{6HX}{a}\left(3G_{5X}+2XG_{5XX}\right)+\frac{6}{Ha}\left(XG_{3X}-G_{4\phi}-2XG_{4\phi X}\right)\\
 & \qquad+\frac{12\phi^{\prime}}{a^{2}}\left(G_{4X}+2XG_{4XX}-G_{5\phi}-XG_{5\phi X}\right)\,,\nonumber \\
M= & \frac{2H^{2}\phi^{\prime}}{a^{3}}\left[3G_{5X}+X\left(7G_{5XX}+2XG_{5XXX}\right)\right]\\
 & \qquad+\frac{1}{Ha^{2}}\left[G_{2X}-2G_{3\phi}+2X\left(G_{2XX}-G_{3\phi X}\right)\right]\nonumber \\
 & \qquad+\frac{6\phi^{\prime}}{a^{3}}\left[G_{3X}-3G_{4\phi X}+X\left(G_{3XX}-2G_{4\phi XX}\right)\right]\nonumber \\
 & \qquad+\frac{6H}{a^{2}}\left[G_{4X}-G_{5\phi}+X\left(8G_{4XX}-5G_{5\phi X}+2X\left(2G_{4XXX}-G_{5\phi XX}\right)\right)\right]\,,\nonumber \\
P= & -2H^{3}X\left[-3G_{5X}+4X\left(2G_{5XX}+XG_{5XXX}\right)\right]\\
 & \qquad-\frac{4H^{2}\phi^{\prime}}{a}\left[-3G_{4X}+3G_{5\phi}+X\left(3G_{4XX}-4G_{5\phi X}+2X\left(3G_{4XXX}-2G_{5\phi XX}\right)\right)\right]\nonumber \\
 & \qquad-\frac{1}{H}\left[G_{2\phi}-2X\left(G_{2\phi X}-G_{3\phi\phi}\right)\right]\nonumber \\
 & \qquad-\frac{2\phi^{\prime}}{a}\left[-G_{2X}+2G_{3\phi}+X\left(G_{2XX}-4G_{3\phi X}+6G_{4\phi\phi X}\right)\right]\nonumber \\
 & \qquad-6H\left[2G_{4\phi}+X\left(-G_{3X}+G_{5\phi\phi}+2X\left(G_{3XX}-4G_{4\phi XX}+G_{5\phi\phi X}\right)\right)\right]\,,\nonumber \\
R= & -\frac{16}{9}H^{3}\phi^{\prime}X\left(2G_{5X}+XG_{5XX}\right)\\
 & \qquad+\frac{4}{9}aH^{2}\left[3G_{4}-7XG_{4X}-2X^{2}G_{4XX}+4XG_{5\phi}-2X\left(5\left(G_{4X}-G_{5\phi}\right)+2X\left(5G_{4XX}-3G_{5\phi X}\right)\right)\right]\nonumber \\
 & \qquad+\frac{1}{9}a\left[-9\rho_{\textrm{m}}+2G_{2}-3p_{\textrm{m}}-4XG_{3\phi}+8XG_{4\phi\phi}-6X\left(G_{2X}-2G_{3\phi}+2G_{4\phi\phi}\right)\right]\nonumber \\
 & \qquad-\frac{4}{9}H\phi^{\prime}\left[3XG_{3X}-4G_{4\phi}-6XG_{4\phi X}+X\left(2G_{3X}-6G_{4\phi X}+G_{5\phi\phi}\right)\right]\,.\nonumber 
\end{align}

\subsection{In terms of densities}

A more familiar description can be achieved by analogy with standard matter fluids. It is possible to define the energy density and pressure of an effective fluid that describes the evolution of the scalar field. The Friedmann equations and the conservation of the energy-momentum tensor read
\begin{align}
& H^{2}=  \rho_{\textrm{m}}+\mathcal{E}\,,\\
&\frac{2H^{\prime}}{3a}+H^{2}=  -p_{\textrm{m}}-\mathcal{P} \label{eq:app_friedmann1}\,,\\
&\mathcal{E}^{\prime}+  3aH\left(\mathcal{E}+\mathcal{P}\right)=0\,,
\end{align}
where $\mathcal{E}$ and $\mathcal{P}$ are the energy density and pressure respectively and are defined as
\begin{align}
\mathcal{E}= & \frac{2H^{3}\phi^{\prime}X}{3a}\left(5G_{5X}+2XG_{5XX}\right)+\frac{1}{3}\left[-G_{2}+2X\left(G_{2X}-G_{3\phi}\right)\right]\\
 & \qquad+\frac{2H\phi^{\prime}}{a}\left[-G_{4\phi}+X\left(G_{3X}-2G_{4\phi X}\right)\right]\nonumber \\
 & \qquad+H^{2}\left[1-2G_{4}+X\left(2\left(4G_{4X}-3G_{5\phi}\right)+4X\left(2G_{4XX}-G_{5\phi X}\right)\right)\right]\,,\nonumber \\
\mathcal{P}= & \frac{2H^{3}\phi^{\prime}X}{3a}\left(G_{5X}+2XG_{5XX}\right)\\
 & \qquad+\frac{2H^{\prime}}{3a}\left[-1+2G_{4}-2X\left(2G_{4X}-G_{5\phi}\right)-\frac{2H\phi^{\prime}X}{a}G_{5X}\right]\nonumber \\
 & \qquad+\frac{1}{3}H^{2}\left[-3\left(1-2G_{4}\right)+2X\left(-2G_{4X}-G_{5\phi}+2X\left(4G_{4XX}-3G_{5\phi X}\right)\right)\right]\nonumber \\
 & \qquad+\frac{2\phi^{\prime\prime}}{3a^{2}}\left[-H^{2}X\left(3G_{5X}+2XG_{5XX}\right)+G_{4\phi}-X\left(G_{3X}-2G_{4\phi X}\right)\right]\nonumber \\
 & \qquad-\frac{4H\phi^{\prime}\phi^{\prime\prime}}{3a^{3}}\left[G_{4X}-G_{5\phi}+X\left(2G_{4XX}-G_{5\phi X}\right)\right]\nonumber \\
 & \qquad+\frac{1}{3}\left[G_{2}-2X\left(G_{3\phi}-2G_{4\phi\phi}\right)\right]+\frac{2H\phi^{\prime}}{3a}\left[G_{4\phi}+X\left(G_{3X}-6G_{4\phi X}+2G_{5\phi\phi}\right)\right]\,.\nonumber 
\end{align}

\subsection{Shift-current}

The equation for the evolution of the scalar field can be rewritten equivalently as
\begin{align}
 & \frac{\mathcal{J}^{\prime}}{a}+3H\mathcal{J}=\mathcal{S}_{\phi}\,,
\end{align}
where $\mathcal{J}$ represents a \emph{shift charge density} describing (on a cosmological background) a Noether current associated with a shift symmetry in the scalar field. On the other hand $\mathcal{S}_{\phi}$ is a source that violates shift-symmetry. They are defined as
\begin{align} \label{eq:field_shift_charge}
\mathcal{J}= & \frac{\phi^{\prime}}{a}\left(G_{2X}-2G_{3\phi}\right)+6HX\left(G_{3X}-2G_{4\phi X}\right)\\
 & \qquad+6\frac{H^{2}\phi^{\prime}}{a}\left[G_{4X}-G_{5\phi}+X\left(2G_{4XX}-G_{5\phi X}\right)\right]+2H^{3}X\left(3G_{5X}+2XG_{5XX}\right)\,,\nonumber \\
\mathcal{S}_{\phi}= & G_{2\phi}-2XG_{3\phi\phi}-\frac{2X}{a^{2}}\left(\phi^{\prime\prime}-aH\phi^{\prime}\right)G_{3\phi X} \label{eq:field_shift_source}
\\
 & \qquad+6\left(2H^{2}+\frac{H^{\prime}}{a}\right)G_{4\phi}+\frac{6H\phi^{\prime\prime}\phi^{\prime}}{a^{3}}G_{4\phi X}+\frac{2H^{3}\phi^{\prime}X}{a}G_{5\phi X}-6H^{2}XG_{5\phi\phi}\,,\nonumber 
\end{align}
and they can be useful to have insights on the dynamics of the scalar field without numerically solving the background equations (cf. Sec.~\ref{sec:background_dynamics}).

\section{Dimensionless Lagrangian Functions for Solver}\label{sec:dimless}

In this release, \hiclass obtains the ability to directly solve for a model from the Lagrangian: the appropriate background equations of motion are computed automatically and then the resulting $\alpha$ functions on these backgrounds are derived.

In this section, we discuss how to bring the (dimensionful) Lagrangian functions defining the particular Horndeski model into their appropriate dimensionless form which can be entered into the code when adding the user's choice of model.

Internally, all the variables in the code are dimensionless. \hiclass expresses all dimensionful output using the fixed lengthscale of $L\equiv 1$~Mpc. In particular, all derivatives are understood to be taken with this lengthscale as the unit. Secondly, all the energy densities and pressures are also given in units of Mpc$^{-2}$ by absorbing a factor $M_\text{Pl}^{-2}=8\pi G_\text{N}$. As already mentioned, in the code the units are such that $M_\text{Pl}^{-2}=1$, here we restore it to make all the units explicit. In particular, the best way to derive the dimensionless form of the Lagrangian functions is to redefine the volume element in the (dimensionless) action:
\begin{equation}
	S = \int d^4x \sqrt{-g} \mathcal{L} = \int \left[ M_\text{Pl}^2 L^{-2} d^4x \right] \sqrt{-g} \frac{\mathcal{L}}{M_\text{Pl}^2L^{-2}}\,.
\end{equation}
We then have made the choice to use $\tilde\phi\equiv\phi/M_\text{Pl}$ as the dimensionless representation of the scalar field value $\phi$. This combined with the lengthscale $L$ fixes the units of the time derivative to give the dimensionless $\tilde\phi' \equiv L\phi'$ and therefore gives the only consistent representation of the dimensionless 
\begin{equation}
\tilde X \equiv X L^2/M_\text{Pl}^2\,.
\end{equation}
This implies that the appropriate dimensionless representation of the Lagrangian for a canonical massive scalar field would be
\begin{equation}
		\tilde{\mathcal{L}} \equiv \frac{\mathcal{L}}{M_\text{Pl}^2L^{-2}} =
		 \tilde{X} - \frac{1}{2}(m L)^2 \tilde\phi^2\,,
\end{equation}
where $m L$ is the mass of the scalar in units of Mpc$^{-1}$. For the purpose of \hiclass we can now self-consistently replace every $\mathcal{L}$ (or in this case, a particular choice of function $G_2(X,\phi)$) with $\tilde{\mathcal{L}}$ and its derivatives $\mathcal{L}_{,X}$ and $\mathcal{L}_{,\phi}$ with the dimensionless derivatives applied to the tilded variables $\tilde{\mathcal{L}}_{,\tilde{\phi}}$ and $\tilde{\mathcal{L}}_{,\tilde{X}}$. We stress that only this particular choice of the units of $X$ and $\phi$ makes the calculation of $X$ from the derivative of $\phi$ in the code correct.

Let us now generalise this to generic choice of the $G_i$ functions. For example, for $G_2$, one has an arbitrary function of dimension 4, which is a series with $\phi/M$ and $X/\Lambda_2^4$ as dimensionless order parameters, e.g.
\begin{equation}
		G_2(X,\phi) = \Lambda_2^4 g_2\left( \frac{X}{\Lambda_2^4},\frac{\phi}{M} \right)\,,
\end{equation}
with some scales $\Lambda_2$ and $M$ typifying the variation in $X$ and $\phi$ and $g_2$ some dimensionless function of two variables. The appropriate dimensionless version to be implemented in \hiclass would then be
\begin{equation}
	\tilde{G}_2(\tilde{X},\tilde\phi) = \frac{\Lambda_2^4}{\Mpl^2 L^{-2}} g_2 \left(\frac{\Mpl^2L^{-2}}{\Lambda_2^4} \tilde{X}, \tilde{\phi}\frac{\Mpl}{M} \right)\,. \label{eq:G2}
\end{equation}
The choice of scales $\Lambda_2$ and $M$ is of course determined by the underlying theory from which a particular model arises. But, from the form of expression \eqref{eq:G2}, it is clear that if the functional form of $G_2$ is to be relevant for dark energy, i.e.\ contribute $\mathcal{O}(H_0)$ modifications to $H$ at late times, then one would expect that the scales $M=\Mpl$ while $\Lambda_2^2 = \Mpl H_0$ (i.e.\ derivatives have typical scales $H_0$). This then simplifies the expression to $\tilde{G}_2 = (H_0L)^2 g_2\left(\tilde{X}/(H_0 L)^{2},\tilde\phi\right)$, with $H_0 L$ the value of the Hubble constant in units of Mpc$^{-1}$.

For the other functions, $G_{3,4,5}$, the appropriate expressions can be obtained by introducing another scale, $\Lambda_3$, related to the typical size of the second derivatives of the scalar, $\Box\phi$. This then leads to the following dimensionless functions,
\begin{align}
	\tilde{G}_{3}(\tilde{X},\tilde{\phi}) &\equiv \frac{\Lambda_2^4}{\Lambda_3^3\Mpl} g_3 \left(\frac{\Mpl^2L^{-2}}{\Lambda_2^4} \tilde{X}, \tilde{\phi}\frac{\Mpl}{M} \right) \,,\label{eq:Gi}\\
	\tilde{G}_4 (\tilde{X},\tilde{\phi}) &\equiv \frac{1}{2} +\frac{\Lambda_2^8}{\Lambda_3^6\Mpl^2} g_4 \left(\frac{\Mpl^2L^{-2}}{\Lambda_2^4} \tilde{X}, \tilde{\phi}\frac{\Mpl}{M} \right)\,, \notag\\
	\tilde{G}_5(\tilde{X},\tilde{\phi}) &\equiv \frac{\Lambda_2^8 }{\Mpl\Lambda_3^9 L^2} g_5\left(\frac{\Mpl^2L^{-2}}{\Lambda_2^4} \tilde{X}, \tilde{\phi}\frac{\Mpl}{M} \right) \,, \notag
\end{align}
with the $g_i$ some dimensionless functions of two variables. Again, there may be a fundamental reason for the choice of some particular $\Lambda_3$, but from the above expressions, one can see that the choice $\Lambda_3^3=\Mpl H_0^2$ gives contributions from $G_i$ functions which are relevant for dark energy when the dimensionless Taylor-expansion coefficients for the $g_i$ are of order one.

When defining a particular Lagrangian-based model in \hiclassx, one needs to enter all necessary derivatives of the functions $G_i$: these should all be understood to be the appropriate derivatives of the $\tilde{G}_i$ with respect to $\tilde{X}$ and $\tilde{\phi}$ once the $G_i$ are brought to the standard forms \eqref{eq:G2} and \eqref{eq:Gi}.

\bibliographystyle{utcaps}

\bibliography{hi_class}

\end{document}